%% file: transfer_matrix_resub.tex
\documentclass[aps, prb, reprint, showpacs]{revtex4-1}
\usepackage{amsmath, amssymb, graphicx, color, mathrsfs, hyperref, array, bbm}

\input{tm_defcommands.tex}

\renewcommand{\emptyset}{\varnothing}
\newcommand{\loopset}{Z_1}
\newcommand{\wt}[1]{\widetilde{#1}}

\newcommand{\vb}{\mathbf{b}}
\newcommand{\vs}{\boldsymbol{\sigma}}
\newcommand{\schur}{\mathcal{S}}

\begin{document}
\title{Of Bulk and Boundaries: \\ Generalized Transfer Matrices for Tight-Binding Models}

\author{Vatsal Dwivedi}
\author{Victor Chua}
\affiliation{Department of Physics and Institute for Condensed Matter Theory, \\ University of Illinois at Urbana-Champaign, IL 61801, USA}

\begin{abstract}
We construct a generalized transfer matrix corresponding to noninteracting tight-binding lattice models, which can subsequently be used to compute the bulk bands as well as the edge states. Crucially, our formalism works even in cases where the hopping matrix is non-invertible. Following Hatsugai [PRL 71, 3697 (1993)], we explicitly construct the energy Riemann surfaces associated with the band structure for a specific class of systems which includes systems like Chern insulator, Dirac semimetal and graphene. The edge states can then be interpreted as non-contractible loops, with the winding number equal to the bulk Chern number. For these systems, the transfer matrix is symplectic, and hence we also describe the windings associated with the edge states on $\Sp(2, \real)$ and interpret the corresponding winding number as a Maslov index. 
\end{abstract}

\maketitle

\section{Introduction}
The topological phases of matter have been a subject of considerable interest\cite{bernevig-hughes_book,shen_book, hasan-kane_TI}. In their simplest form, for noninteracting periodic systems, they are characterized by a topological invariant on the Brillouin zone. For instance, the two dimensional topological states may have a nontrivial Chern number of the $\U(1)$ bundle of the phase of the Bloch eigenstates over the Brillouin zone\cite{tknn}.                                                                                                                                                                                                                                                                                                                                                                                                                                                                                                                                                                                                                                                                   

One of the remarkable features of the topological phases is the existence of boundary states -- modes localized on the surface/edge at energies that reside in the bulk band gap, which cannot be gapped out by local perturbations. For topological phases that are insulating in the bulk, the existence of these modes is related to the fact that the topological phase is topologically different from vacuum, a trivial insulator, and hence they cannot be connected without closing the gap.

The topological phases are commonly described by noninteracting lattice Hamiltonians, which can be studied in the framework of Bloch theory (also known as Floquet theory in the differential equation literature\cite{eastham_spectral,eastham_periodic_operators}). But one of the central tenets of Bloch theory is that we only consider eigenstates to the Hamiltonian that are translation invariant(up to a phase), which is desirable if the system in question is truly periodic, so that the (quasi-)momentum $\vk$ is well defined. However, the presence of an edge naturally breaks that symmetry. Can we still account for the edge modes in a Bloch-like formalism? As it turns out, the answer is yes, if we allow the quasi-momentum to be complex. 

Let $\psi_{\vn}$ be a wavefunction for lattice sites indexed by $\vn = \left( n_1, n_2, \dots n_d \right) \in \intg^d$. Then, an imaginary part of the quasi-momentum, Im$(k_i) = \kappa_i \neq 0$ leads to $\psi_\vn \sim e^{\pm n_i \kappa_i}$, a growing or decaying exponential along the direction $i$, which is not normalizable in the case of an infinite system. However, an edge state can be naturally thought of as a decaying exponential (indeed, we can calculate the \emph{penetration depth} of the edge states), which corresponds to a quasi-momentum $\vk$ with a nonzero imaginary part.

A system with boundary can be naturally described in the language of transfer matrices\cite{hatsugai_cbs, hatsugai_cbs_PRL, dh_lee_cbs, schulman_cbs, tauber-delplace_tm}. By assuming the system to be periodic in directions parallel to the edge/surface, we can reduce it to a family of quasi-1D system\cite{vasudha-roger_tm} parametrized by the transverse quasi-momentum $\vk_\perp$, for which a transfer matrix gives the wavefunction of a block in terms of previous block(s). The eigenvalues of the transfer matrix decide whether the state is periodic or decaying: an eigenvalue with magnitude unity corresponds to Bloch state, while the magnitude being less(greater) than unity corresponds to a decaying(growing) exponential as $n \to + \infty$.

Transfer matrices have been studied in diverse contexts, for instance, electronic band structure\cite{hatsugai_cbs, schulman_cbs, dh_lee_cbs, avila_edge2d, graf-agazzi_hofstadter}, disorder\cite{mackinnon1983scaling} conductivity\cite{tm_cond}, Majorana fermions\cite{suraj_quench} and wave motion in electromechanical systems\cite{stephen_tm_mech}. They have also been studied by mathematicians as monodromy matrices under the banner of Floquet theory\cite{eastham_spectral, eastham_periodic_operators, yakubovich_periodic, jared_defect_modes}. In condensed matter, they have been used to compute the $\intg_2$ invariants for time-reversal invariant systems\cite{avila_edge2d}. However, all of these constructions have been limited to very specific models and/or invertible hopping matrices. 

As we will see, the perspective of transfer matrices and a complex k-space is not a mere curiosity or a tool to simplify calculations; instead, it offers interesting insights into the geometry associated with the edge states of the system. For one, as the eigenvalue condition for the transfer matrix, expressed as the characteristic polynomial, is algebraic in energy $\ve$ and $\bze = e^{i\vk}$, one can associate algebraic curves with the characteristic polynomials. A natural thing to do is to complexify $\ve$, as the algebraic equation always has roots in the complex plane, which follows from the fundamental theorem of algebra. The characteristic polynomial defines an algebraic variety of codimension 1 on $(\ve, \bze) \in \cmplx^{d+1}$ for a $d$-dimensional system, often termed as a Bloch variety\cite{algebraic_fermi_curves}. We shall not delve much into this picture, however, algebraic curves in complex spaces can also be naturally thought of as Riemann surfaces, a perspective that turns out to be particularly useful, which we shall consider in this article. 

One of the first steps in that direction was taken by Y Hatsugai\cite{hatsugai_cbs_PRL, hatsugai_cbs}, who showed that for the Hofstadter model with flux $\phi = p/q$ per plaquette, the edge states correspond to nontrivial windings around the holes of the $\ve$-Riemann surface, which is a 2-dimensional surface of genus $q-1$, where $q$ is the periodicity of the lattice in presence of the magnetic field, or, equivalently, the number of bulk bands (hence $q-1$ being the number of band gaps). Furthermore, he proves that the winding number so associated with the edge states is equal to the bulk Chern number. The Riemann surface picture was first introduced in condensed matter physics by W Kohn\cite{kohn_cbs}, however, it has been investigated in substantial detail in mathematics literature\cite{mckean-moerbeke_hilleqn}.

Substantial progress has been made since, with regards to the bulk-boundary correspondence. In particular, using the methods of non-commutative geometry, the equality of the Chern number and the Hall conductivity at finite disorder have been rigorously addressed within the mathematical physics literature\cite{bellissard-baldes_noncomm_IQHE,baldes_disorder_IQHE}, and with generalizations to time-reversal invariant topological insulators\cite{prodan_disorder_TI}. Complementary to this are approaches based on Green's functions\cite{volovik_book, essin-gurarie_bulk_bdry}, which have also demonstrated the bulk-boundary correspondence from a field theoretic perspective. In addition, aspects of this correspondence have also been discussed from the viewpoint of quantum transport using S-matrices\cite{fulga_scattering,graf-ortelli_pumping}. 

In this article we seek to study the bulk and edge spectra of and the complex geometry associated with generic tight-binding Hamiltonians using transfer matrices, as a continuation of Hatsugai's analysis beyond the Hofstadter model. We work out a general construction of transfer matrices for quasi-1D systems, including the cases when the hopping matrices are singular. The size of the transfer matrices turns out to be twice the rank of the hopping matrix. We work out analytic computations in some detail for various systems where the hopping matrix is of rank 1, e.g, Hofstadter model, Chern insulator, Dirac semimetal and graphene, as well as for a model with the a rank 2 hopping matrix, viz, the topological crystalline insulator model proposed by Fu\cite{fu_TCI}. In all these case, we derive explicit analytic expressions for the bulk band edges as well as the edge state spectra. 

For the rank 1 systems, we also describe the topological winding associated with the edge states on the energy Riemann surface, following Hatsugai. The case of Chern insulator is particularly nice, as the energy Riemann surface turns out to be a 2-torus, and we work out Hatsugai's construction explicitly using elliptic functions. Furthermore, as the transfer matrix for these models is  $2 \times 2$ real symplectic matrix, and the associated Lie group $\Sp(2, \real)$ is homeomorphic to a solid 2-torus, we plot the transfer matrix as a function of the transverse momentum for a given edge spectrum using an explicit parametrization of $\Sp(2, \real)$. Using the fact that $\pi_1(\Sp(2, \real)) \cong \intg$, we identify the corresponding winding number as a Maslov index, which, in all the cases discussed, turns out to be equal to the bulk Chern number. 

Finally, we use our transfer matrix formalism to study Chern insulator on a rectangular geometry in presence of diagonal disorder, where we marry our general transfer matrix formalism with the conventional numerical methods\cite{kramer1993localization} to compute localization lengths and the their scaling. We then demonstrate the existence of edge states for a disordered Chern insulator.

The rest of this article is organized as follows: in \S \ref{sec:tmat}, we describe our general construction of the transfer matrix and discuss their properties and applications. In \S \ref{sec:r1}, we describe the computations for rank 1 systems, taking Chern insulator as the prototypical model. In \S \ref{sec:winding}, we construct the $\ve$-Riemann sheet and describe the windings associated with edge states. In \S \ref{sec:r2}, we analytically compute the transfer matrix for a rank 2 system, the TCI. In \S \ref{sec:dis}, we apply our formalism to Chern insulator with diagonal disorder. Finally, we conclude and make general comments in \S \ref{sec:conc}. Details of some of the lengthy calculations and mathematical facts relevant to the work are relegated to the appendices.


\section{Transfer matrices}   \label{sec:tmat}
Transfer matrices arise naturally in discrete calculus as a representation of finite order linear difference equations, where it is an operator that implements a first order shift on a block. In general, a transfer matrix will depend on the independent variable; however, if the system is periodic, the transfer matrix that translates by a single period acquires a special significance, as its repeated application can propagate a solution as far as one wishes. This matrix is often known as the monodromy matrix in dynamical systems literature\cite{eastham_spectral, eastham_periodic_operators}. 

As noninteracting lattice models are essentially composed of hopping (i.e, shift) operators, which act on the wavefunctions, the Schr\"odinger equation can alternatively be written as difference equations (or recursion relations) by action on a 1-particle state, as we demonstrate in the following. 

\subsection{Outline for tight-binding lattice Hamiltonians}   \label{sec:tmat_lattice}
The tight binding lattice Hamiltonians, in presence of (discrete) translation symmetry, are diagonal in the momentum basis, thereby reducing the computation of the bulk spectrum to the diagonalization of a finite-dimensional, momentum-dependent Bloch Hamiltonian. However, as we are interested in the edge states, the translation symmetry is naturally broken in the direction normal to the edge, as the system is finite in that direction. We shall assume that for a system in $d$ space dimensions, we have periodic boundary conditions(PBC) along $d-1$ directions parallel to the edge, so that the corresponding quasi-momentum $\vk_\perp \in \torus^{d-1}$ (the $d-1$ dimensional torus) is a good quantum number\cite{vasudha-roger_tm, hughes-srinidhi_emres}, which reduces a given $d$-dimensional system into a family of 1D system parametrized by $\vk_\perp$. 

Consider, then, the tight-binding Hamiltonian of such an arbitrary 1D model parametrized by the transverse quasi-momentum $\vk_\perp$, with finite range hopping along the finite direction. In the most general form, we can write such a Hamiltonian as
\begin{align}
  \hlt = & \;  \sum_{n=0}^N \sum_{\alpha, \beta = 1}^\ndof \sum_{\ell=0}^{R} \left[ \cd_{n+\ell, \alpha} t_{\ell, \alpha\beta}^{\phantom{\dagger}} c_{n,\beta}^{\phantom{\dagger}} + \text{h.c.} \right] \nonumber \\ 
  = & \; \sum_{n=0}^N \sum_{\ell=0}^{R} \left[ \vcd_{n+\ell} \mathbf{t}_{\ell} \vc_{n} + \text{h.c.} \right],
\end{align}
where $R$ is the range of the hopping and we have $\ndof$ internal degrees of freedom (spin/ orbital/ sublattice) per site of the lattice. In the second line, we have bundled up the creation/annihilation matrices corresponding to $q$ orbitals in the $q$-vectors $\vcd / \vc$, while $\mathbf{t}_\ell$ is the corresponding hopping matrix with $\ell^\text{th}$ nearest neighbors. We shall suppress the explicit dependence on $\vk_\perp$ in the following equations to avoid notational clutter; however, all parameters should be assumed to depend on $\vk_\perp$, unless stated otherwise.

\begin{figure}
\centering 
\includegraphics[width=\columnwidth]{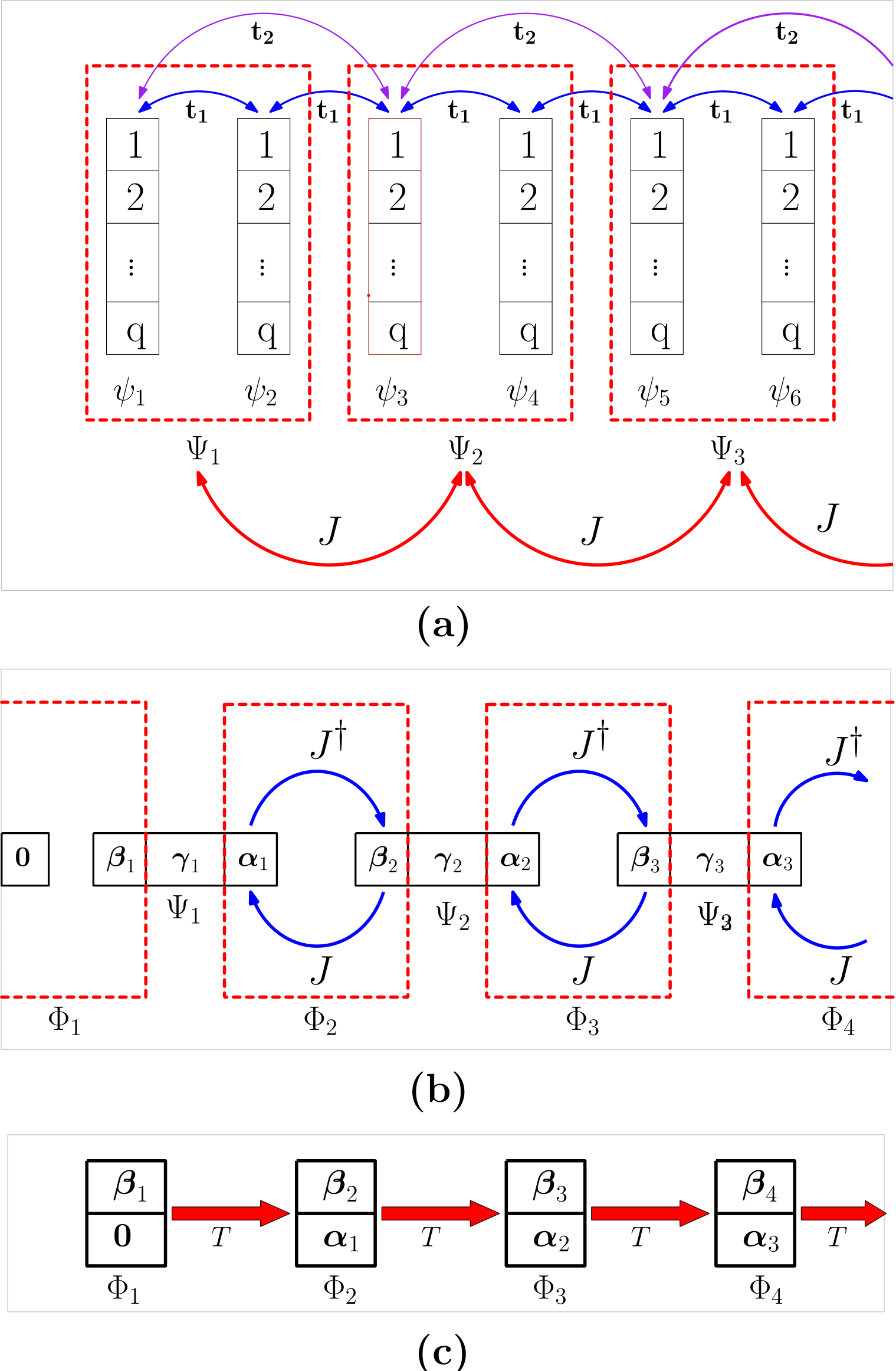}
\caption{(a) A schematic depiction of the recursion relation, with $q$ internal degrees of freedom, range of interaction $R = 2$ and Dirichlet boundary conditions at the left edge. We can form blocks (supercells) of such \emph{sites} with 2 sites each, so the there is only nearest neighbor hopping between them. (b) A simplified depiction of the reduced recursion relation, with $\bal$, $\bbe$, $\bga$ corresponding to the coefficients of $V$, $W$ and $X$ subspaces(introduced in \S \ref{sec:tmat_general}), respectively. (c) We club together $\bbe_n$ with $\bal_{n-1}$ to obtain $\Phi_n$, which is translated by one step using the transfer matrix.}    \label{fig:sch}
\end{figure}

By considering the action of this Hamiltonian on a 1-particle state
\beq \ket{\psi} = \sum_{n=0}^N \psi_n \vcd_n \ket{\Omega}, \quad \eeq
where $\ket{\Omega}$ is the fermionic vacuum state and $\psi_n \in \cmplx^q$ is the wavefunction for each physical site, the eigenvalue problem $\hlt \ket{\psi} = \ve \ket{\psi}$ can be written as a recursion relation in $\psi_n$, as
\beq \sum_{\ell=0}^{R} \left( \mathbf{t}_\ell \psi_{n+\ell} + \mathbf{t}_\ell^\dagger \psi_{n-\ell} \right)  = \ve \, \psi_n. \label{eq:recur_mat} \eeq
We now construct blocks consisting of these \emph{sites}, so that the system is periodic in these blocks and the hopping between such blocks is restricted to nearest neighbor\cite{dh_lee_cbs}, as shown in Fig. \ref{fig:sch}). These blocks form the sites of a \emph{superlattice}. We shall hereafter refer to those blocks as \emph{supercells}. Note that this is always possible as the hopping has a finite range, hence we can always choose a supercell consisting of $R$ sites. In terms of these supercells, each containing $\nuc = \ndof R$ degrees of freedom, the recursion relation becomes
\beq J \Psi_{n+1} + M \Psi_n + J^\dagger \Psi_{n-1} = \ve \Psi. \eeq
Here, $J$ is the \emph{hopping matrix} connecting nearest neighbor supercells and $M$ is the \emph{on-site matrix}, which encodes the hopping between degrees of freedom inside the supercell as well as the on-site energies. The wavefunction for a supercell, $\Psi_n$, is defined as 
\beq 
\Psi_n = \left( \begin{array}{c} \psi_n \\ \psi_{n+1} \\ \vdots \\ \psi_{n+R-1}\end{array} \right)  \in \cmplx^\nuc. 
\eeq 
We shall denote the standard basis of $\cmplx^\nuc$ with $\uvec_i, i = 1, \dots \nuc $, where $\left( \uvec_{i} \right)_j = \delta_{ij}$.  

For a nonsingular $J$, the transfer matrix construction works by noticing that\cite{avila_edge2d}
\beq \Psi_{n+1} = J^{-1} (\ve \id - M) \Psi_n - J^{-1} J^\dagger \Psi_{n-1} \eeq
can be rewritten as
\begin{align}
  \left( \begin{array}{c} \Psi_{n+1} \\ \Psi_n \end{array} \right) = & \; \left( \begin{array}{cc} 
  J^{-1} (\ve \id - M) \quad &  - J^{-1} J^\dagger \\ 
  \id \quad & 0 \end{array} \right) 
  \left( \begin{array}{c} \Psi_{n} \\ \Psi_{n-1} \end{array} \right)  \nonumber \\ 
  \equiv & \; T  \left( \begin{array}{c} \Psi_{n} \\ \Psi_{n-1} \end{array} \right). 
\end{align}
Hence, we have a $2 \nuc\times 2 \nuc$ transfer matrix $T$. However, this does not work for a singular $J$, which is often the case. 

What exactly does $\rank(J)$ mean? Think of the $\nuc$ degrees of freedom inside each supercell as $\nuc$ sites\footnote{This is the opposite of the traditional way of doing things, where sites inside a supercell are effectively treated as \emph{orbitals}.}. Then, $\rank(J)$ is the number of linearly independent rows in $J$, and hence the number of degrees of freedom that enter in the recursion relation, when expressed in a suitable basis. In more physical terms, $\rank(J)$ denotes the number of \emph{bonds} between adjacent supercells. Hence, physically, the singularity of $J$ means that there are sites in the supercell from which one cannot hop directly to a site in another supercell. In other words, if the $\nuc$ degrees of freedom in a supercell are thought of as nodes of a graph, where $J$ and $M$ encode the connectivity of the graph, then there are nodes in a supercell that are not connected to any nodes in other supercells, and $\rank(J)$ is the number of links between the supercells. We seek to compute a transfer matrix for singular $J$, where we, in some sense, \emph{mod out} the redundant degrees of freedom, thereby inverting $J$ on a reduced subspace to get a reduced transfer matrix.

We shall seek to compute the transfer matrix in a basis-independent fashion, i.e., without referring to the explicit forms of the $J$ and $M$ matrices. However, we present an explicit calculation for the case of Chern insulator in Appendix \ref{app:chern} which motivated us for the following general construction.

\subsection{Constructing the transfer matrix}    \label{sec:tmat_general}
We begin with the recursion relation 
\beq J \Psi_{n+1} + J^\dagger \Psi_{n-1}  = (\ve \id - M) \Psi_n. \label{eq:wf_recur} \eeq 

Let $\rank(J)=r$. We will see that the corresponding transfer matrix will be $2r \times 2r$. Indeed, if $J$ had full rank ($\rank(J) = \nuc$), we could have inverted it to get a $2\nuc \times 2\nuc$ transfer matrix, as computed in \S \ref{sec:tmat_lattice}. In the following, we shall also assume a big enough supercell that $J$ is nilpotent of degree 2, i.e, $J^2 = 0$, so that $r  \leq \nuc / 2$. Physically, for $\nuc > 2$, this simply means that in a given supercell, the nodes in a supercell that are connected to the right neighboring supercell and the left neighboring supercell are not directly connected to each other.

Now consider $M$. We note that $\green = (\ve \id - M)^{-1}$ is the resolvent (or the Green's function) of a single supercell. Clearly, $\ve \id - M$ is singular when $\ve$ is an eigenvalue of $M$. What does that mean? Consider a system with the uncoupled $\nuc$-degrees-of-freedom supercells, obtaining by tuning to $J = 0$ in the recursion relation of eq. (\ref{eq:wf_recur}). The corresponding spectrum consists of $\nuc$ degenerate levels. As we turn on the hopping $J$, these degenerate levels broaden into bands. Hence, the eigenvalues of $M$ can be interpreted as the centers of the bands. Since we are primarily concerned with the band gaps and the edge states therein, we can take $\ve \id - M $ to be nonsingular as far as we do not venture deep inside the bands\footnote{This breaks down if the bandwidth turns out to be zero, i.e, when the band pinches to a point(for instance, in case of graphene, or Chern insulator with $m = 1$). However, we can get around that using the well known trick of adding a small imaginary part to the Green's function in order to move the poles off the real line.}.

We perform a reduced singular value decomposition\cite{strang_book} (SVD) of $J$, 
\beq J = V \cdot \Xi \cdot W^\dagger, \eeq 
where 
\begin{align}
  V = & \; \left( \vv_1, \vv_2, \dots \vv_r \right)_{\nuc \times r}, \nonumber \\ 
  W = & \; \left( \vw_1, \vw_2, \dots \vw_r \right)_{\nuc \times r}, \nonumber \\
  \Xi = & \; \text{diag}\{ \xi_1, \xi_2, \dots, \xi_r \}_{r \times r},
\end{align}
with 
\beq V^\dagger \cdot V = W^\dagger \cdot W = \id , \quad  V^\dagger \cdot W = 0.   \label{eq:svd_prop}  \eeq
The first two expressions follow from the definition of SVD, while $J^2 = 0$ implies the third. Hence, $J^2 = 0$ is required to ensure that the $V$ and $W$ subspaces are orthogonal and the corresponding coefficients can be extracted by taking inner products. 

The SVD can equivalently be written as
\beq J = \sum_{i = 1}^r \xi_i \vv_i \otimes \vw_i \eeq
with
\beq \langle \vv_i, \vv_j \rangle = \langle \vw_i, \vw_j \rangle = \delta_{ij} , \quad  \langle \vv_i, \vw_j \rangle = 0.  \eeq
We shall hereby refer to these vector pairs $(\vv_i, \vw_i)$ as \emph{channels}. As we can still change the phases of $\vv$ and $\vw$ without violating the orthonormality, we choose their phases such that all the singular values are positive, i.e, $\xi_i > 0 \; \forall \, i$. Clearly, $\Xi^\dagger = \Xi$. 

Now, \emph{morally speaking}, we claim that the only directions in $\cmplx^\nuc$ relevant for the problem are $\vv_i$'s and $\vw_i$'s, i.e, span$\{V\}$ and span$\{W\}$. Take a basis of $\cmplx^\nuc$ as $\{\vv_i, \vw_i, \vx_j\}$, where $i = 1, \dots r, \; j = 1, \dots \nuc-2r$, and expand $\Psi_n$ as 
\beq \Psi_n = \sum_{i=1}^{r} \left( \alpha_{n, i} \vv_i + \beta_{n, i} \vw_i \right)  + \sum_{j=1}^{\nuc-2r} \gamma_{n, j} \vx_j,  \eeq
with $\alpha_{n,i}, \beta_{n,i}, \gamma_{n, i} \in \cmplx$, or, equivalently, 
\beq \Psi_n = V \bal_n + W \bbe_n + X \bga_n, \eeq
with $ \bal_n, \bbe_n \in \cmplx^r, \; \bga_n \in \cmplx^{\nuc-2r}$. We have defined $X$ analogous to $V$ and $W$, so that 
\beq V^\dagger \cdot X = W^\dagger \cdot X = 0, \quad X^\dagger \cdot X = \id. \label{eq:svd_xprop}  \eeq
Also, 
\beq \bal_n = \left( \alpha_{n, 1},  \alpha_{n, 2}, \dots  \alpha_{n, r} \right), \eeq
with $\bbe_n$ and $\bga_n$ defined in a similar fashion. 

We can rewrite the recursion relation in \eq{eq:wf_recur}, in terms of the Green's function $\green= (\ve \id - M)^{-1}$, as 
\beq \Psi_n = \green\cdot J \; \Psi_{n+1} + \green\cdot J^\dagger \; \Psi_{n-1}. \label{eq:recur_green} \eeq
But
\beq J \Psi_n = V \cdot \Xi \; \bbe_n , \quad J^\dagger \Psi_n =  W \cdot \Xi \; \bal_n, \eeq 
which follows from the SVD, \eq{eq:svd_prop} and \eq{eq:svd_xprop}. 
We can now premultiply \eq{eq:recur_green} by $V^\dagger$, $W^\dagger$ and $X^\dagger$ to extract the coefficients $\bal_n$, $\bbe_n$ and $\bga_n$, respectively. In order to simplify notation, we denote the restriction of $\green$ to $V$ and $W$ subspaces by $\green_{vv} = V^\dagger \cdot \green\cdot V$, $\green_{vw} = W^\dagger \cdot \green\cdot V$, etc\footnote{Note that the order of $V$ and $W$ in subscript is opposite to that in the expression.}. Then,  
\begin{align} 
 \bal_n = & \; \green_{vv} \cdot \Xi \; \bbe_{n+1} + \green_{wv} \cdot \Xi \; \bal_{n-1}, \nonumber \\
 \bbe_n = & \; \green_{vw} \cdot \Xi \; \bbe_{n+1} + \green_{ww} \cdot \Xi \; \bal_{n-1}, \nonumber \\
 \bga_n = & \; \green_{vx} \cdot \Xi \; \bbe_{n+1} + \green_{wx} \cdot \Xi \; \bal_{n-1},  \label{eq:recur}
\end{align}
where the $\green_{ab}, \; a, b \in \{v, w\}$ is a $r \times r$ matrix. After some matrix gymnastics (see Appendix \ref{app:math_matrix}), the first two equations can be reorganized as 
\beq 
\Phi_{n+1} = T \Phi_n, \quad \Phi_n \equiv \left( \begin{array}{c} \bbe_{n} \\ \bal_{n-1} \end{array} \right), 
\label{eq:TMM}
\eeq
with
\begin{align}
 T = & \;  - \left( \begin{array}{cc} \green_{vv} \cdot \Xi \quad\quad & -\id \\ \green_{vw} \cdot \Xi \quad\quad & 0 \end{array} \right)^{-1} \left( \begin{array}{cc} 0 & \quad \green_{wv} \cdot \Xi \\ -\id & \quad \green_{ww}  \cdot \Xi \end{array} \right) \nonumber \\ 
 = & \;  \left( \begin{array}{cc} \Xi^{-1} \cdot \green_{vw}^{-1} \quad & - \Xi^{-1}  \cdot \green_{vw}^{-1} \cdot \green_{ww} \cdot \Xi \\ \green_{vv} \cdot \green_{vw}^{-1} \quad & \left( \green_{wv} - \green_{vv} \cdot \green_{vw}^{-1} \cdot \green_{ww} \right) \cdot \Xi \end{array} \right). 
 \label{eq:tmat_def}
\end{align}
Hence, we have managed to construct a closed form expression for a  $2r \times 2r$ transfer matrix explicitly for the given recursion relation. This is one of our central results.

Defining $\gralt_{ab} = \green_{ab} \cdot \Xi$, we can also express this result as 
\beq T =  \left( \begin{array}{cc} \gralt_{vw}^{-1} \quad\quad & \gralt_{vw}^{-1} \cdot \gralt_{ww} \\ \gralt_{vv} \cdot \gralt_{vw}^{-1} \quad\quad &  \gralt_{wv} - \gralt_{vv} \cdot \gralt_{vw}^{-1} \cdot \gralt_{ww} \end{array} \right). \eeq 
This expression is somewhat cleaner, but it obscures the different physical significance associated with $\green$ and $\Xi$, as well as properties of $\green$, which we now state. As the Green's function is Hermitian, i.e, $\green^\dagger = \green$, we have
\beq \green_{vv}^\dagger = \green_{vv}, \quad \green_{ww}^\dagger = \green_{ww}, \quad \green_{vw}^\dagger = \green_{wv}. \label{eq:green_prop} \eeq 
Using these and \eq{eq:det_block}, an explicit computation shows that
\beq \det T = \det \left( \green_{vw}^{-1} \right) \det \left( \green_{wv} \right) = \frac{\left( \det   \green_{vw} \right)^*}{\det \green_{vw} },  \eeq
which we can write as
\beq \det T = e^{-2i \theta} \in U(1), \quad \theta = \arg \left( \det \green_{vw} \right).  \eeq
However, we can gauge this phase away by the gauge transform
\beq \Phi_n \to e^{i n \theta / r} \Phi_n , \quad T \to e^{i \theta / r} T. \label{eq:gaugetr} \eeq 
In the following, whenever we refer to the transfer matrix, we shall assume that we have gauged away the phase of the determinant of $T$ so that $\det T  = 1$. 

\subsection{Properties}    \label{sec:tmat_prop}
Before we go on to compute physically relevant quantities from the transfer matrix, we discuss a few features of our construction:
\begin{itemize}
  \item[a)] The transfer matrix propagates $\bal$ and $\bbe$ degrees of freedom. Given one of the $\Phi$'s, say, $\Phi_m$, we can compute $\Phi_n$ for all $n$, and, using the expression for $\bga_n$ in eq. (\ref{eq:recur}), compute the wavefunction $\Psi_n \, \forall \, n$. Furthermore, as $T$ is nonsingular by construction, we can also use $T^{-1}$ to propagate $\Phi_n$ backwards. 

  \item[b)] The transfer matrix is basis independent, as we have never referred to the explicit form of the $J$ and $M$ matrices. It reduces the computation of transfer matrix for a system to the identification of the $J$ and $M$ matrices, as everything else can be mechanized. We shall illustrate that with examples later on (see Table \ref{tb:eg}). 
 
  \item[c)] The size and spectral properties of the transfer matrix are independent of the size of the supercell chosen, once it is above a certain size. Hence, we can define a \emph{minimal supercell}, which is a block consisting of the minimum number of sites so that the hopping between the supercells is nearest neighbor and the corresponding hopping matrix is nilpotent.  In Appendix \ref{app:sqr}, we show that if we take a supercell that is $m$ times the minimal supercell, the transfer matrix is simply exponentiated by $m$, i.e, $T \to T^{m}$, but its size, which is twice the rank of the hopping matrix, stays invariant under this operation. But as in computing the band structures, we are concerned only with the behavior of $T^n$ for large $n$(See \S \ref{sec:tmat_appl}), the band structure, as expected, stays invariant under such a transformation. Hence, we can always make the supercell bigger than the minimal supercell, while leaving the bands and edge states invariant. We shall use this property in certain proofs. 

  \item[d)] As $\green_{ab}, \; a, b \in \{v, w\}$ are simply restrictions of the Green's functions, they are propagators connecting the $a$ and $b$ degrees of freedom for each supercell, while $\Xi$ encodes the tunneling probabilities, or the relative strength of each channel. In fact, the recursion equation in terms of $\bal$ and $\bbe$ (\eq{eq:recur}) has a simple diagrammatic interpretation as superpositions of possible nearest neighbor hopping processes. This is illustrated in Fig.\,\ref{fig:hopping}, where the Green's functions $\mathcal{G}_{ab}$ express the propagation within a block and $\Xi$ the tunneling between blocks. The transfer matrix equation (\ref{eq:TMM}) can then be seen as an equation of constraint that respects these hopping processes.

\begin{figure}
	\includegraphics[width=0.95\columnwidth]{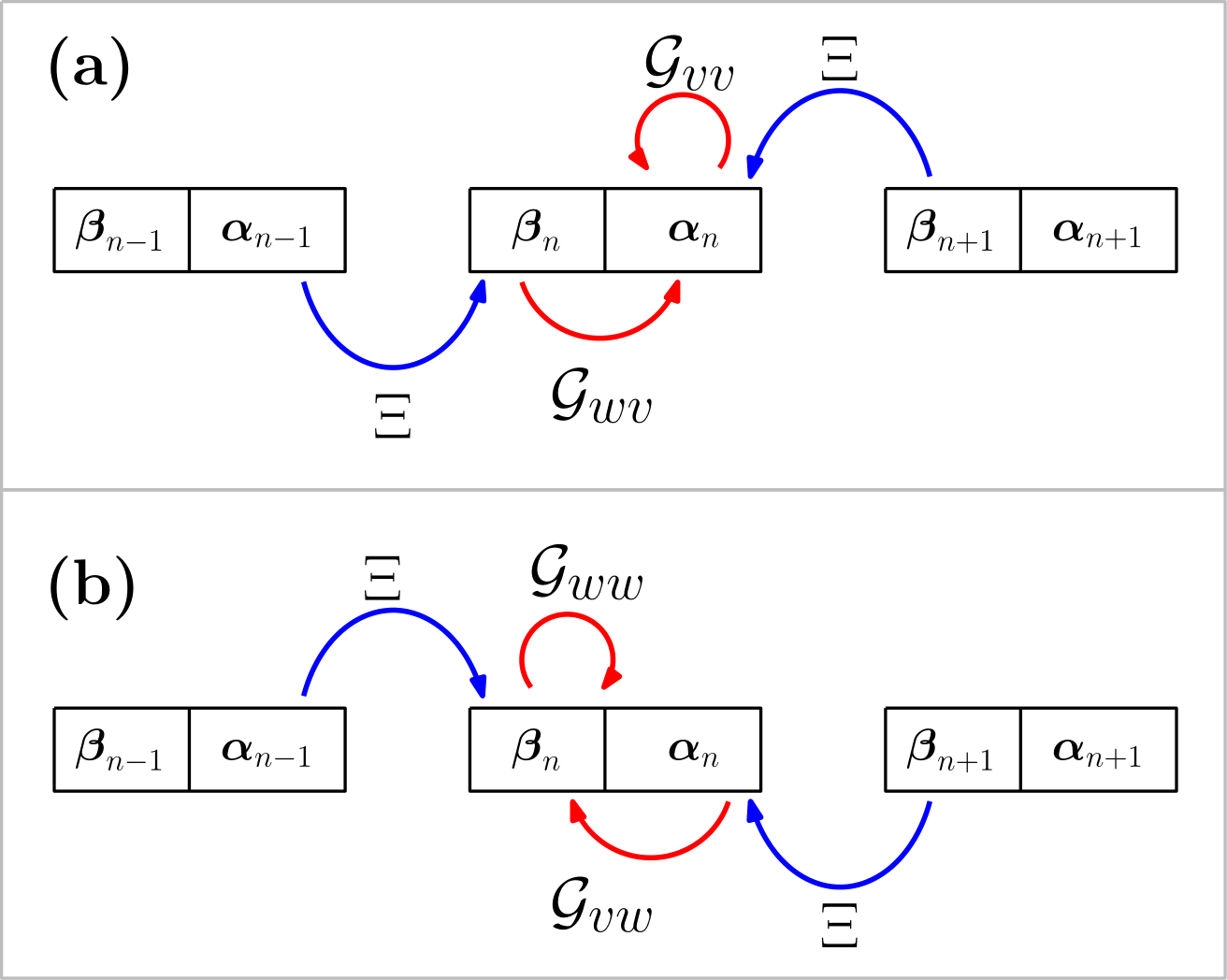}
	\caption{Diagrammatic representation of the recursion relations (\eq{eq:recur}) for (a) $\bal_{n}$ and  (b) $\bbe_n$.}
	\label{fig:hopping}
\end{figure}

  \item[e)] In Floquet theory for a continuous independent variable, the monodromy matrix is symplectic if the system is Hamiltonian\cite{jared_mod_ins}. Is that also true for the discrete case? An explicit computation using eq. (\ref{eq:green_prop}) shows that 
  \beq T^\dagger \cdot \symp \cdot T = \symp, \quad \symp = \left( \begin{array}{cc} 0 & 1 \\ -1 & 0 \end{array} \right), \eeq 
  if 
  \beq 
    [\green_{ab}, \Xi] = 0, \quad a, b \in \{ v, w \}.  \label{eq:symp_cond}
  \eeq
  Physically, this condition implies that the various channels that connect the nearest neighbor supercells need to be independent, so that the order of tunneling ($\Xi$) and propagation ($\green$) is irrelevant. We term this condition as $T$ being $\symp$-unitary or complex-symplectic ($T \in \Sp(2r, \cmplx)$). The spectral properties of $\symp$-unitary operators have been studied in great detail in the mathematics literature\cite{baldes_fredholm}. Furthermore, if $T$ is real, we say that $T$ is symplectic ($T \in \Sp(2r, \real)))$.  In the discussion on bulk bands, we show that if the transfer matrix is symplectic, it can effectively be decomposed into a set of chains, one corresponding to each channel. The conditions on $\green_{ab}$ obtained above are physical manifestations of that fact. As $\Xi$ is, by definition, a diagonal matrix, in order for it to commute with another matrix $A$,  $A$, in general, must also be diagonal\footnote{This breaks down if two or more diagonal entries of $\Xi$ are equal, as $\Xi$ then becomes proportional to identity in that subspace, so that $A$ restricted to that subspace can be anything.}. Hence, for $T$ to be symplectic, $\green_{vv}$ and $\green_{vw}$ must also be diagonal. 
  
  \item[f)] Recall that a complex square matrix A is termed \emph{normal} if it commutes with its adjoint, i.e, if 
  \beq A^\dagger A = A A^\dagger. \eeq
  The matrix $A$ is diagonalizable by a unitary matrix if and only if it is normal. In other words, the eigenvectors of a matrix form an orthonormal basis if and only if it is normal, a condition often ignored in physics literature. However, as it turns out, the transfer matrices are almost never normal. Hence, in the subsequent arguments, we shall not assume that $T$ is normal in general, which, naturally, makes them somewhat more involved. 

\end{itemize}

\subsection{Using the transfer matrix}     \label{sec:tmat_appl}
We now discuss the computation and interpretation of spectra from the transfer matrices.

\subsubsection{Bulk bands} \label{sec:tmat_appl_bulk}
The transfer matrix can be used to propagate a state spatially into the system. The eigenstates with an eigenvalue $\rho \in \cmplx$ propagates indefinitely if $|\rho| = 1$, i.e, the $\rho$ lies on the unit circle $S^1 \equiv \{z\in \mathbb{C}\mid\, |z|=1 \}$ in the complex plane, while it grows/decays as $n \to \infty$ is $\rho$ lies outside/inside the unit circle. Hence, a given $(\ve, \vk_\perp)$ lies in the bulk band if all eigenvalues of $T(\ve, \vk_\perp)$ lie on the unit circle, while it lies in the gap if all of them lie off the unit circle. 

Formally, let $\sigma \left[T(\ve, \vk_\perp)\right]\subset \cmplx$  be the spectrum of $T$ for a point $(\ve, \vk_\perp) \in \real \times \torus^{d-1}$. We define the \emph{bulk band}, $\band \subset\real \times \torus^{d-1}$, as
\beq 
\band = \left\{ (\ve, \vk_\perp) \, | \, \sigma\left[ T(\ve, \vk_\perp) \right] \subset S^1 \right\}, 
\label{eq:band_def} \eeq 
and the \emph{bulk gap} as
\beq 
\gap = \left\{ (\ve, \vk_\perp) \, | \, \sigma\left[ T(\ve, \vk_\perp) \right] \subset  \mathbb{C} \backslash S^1 \right\}. 
\label{eq:gap_def} \eeq 
For $r>1$, the possibility exists that there can be points $(\ve, \vk_\perp)$ for which some eigenvalues are on and some off the unit circle.  We shall term such points \emph{partial gaps}, $\pgap$, defined as
\begin{align}
\pgap = \left( \real \times \torus^{d-1} \right) \backslash \left( \gap \cup \band \right). \label{eq:pgap_def}
\end{align} 
By construction, each $(\ve, \vk_\perp)$ falls in one of these sets. 

To compute the bulk bands, one needs to compute the eigenvalues of the transfer matrix. This can always be done numerically in a given case, however, if the transfer matrix is symplectic, its characteristic polynomial has further structure(see Appendix \ref{app:tmat}) which allows us to compute the eigenvalues analytically. 

Let us start off with $r = 1$, where $\det T = 1$ implies that $T$ is symplectic. It also implies that the product of eigenvalues is unity, so that the eigenvalues are reciprocals of each other, so that 
\beq \Delta \equiv \tr T = \rho + \rho^{-1}, \eeq
which can be solved to get
\beq \rho_{\pm} = \frac{1}{2} \left[ \Delta \pm \sqrt{\Delta^2 - 4} \right]. \eeq 
Hence, either both the eigenvalues are on the unit circle or both on the real line. In turn, a given $(\ve, \vk_\perp)$ either belongs to $\gap$ or $\band$, so that $\pgap = \emptyset$. 

For $r > 1$, if the transfer matrix is symplectic, the eigenvalues always come in reciprocal pairs, i.e, if $\rho_i$ is an eigenvalue, so is $\rho_i^{-1}$. Hence, given a $2r \times 2r$ transfer matrix, we construct the $r$ \emph{Floquet discriminants} using the traces of powers of $T$ (see Appendix \ref{app:tmat} for details), where 
\beq \Delta_i = \rho_i + \rho_i^{-1}, \quad i = 1, 2, \dots r, \eeq 
so that the eigenvalues of the transfer matrix are  
\beq \rho_{i, \pm} = \frac{1}{2} \left[ \Delta_i \pm \sqrt{\Delta_i^2 - 4} \right]. \eeq 
For instance, for $r = 2$, the explicit expression for the Floquet discriminants is
\beq \Delta_{\pm} = \frac{1}{2} \left[ \tr T \pm \sqrt{2 \tr T^2 - (\tr T)^2 + 8} \right].  \eeq
Hence, if the transfer matrix is symplectic, we can essentially decompose it into a set of $r=1$ systems, which are independent of each other! 

From the expression for the eigenvalues, we deduce that we have an oscillating state for $|\Delta_i| \leq 2$ and a growing/decaying state for $|\Delta_i|>2$. Hence, we can alternatively define the bulk band and the band-gap as 
\begin{align}
 \gap = & \;  \{(\ve, \vk_\perp) \; | \; |\Delta_i(\ve, \vk_\perp)| > 2 \; \forall \; i = 1, \dots r\}, \nonumber \\ 
 \band = & \;  \{(\ve, \vk_\perp) \; | \; |\Delta_i(\ve, \vk_\perp)| \leq 2 \; \forall \; i = 1, \dots r \}.
\end{align}

Furthermore, the band edges are simply given by the conditions $|\Delta_i| = 2, \; i = 1, 2, \dots r$. Hence, we can simply solve this conditions for $\ve(\vk_\perp)$, numerically if needed, to compute the band edges, without having to diagonalize $T$ for all possible $(\ve, \vk_\perp)$, which one would need to do in general.   

\subsubsection{Decay conditions}
The edge states are typically the states that reside outside the bulk bands, $(\ve, \vk_\perp) \in \gap \cup \pgap$, which implies that they are growing/decaying as $n \to \pm \infty$. In order to be normalizable, they are taken to be decaying into the bulk away from the edges. Typically, one is interested in the existence of these states, and, should they exist, in the \emph{edge spectrum}, i.e, the energy of the edge state, $\ve_{edge}(\vk_\perp)$, as a function of the transverse momentum $\vk_\perp$. 

Given $\Phi_k$ for an arbitrary site $k$, we can use the transfer matrix to compute $\Phi_{n+k} = T^n \Phi$. Hence, given a $\Phi \in \cmplx^{2r}$, we are concerned with the asymptotics of $|T^n \Phi|$ for $n\to\pm\infty$, where $|.|$ is the vector norm over $\cmplx^{2r}$. A $\Phi$ can be a legitimate \emph{left} edge state if $|T^n \Phi| \to 0$ as $n \to \infty$. Similarly, $\Phi$ can be a legitimate \emph{right} edge state if $|T^n \Phi| \to 0$ as $n \to -\infty$. In this subsection, we seek the conditions imposed on $\Phi$ by the transfer matrix (i.e, the bulk) for it to be a legitimate decaying edge state, while we defer the implications of the boundary condition to the next subsection. In the following, we consider the left edge, the arguments for the right edge being their exact analogues.

Consider first the case when $T$ is \emph{normal} and satisfies the conventional eigenvalue equation
\beq T \varphi_s = \rho_s \varphi_s, \quad \rho_s \neq 0 \, \forall \, s = 1,\dots 2r. \eeq
Now, span$\{\varphi_s\} = \cmplx^{2r}$, so that any state $\Phi \in \cmplx^{2r}$ can be written as a linear combination of the eigenvectors of the transfer matrix, and
\beq \Phi = \sum_{s = 1}^{2r} \alpha_s \varphi_s \implies T^n \Phi = \sum_{s = 1}^{2r} \rho_s^n \alpha_s \varphi_s. \eeq
We deduce that $|T^n \Phi|$ decays as $n \to \infty$ only if the contribution of the growing eigenvalue is 0, i.e, 
\beq |\rho_s|>1 \implies \alpha_s = 0. \eeq 
This restricts $\Phi$ to a subspace of $\cmplx^{2r}$ corresponding to $|\rho_s| \leq 1$.

However in general, $T$ is \emph{not} a normal matrix and hence it cannot have an eigenvalue equation in the usual sense and cannot be diagonalized by a unitary matrix, a fact that is often overlooked in the physics literature. Nevertheless, it can always be brought to a Jordan canonical form \cite{kato_book}. The behavior of $|T^n\Phi|$ is still dictated by the \emph{generalized} eigenvalues of $T$, so that to ensure exponential spatial decay of $|T^n\Phi|$ as $n \to \infty$, we now require that $\Phi$ contain no generalized eigenvectors with eigenvalues $|\rho_s|>1$. Similarly, for the right edge, we want $|T^n\Phi|$ to decay as $n \to -\infty$. Hence, the corresponding condition demands that $\Phi$ contain no generalized eigenvectors with eigenvalues $|\rho_s|<1$.

In order to obtain a precise mathematical condition for the $|\rho_s| \lessgtr 1$ subspaces, we express $T$ in its Jordan canonical form \cite{kato_book} 
\beq 
T = \sum_{s} [\rho_s \proj_s + \nilp_s], 
\label{eq:jordan}
\eeq 
where the sum extends over all generalized eigenvalues $\rho_s$, $\proj_s$ project in the eigenspace of $\rho_s$ and $\nilp_s$ are nilpotent matrices. However, it remains true that the determinant of $T$ is equal to the product of its generalized eigenvalues $\prod_s \rho_s=\det\,T$, as can be seen by applying a similarity transform to (\ref{eq:jordan}). We define the projector to the $|\rho_s| < 1$ subspace as 
\beq \proj_< \equiv \sum_{|\rho_s|<1} \proj_s, \eeq
and similarly, $\proj_0$ and $\proj_>$ for $|\rho_s| = 1$ and $|\rho_s| > 1$, respectively. Clearly, 
\beq \proj_< + \proj_0 + \proj_> = \id. \eeq 
Then a sufficient condition for $\Phi \in \cmplx^{2r}$ to be a \emph{left} edge state is 
\begin{align}
\proj_< \Phi = \Phi  \label{eq:dcond_L}
\end{align}
and similarly for a \emph{right} edge,
\begin{align}
\proj_> \Phi = \Phi.\label{eq:dcond_R}
\end{align}
A rigorous proof of this statement is provided in Appendix \ref{app:math_op}. We shall term these \emph{decay conditions}.

\subsubsection{Boundary conditions}
We now discuss the boundary conditions required to compute the physical edge states of the system, as observed in an exact diagonalization of the lattice models on finite size lattices. Most of the following is a restatement of the results by Lee and Joannopoulos\cite{dh_lee_cbs} in our formalism, which, we believe, is more general. In the following, we shall only consider the system that is terminated abruptly at layer 0 and $N$, hereafter termed a \emph{hard boundary} condition. 

We mention in passing that as the edge state spectrum is strongly dependent on the boundary conditions, it can get modified quite drastically by local terms at the boundary, an effect commonly known as \emph{edge reconstruction}. In order to consider the most general case, we should take a Hamiltonian $\tilde{\hlt} = \hlt + \delta \hlt$, where $\delta \hlt$ is an operator localized at the edge, which can account for the edge reconstruction, for instance, due to an impurity\cite{zeng_graphene_edge} or lattice deformation\cite{kim_graphene_edge}. Such a boundary condition imposes additional conditions\cite{dh_lee_cbs, tsymbal_TI_edge, yamakage_TI_edge} on the eigenvectors of the transfer matrix. As our purpose in this article is to expound the geometry and topology associated with the band structure which is independent of such local deformations, we shall not discuss such cases in detail. 

For concreteness, we consider just a left edge since the right edge is analogous. The \emph{hard boundary} condition\cite{hatsugai_cbs_PRL} is the simplest Dirichlet boundary condition at an edge, whereby we simply demand that $\Psi_{0}=(0,0)$, which leads to $\alpha_{0}=0$. Hence, any initial state vector 
\begin{align}
\Phi_1 \equiv \begin{pmatrix} \bbe_1 \\ 0 \end{pmatrix}
\label{eq:phi1_exp}\end{align}
will satisfy this Dirichlet boundary condition on the left edge. Similarly, on the right edge, as $\Psi_{N+1}=(0,0)^T$, we have
\begin{align}
\Phi_N \equiv \begin{pmatrix} 0 \\ \bal_N \end{pmatrix}.
\label{eq:phiN_exp}\end{align}
Note that $\bbe_1,\bal_1 \in \mathbb{C}^{r}$ are still undetermined for $r>1$ on their respective edges. We shall use the decay conditions to fix these in the next section. 

Formally, we can also define projectors to write the boundary condition in a way similar to the decay conditions. We begin by defining the $2r \times r$ matrices 
\beq \proja = \left( \begin{array}{c} 0_{r\times r} \\ \id_{r\times r} \end{array} \right), \quad \projb = \left( \begin{array}{c} \id_{r\times r} \\ 0_{r\times r} \end{array} \right),  \eeq  
as the injectors into the $\bbe$  and $\bal$ subspaces, respectively. In terms of these operators, the Dirichlet condition on the left edge is equivalent to the statement that $\Phi \in \text{range} (\projb)$, while the right edge is equivalent to $\Phi \in \text{range}(\proja)$. Finally, define the projectors 
\beq \proj_R = \proja \proja^\dagger, \quad \proj_L = \projb \projb^\dagger. \eeq 
Then a sufficient condition for $\Phi \in \cmplx^{2r}$ to be a \emph{left} edge state is 
\beq \proj_L \Phi = \Phi, \label{eq:bcond_L} \eeq
while for the right edge, we have
\beq \proj_R \Phi = \Phi. \label{eq:bcond_R} \eeq
These are our \emph{boundary conditions}.

\subsubsection{Physical edge states}
We have obtained two sets of conditions, viz, the decay conditions and the boundary conditions, that we need to solve simultaneously in order to obtain the physical edge states. However, before we attempt to do so, we can ask a somewhat perverse question, which turns out to have important consequences: What if we chose the \emph{wrong} decay condition for a given boundary? We tabulate the situation as follows:
\begin{table}[ht!]
  \vspace{0.1in}
  \renewcommand{\arraystretch}{1.5}
  \setlength{\tabcolsep}{20pt}
  \begin{tabular}{|c|ll|}
    \hline 
     & $\proj_< \Phi = \Phi$ & $\proj_> \Phi = \Phi$ \\ 
     \hline
     $\proj_L \Phi = \Phi$ & Left edge  & \emph{Unphysical} \\ 
     $\proj_R \Phi = \Phi$ & \emph{Unphysical} & Right edge \\ 
    \hline  
  \end{tabular}
  \caption{Boundary(rows) vs decay(column) conditions.}   \label{tb:edge}
\end{table}

The \emph{wrong} choice of decay condition implies that the corresponding state grows (instead of decaying) exponentially in the bulk, and is hence not normalizable and unphysical. However, we shall see that in order to account for all the windings corresponding to the edge state, we shall need to take the unphysical states into account. Furthermore, these should not be thought of as a complete fantasy, as they can be revealed by changing the boundary condition, as we shall demonstrate explicitly in \S \ref{sec:r1_CI} 

At this point, we can compute the physical edge states by solving the decay conditions and the boundary conditions simultaneously. For instance, for the left edge state, we seek to simultaneously solve 
\beq \proj_< \Phi = \Phi = \proj_L \Phi,  \label{eq:cond_L1} \eeq 
or, alternatively, 
\beq \proj_< \Phi_1 = \Phi_1; \quad \Phi_1 = \left( \begin{array}{c} \bbe_1 \\ 0 \end{array}  \right). \label{eq:cond_L2} \eeq 
Note that as $\rank(\proj_<) \leq r$, this is a homogeneous linear system of up to $r$ equations for the $r$ variables, viz, the coefficients of $\bbe_1$. But for a nontrivial state, we demand that $\bbe_1 \neq 0$, from which we can obtain a Cramer's condition, which can be numerically solved to obtain the physical edge spectrum.

In the following, we also analytically construct a closed form expression combining the decay and the boundary conditions for the case when there are an equal number $(=r)$ of eigenvalues are 
inside and outside the unit circle in the complex plane, which corresponds to an $(\ve, \vk_\perp) \in \gap$, i.e, in the bulk gap. This implies that $\Tr{\proj_<} = \Tr{\proj_>} = r$, so that $\proj_< + \proj_> = \id$ and
\beq \proj_< \Phi_1 = \Phi_1 \implies \proj_> \Phi_1 = 0.  \label{eq:cond_L3} \eeq

We seek to represent $\proj_>$ in terms of the (generalized) eigenvectors of $T$. Let $\rho_i \in \cmplx$ be the generalized eigenvalues of $T$ with corresponding left and right generalized eigenvectors being $\phi_i$'s and $\varphi_i$'s. Furthermore, let us assume that that $\rho_i$ lies outside the unit circle for $i = 1, \ldots r$ while it lies inside the the unit circle for $i = r+1, \ldots 2r$. Then, we define the left and right subspaces corresponding to $\proj_>$ as  
\beq \mathcal{L}_> = (\phi_1,\ldots,\phi_r), \quad \mathcal{R}_> = (\varphi_1,\ldots,\varphi_r), \label{eq:defLR} \eeq 
where $\mathcal{L}_>, \mathcal{R}_> \in \cmplx^{2r\times r}$ span the co-kernel and range of $\proj_>$, respectively.

If $T$ were normal, i.e, diagonalizable by a unitary transform, then the right eigenvectors $\varphi_i$'s form an orthonormal basis of $\cmplx^{2r}$. As $\proj_>$ projects along a subset of these eigenvectors, it is an orthogonal projection, which can be written as
\beq \proj_> = \varphi_1 \varphi_1^\dagger + \varphi_2 \varphi_2^\dagger + \dots + \varphi_r \varphi_r^\dagger = \mathcal{R}_> \mathcal{R}_>^\dagger. \eeq 
Alternatively, in terms of the left eigenvectors, we can also write $\proj_> = \mathcal{L}_> \mathcal{L}_>^\dagger$.

In general, the analogue of this expression is the non-orthogonal representation\cite{meyer2000matrix} of $\proj_>$ 
\begin{align}
\proj_> = \mathcal{R}_>(\mathcal{L}_>^\dagger \mathcal{R}_> )^{-1}\mathcal{L}_>^\dagger. 
\end{align}
Hence, the decay condition $\proj_> \Phi_1 = 0$ (\eq{eq:cond_L3}) implies $\mathcal{L}^\dagger_> \Phi_1 = 0$, which, using \eq{eq:defLR}, can be written explicitly as
\begin{align}
\sum_{j=1}^{r} (\phi^*_j)_i (\bbe_1)_j &= 0, \quad i=0,\ldots,r,
\label{eq:e_Beta2}
\end{align}   
which constitutes $r$ linear equations for $r$ variables $(\bbe_1)_j$. Note that $\bbe_1$ is unique up to a non-zero complex scalar since the right-hand sides are all zero. Thus the space of unique solutions really is the complex projective $\mathbb{CP}^{r-1}$ valued. The equations (\ref{eq:e_Beta2}) have a nontrivial solution if and only if 
\begin{align}
\det \left[ \mathcal{L}_>^\dagger \projb \right] = 0. 
\label{eq:eQ}
\end{align}
which is essentially  a Cramer's condition. The analogous right edge conditions reads as
\begin{align}
\det\left[ \mathcal{R}_<^\dagger \mathcal{Q}_\alpha \right] =0 
\end{align} 
These conditions incorporate both the boundary and decay conditions and can be solved numerically to obtain $\ve$ as a function of $\vk_\perp$ to obtain the edge spectrum, $\ve_{edge}(\vk_\perp)$.  

Equation (\ref{eq:eQ}) is very convenient for numerical computations, but we also present an alternative characterization which is more explicit in terms of $T$'s projection. The general spectral decomposition of the resolvent of $T$\cite{kato_book} yields 
\begin{align}
\proj_> &= \oint_{|z|=1} \frac{dz}{i2\pi} (z-T^{-1})^{-1} \nonumber \\
&= \oint_{|z|=1} \frac{dz}{i2\pi} T (zT -\mathbb{I})^{-1} 
\label{eq:P_more}.
\end{align}
Essentially, we use the fact that the integrand has poles whenever $z$ equals an eigenvalue $\rho_s$ of $T$ so that $|\rho_s| > 1$. 

Now, in the simpler case of a normal $T$, we have $\proj_>= \mathcal{L}_> \mathcal{L}_>^\dagger = \mathcal{R}_> \mathcal{R}_>^\dagger$, so that
\beq 
\det \left[ \projb^\dagger \proj_> \projb \right] = \det \left(\projb^\dagger \mathcal{L}_>\right) \det\left(\mathcal{L}_>^\dagger \projb\right) = 0.  \label{eq:detcond2}
\eeq  
Substituting the integral representation of $\proj_>$ from \eq{eq:P_more}, we get
\begin{align}
\det\left[ \oint_{|z|=1} dz  \left[ T(\ve) (z T(\ve)-\mathbb{I})^{-1} \right]_{\beta\beta}  \right] = 0,
\end{align}
where $[\ast]_{\beta\beta}$ denotes the $r\times r$ sub-matrix of the argument and we have expressed the $\ve$ dependence of $T$ explicitly. Such an equation, though impractical for numerical computations, make explicit the analytic properties of an edge dispersion $\ve({\bf k}_\perp)$ in open neighborhoods where it exists as a solution. 

In the most general case where $\proj_>$ is oblique(non-orthogonal), the analogue of \eq{eq:detcond2} is  
\begin{align}
\det \left[ [\proj_>^\dagger \proj_>]_{\beta\beta} \right] \equiv \det [ \projb^\dagger \proj_>^\dagger \proj_> \projb ] = 0
\end{align} 
where $\proj_>$ is still given by the integral equation (\ref{eq:P_more}). 

\section{The case of $r = 1$}    \label{sec:r1}
Now that we have a hammer, we look for a nail. The simplest nontrivial case for our formalism corresponds to $r = 1$. In the following, we shall see that this case offers further simplifications, as well as additional structure that is not present, or at least not immediately obvious, in the higher rank cases. 

\subsection{Transfer matrix}    \label{sec:r1_tm}
Let us start with an explicit calculation of the transfer matrix using \eq{eq:tmat_def}. For $r = 1$, $\Xi$ is a $1 \times 1$ matrix, i.e, a number, which we can set to $1$ by a suitable normalization of the recursion relation\footnote{This also rescales $\ve$, but that is analogous to writing the energy in \emph{units} of the hopping energy, a practice that is quite standard.}.  
We write the $2 \times 2$ transfer matrix as  
\beq T = \frac{1}{|\green_{vw}|} \left( \begin{array}{cc} 1 & \quad  -\green_{ww} \\ \green_{vv} & \quad  -\det\left(\left. \green\right|_{\text{span}(\vv, \vw)}\right) \end{array} \right). \label{eq:rank1_tm} \eeq 
where we have defined the restricted determinant as
\[ \det\left(\left. \green\right|_{\text{span}(\vv, \vw)}\right)  = \left|  \begin{array}{cc} \green_{vv} & \green_{vw} \\ \green_{wv} & \green_{ww} \end{array} \right| = \green_{vv}\green_{ww} - \green_{vw}\green_{wv}. \] 
The prefactor becomes $|\green_{vw}|$ after we gauge away the phase of $T$ by the gauge transform from \eq{eq:gaugetr}, as
\beq T \to e^{i \theta} T = e^{i \arg \left(\green_{vw}\right)} T. \eeq
Also, the conditions on the Green's function in eq. \ref{eq:green_prop} reduce to 
\beq \green_{vv}, \; \green_{ww} \in \real, \quad \green_{vw}^* = \green_{wv}. \eeq 
As $T$ is real and $\det T = 1$, $T \in \Sp(2, \real) \cong SL(2, \real)$. Hence, all transfer matrices for $r=1$ are symplectic, by construction\footnote{This may not be true for higher ranks, as $\Sp(n, \real) \subset SL(n, \real)$ is a proper subset for $n > 2$.}.   

We can write out the Floquet discriminant, the trace of the transfer matrix, as 
\beq \Delta(\ve) = \frac{1}{|\green_{vw}|} \left[ 1 -  \det\left(\left. \green\right|_{\text{span}(\vv, \vw)}\right) \right]. \eeq 
The band edges are given by $\Delta(\ve, \vk_\perp) = \pm 2$, which can be used to solve for $\ve(\vk_\perp)$, at least locally. Note that $\ve$ enters the calculation only as $\green = (\ve \id - M)^{-1}$, which is a rational function of $\ve$, so that solving for the band edges is equivalent to finding the zeros of a polynomial in $\ve$. 

The case of the edge states is also particularly simple for $r = 1$.  As the subspaces corresponding to $|\rho_s| \lessgtr 1$ are either $\emptyset$ or 1-dimensional, the decay condition requires that $\Phi_1$ be an eigenvector:
\beq \proj_< \Phi_1 = \Phi_1 \implies \Phi_1 \propto \varphi_1, \eeq
where $\varphi_1$ is the eigenvector of $T$ corresponding to the eigenvalue $|\rho_1| < 1$, i.e, in the bulk gap. We get the analogous condition for the right edge. 

We start off with a somewhat less restrictive condition which can be represented in a neat geometric way. We demand simply that $\Phi_1$, $\Phi_N$ be eigenvectors of $T$ (hereby referred to as the \emph{eigenvalue condition}). We note that for any antisymmetric $\symp\in \mathbb{R}^{2\times 2}$ and $\varphi \in \cmplx^2$, we have
\beq  \varphi^T \, \symp \; \varphi  = 0. \eeq 
The eigenvalue condition ($T \varphi \propto \varphi$) can then be equivalently expressed as
\beq f(\ve, \vk_\perp) \equiv \varphi^T \; \symp \cdot T(\ve, \vk_\perp) \; \varphi  = 0. \label{eq:evans} \eeq 
In dynamical systems literature, the function $f$ is often referred to as the Evans function\cite{jared_defect_modes}. This is equivalent to the statement that $\varphi$ satisfies \emph{either} the left \emph{or} the right decay condition, i.e, it lies entirely in the growing or the decaying subspace. We can later check whether these states are physical by computing the corresponding eigenvalues.


The hard boundary condition (\eq{eq:phi1_exp} and (\ref{eq:phiN_exp})) implies that
\beq \Phi_1 = \left( \begin{array}{c} 1 \\ 0 \end{array} \right), \quad \Phi_N =  \left( \begin{array}{c} 0 \\ 1 \end{array} \right), 
\label{eq:rank1_bdry} \eeq
where we have exercised our right to scale $\Phi_{1,N}$ by an arbitrary complex number. This then suggests that we choose
\beq 
\symp = \left( \begin{array}{cc} 0 & 1 \\ -1 & 0 \end{array} \right) 
\label{eq:symp}\eeq 
such that $\symp$ is the symplectic form with respect to the $\bbe$ and $\bal$ subspaces. Such a choice of $\symp$ automatically incorporates the hard boundary data into the Evans function.

We term a solution $\ve_{edge}(\vk_\perp)$ of the equations (\ref{eq:evans}),(\ref{eq:rank1_bdry}),(\ref{eq:symp}) for either $\varphi= \Phi_{1,N}$ as the \emph{edge spectrum}.  Note that this includes the physical as well as the unphysical edge states, as defined in \S \ref{sec:tmat_appl}. These conditions describe a curve in the $(\ve, \vk_\perp)$ space, which has an associated winding number.

\subsection{Hofstadter model} \label{sec:r1_HS}
We start off by repeating Hatsugai's\cite{hatsugai_cbs} calculation in our formalism. The Hofstadter Hamiltonian, after a partial Fourier transform along $y$, the direction with PBC, is given by 
\beq \hlt = - \sum_{n} \left[ \cd_n c_{n+1} + \cd_{n+1} c_n + 2 \cos (k_y - 2 \pi n \phi) \cd_n c_n 	\right] \eeq 
where $\phi = p/q, \; p, q \in \intg^+$. The system is periodic with period $q$, and we get a gapped system with edge states for odd $q$. We club together $q$ physical sites to make a supercell, so that $J$ has all entries equal to zero except $J_{1q} = 1$, while $M$ has $2 \cos (k_y - 2 \pi n \phi)$ as its diagonal entries while it has 1's on the first diagonal. For instance, for the simplest nontrivial case of $\phi = 1/3$, these matrices are 
\beq J = \left( \begin{array}{ccc}  0 & 0 & 1 \\    0 & 0 & 0 \\   0 & 0 & 0 \\ \end{array} \right) \eeq
and 
\beq M = 2\left(\begin{array}{ccc}  \cos \left(k_y-\frac{2\pi }{3}\right) & 1 & 0 \\  1 & \cos \left(k_y+\frac{2\pi }{3}\right) & 1 \\  0 & 1 &  \cos (k_y) \\ \end{array} \right) \eeq 
Going through the machinery above, we obtain the Floquet discriminant as 
\beq \Delta(\ve, k_y) = \ve^3 - 6 \ve - 2 \cos(3 k_y). \eeq 

\begin{figure}
  \centering  
  \includegraphics[width=\columnwidth]{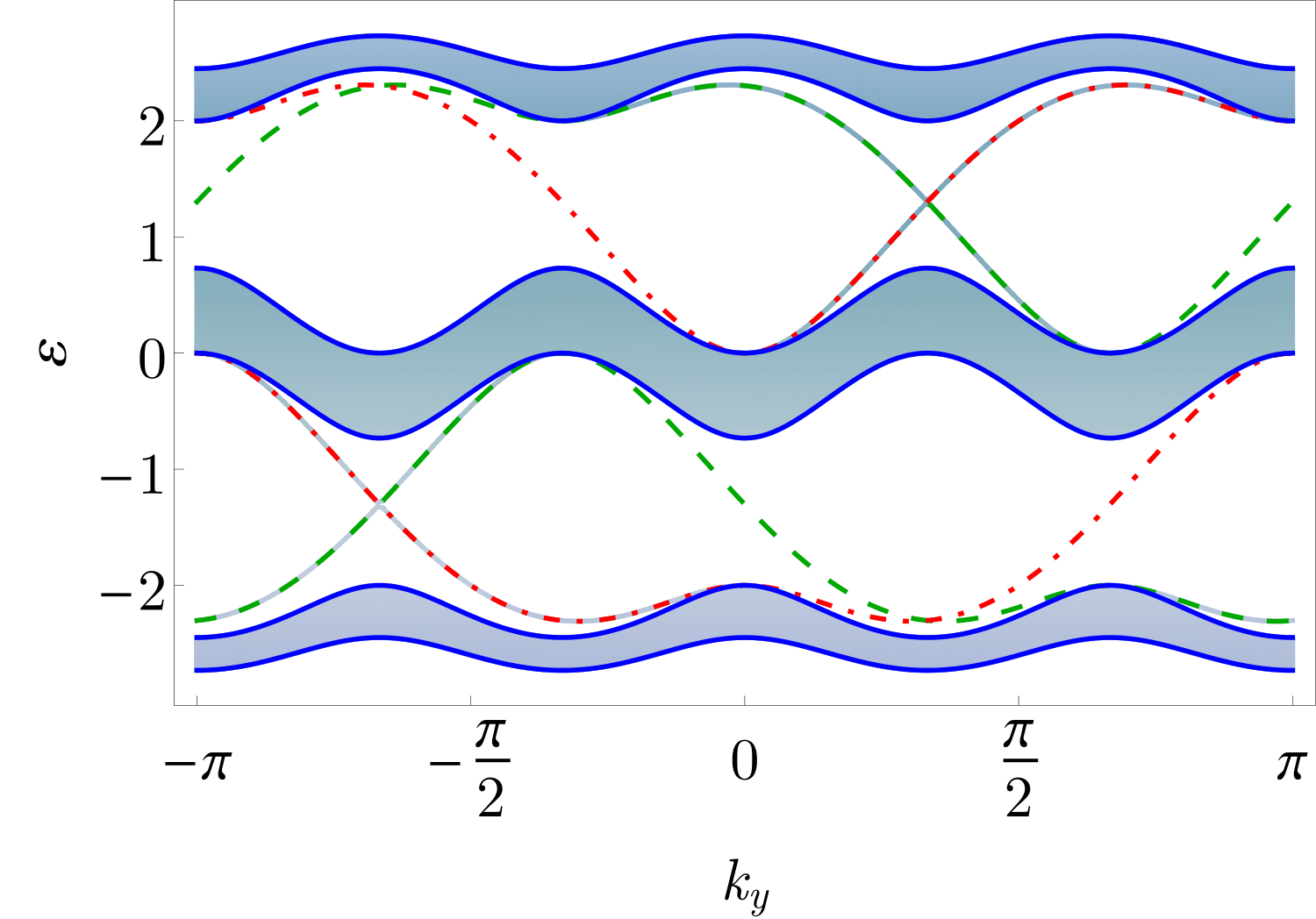}  
  \phantom{blah} 
  \includegraphics[width=\columnwidth]{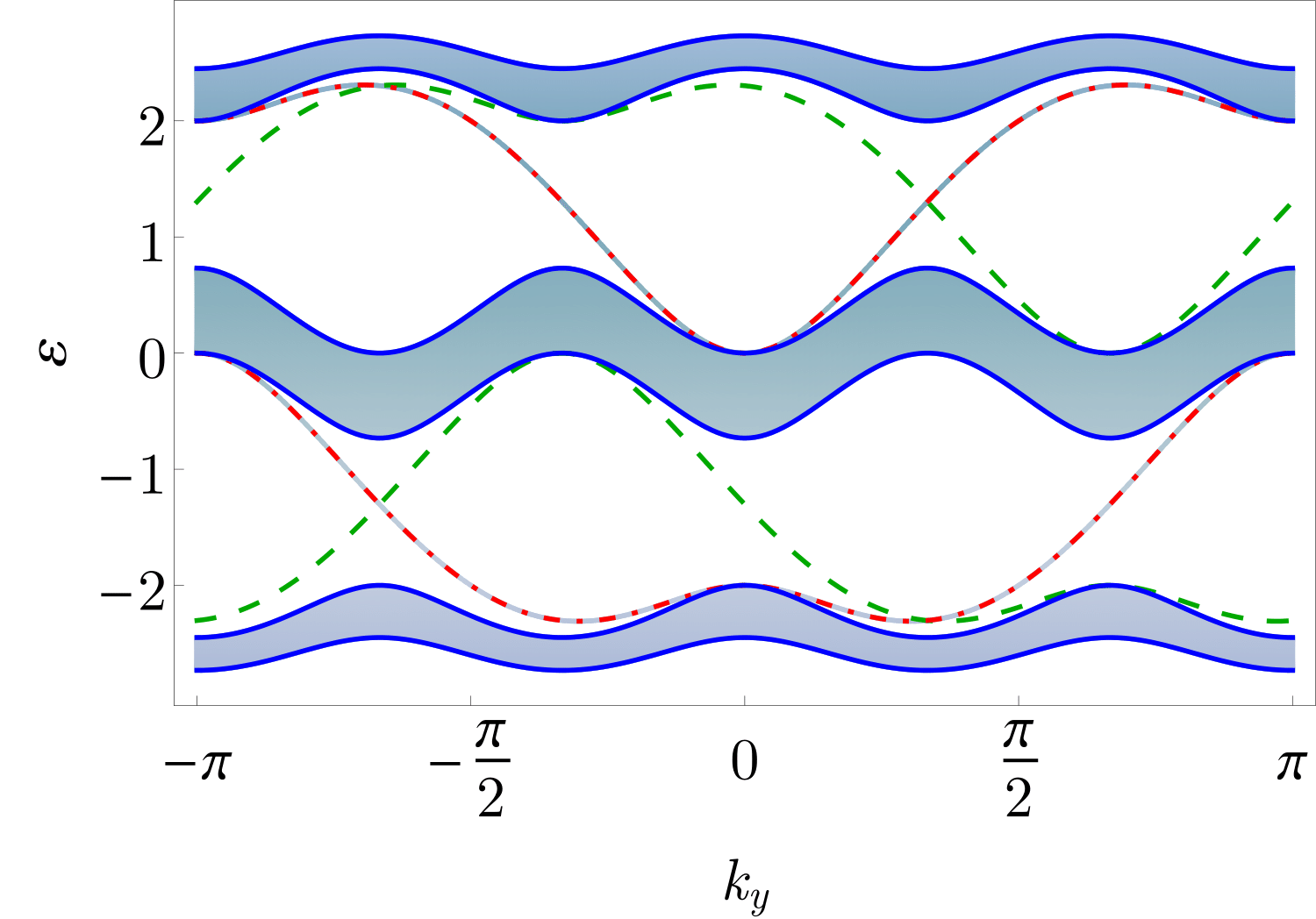}  
  \caption{(color online) The spectrum of Hofstadter model, with the band edges (dark blue) computed using the transfer matrix formalism and the left and right edge state dispersion (dashed and dashed-dot) from the Evans equation (\ref{eq:evans}), overlaid on the spectrum computed using exact diagonalization for a (top) commensurate and (bottom) incommensurate system. Note that in the latter case, the edge states seen in exact diagonalization exactly follow the winding right edge state obtained from the transfer matrix.}    
  \label{fig:HS_spectrum}
\end{figure}

In general, for arbitrary $q$, $\Delta$ is a polynomial in $\ve$ of order $q$. The edge state calculation is identical to Hatsugai's, so we shall not discuss it in any detail. However, we emphasize his remark  that if the total number of sites is \emph{commensurate} with the flux $\phi$, i.e, a multiple of $q$, then for a given $k_y$, we either get edge states on both left and right edges, or no edge states at all. In order to have an edge state for all $k_y$, which will have an associated winding, we need to consider a system with the number of sites \emph{incommensurate} with the flux\footnote{Our usage of the words ``commensurate'' and ``incommensurate'' seems to be opposite that of Hatsugai!} (see fig. \ref{fig:HS_spectrum}). 

In our picture, for the latter case, the number of supercells is not an integer. This makes physical sense for a Hofstadter model as the $\nuc$ degrees of freedoms per supercell are physical sites for the Hofstadter model, so that we can remove those sites. In general, the degrees of freedom inside a supercell are not physical sites. However, we shall see that the number of supercells being fractional still formally makes sense, and hence we can contrive a (potentially unphysical) boundary conditions for those cases which will exhibit the winding of the edge states. We shall hereafter use the word \emph{incommensurate} (with the superlattice) to refer to the cases where the number of supercells is not an integer.

\subsection{Natural basis  and ``unfolding''}   \label{sec:r1_basis}
Hatsugai's calculation of the transfer matrix worked because of the fact that the system had nearest neighbor hopping. Before we proceed to further examples, we stop to consider the implications of nearest neighbor hopping inside a single supercell, which implies, in our notation, that $M$ is tridiagonal, and $J = \uvec_1 \otimes \uvec_\nuc$, as in the Hofstadter model. Explicitly, if
\beq J = \left( \begin{array}{cccc} 
0 & 0 & \ldots & 1 \\ 
0 & 0 & \ldots & 0 \\
\vdots & \vdots & \ddots & \vdots \\ 
0 & 0 & \ldots & 0 
\end{array} \right), \quad 
M = \left( \begin{array}{cccc} 
\mu_1 & \tau_1 & \ldots & 0 \\ 
\tau_1 & \mu_2 & \ldots & 0 \\
\vdots & \vdots & \ddots & \vdots \\ 
0 & 0 & \ldots & \mu_\nuc
\end{array} \right), \label{eq:cond_trid}  \eeq 
with $\mu_n, \tau_n \in \real \; \forall \, n = 1, 2, \dots \nuc$, where we have defined $\tau_\nuc = 1$, then we can write the recursion relation as 
\beq \tau_n \phi_{n+1} + \mu_n \phi_n + \tau_n \phi_{n-1} = \ve \phi_n, \eeq 
where $\tau_n$ and $\mu_n$ are periodic with period $\nuc$. Following Hatsugai and others\cite{dh_lee_cbs, schulman_cbs}, we can compute the transfer matrix as
\beq T = \prod_{n = 1}^\nuc \tmat_n, \quad \tmat_n = \left( \begin{array}{cc} -\frac{1}{\tau_n} (\ve - \nu_n) \quad \ & -1 \\ 1 \quad & 0 \end{array} \right), \eeq 
where $\tmat_n$ is the transfer matrix from site $n$ to site $n+1$ with periodicity $\tmat_{n+\nuc} = \tmat_n$. This construction always results in the transfer matrix being polynomial in $\ve$, as it simply involves a product of matrices linear in $\ve$. Subsequently, the Floquet discriminant, $\Delta = \tr T$ is a polynomial in $\ve$, a fact which we shall use when discussing the winding in \S \ref{sec:winding}. 

Now, an interesting aspect of our computation of the transfer matrix is that it is basis independent, as we have not used the explicit form of $J$ or $M$ anywhere in this calculation. Can we choose a basis for our system where $J$ and $M$ are of the form in \eq{eq:cond_trid}? Let us assume so, and let such a basis of $\cmplx^\nuc$ be $\{ \basis_i , \; i = 1, \dots \nuc\}$. 

We start off by noting that $J$ a natural orthonormal basis for $\cmplx^\nuc$, viz, 
\beq \cmplx^\nuc = \text{span}\{\vv, \vx_j, \vw\}, \quad j = 2, \dots \nuc - 1. \eeq  
As $J = \vv \cdot \vw^\dagger$, in this basis, $\langle \vv J \vw \rangle = 1$, and all other matrix elements of $J$ are zero, which is what we demand in \eq{eq:cond_trid}. Hence, we set $\basis_1 = \vv$ and $\basis_\nuc = \vw$. Now we can represent $M$ in this basis as 
\beq M = \left( \begin{array}{ccc} 
\vv^\dagger M \vv & \vv^\dagger  M X & \vv^\dagger M \vw \\ 
X^\dagger M  \vv & X^\dagger M X & X^\dagger M  \vw \\ 
\vw^\dagger M \vv & \vw^\dagger  M X & \vw^\dagger M \vw  
\end{array}  \right). \label{eq:M_vw}   \eeq
Note that we still have freedom to choose $\{ \basis_j \}$ as linear combinations of $\vx_j, \; j = 2, \dots \nuc - 1$. We seek to turn $M$ tridiagonal by this freedom. Any such choice would correspond to a unitary transform of $M$ as defined in \eq{eq:M_vw} only in the subspace spanned by $\vx_j$'s, i.e, $M \to \uop \cdot M \cdot \uop^\dagger$, where 
\beq \uop = \left( \begin{array}{ccc} 1 & 0 & 0 \\ 0 & \uop_X & 0 \\ 0 & 0 & 1 \end{array}  \right) \cdot \uop_J, \eeq 
where $\uop_J$ simply sets $\basis_1 = \vv$ and $\basis_\nuc = \vw$, while leaving the $X$ subspace invariant. We seek a suitable choice of $\uop_X \in U(\nuc-2)$ which tridiagonalizes $M$. For $\uop \cdot M \cdot \uop^\dagger$ to be real tridiagonal, we need that 
\begin{align}
 &  \vw^\dagger M \vv = 0, \nonumber \\
 & \uop_X \left( X^\dagger M X \right) \uop_X^\dagger \text{ is tridiagonal}, \nonumber \\ 
 & \uop_X \left( X^\dagger M  \vv \right) \propto \left( 1, 0 \ldots 0 \right)^T \in \real^{\nuc - 2} \nonumber \\ 
 & \uop_X \left( X^\dagger M  \vw \right)  \propto \left( 0, 0 \ldots 1\right)^T \in \real^{\nuc - 2}. 
\end{align}
The first condition simply means that the degrees of freedom in a supercell connected to the next and the previous supercells are not directly connected to each other (except for when $\nuc = 2$). This can always be arranged by taking a big enough supercell. 

For the second condition, we note that any Hermitian matrix can be reduced to a real tridiagonal form using the Lanczos/Householder algorithm\cite{press-flannery_book, burden-faires_book}. We choose $\uop_X$ to be the (non-unique) unitary matrix that tridiagonalizes the Hermitian matrix $X^\dagger M X$. 

Finally, the question of tridiagonalizing $M$ has reduced to the question of satisfying the conditions for rotations of $X^\dagger M \vv$ and $X^\dagger M \vw$, which should be checked explicitly for a given case, employing the nonuniqueness of $\uop_X$ for tridiagonalization.  

If such a unitary transform $\uop$ does exist, we shall refer to such a transformation as \emph{unfolding} the model to a 1D chain. A quick survey of the matrix $M$ in this basis reveals various restrictions on the transfer matrices. For instance, if $\vv^\dagger M \vv$ and $\vw^\dagger M \vw$ are real, it immediately follows that the entries of the transfer matrix are real. Furthermore, we can often glean information about the edge states by looking at the hopping of the resulting 1D chain. We discuss that explicitly in \S \ref{sec:r1_CI_edge}

\subsection{Chern insulator}     \label{sec:r1_CI}
Finally we proceed to the case of Chern insulator, where we compute the transfer matrix. This model turns out to be the \emph{drosophila melanogaster} (fruit fly) of topological states, as we shall show in the following calculations.  Furthermore, as the transfer matrix is quadratic in $\ve$, we can carry out much of the computation analytically. 

\subsubsection{Transfer matrix}
Consider the 2D lattice Hamiltonian
\beq \hlt =  \sin k_x \sigma^x +  \sin k_y \sigma^y +  (2-m-\cos k_x - \cos k_y) \sigma^z \eeq
with an edge along the $x$ axis (See Appendix \ref{app:chern} for details of the model as well as a direct computation of the transfer matrix). We begin with the identification
\beq J =  \frac{1}{2i} \left( \sigma^x  - i \sigma^z \right), \quad M =  \sin k_y \sigma^y + \Lambda(k_y) \sigma^z,  \eeq
with $\Lambda(k_y) = 2 - m - \cos k_y$, and compute
\beq T =  \frac{1}{|\Lambda(k_y)|} \left( \begin{array}{cc} -\ve^2 + \Lambda^2(k_y) + \sin^2 k_y \quad &  \ve - \sin k_y \\ -(\ve + \sin k_y) \quad & 1 \end{array} \right). \eeq 
Note that this is not identical to the transfer matrix obtained in \eq{eq:CI_tmat}, but is related by a similarity transform, as they both have the same determinant and trace, 
\beq \Delta(\ve, k_y) =\frac{1 - \ve^2 + \Lambda^2(k_y) + \sin^2 k_y}{|\Lambda(k_y)|}.  \label{eq:CI_trace} \eeq
This is the main advantage of calculating in a basis-independent fashion: the transfer matrix itself depends on the choice of basis, but the Floquet discriminant, which governs the bulk properties, is a basis-independent quantity. 

\begin{figure}
  \centering
  \includegraphics[width=\columnwidth]{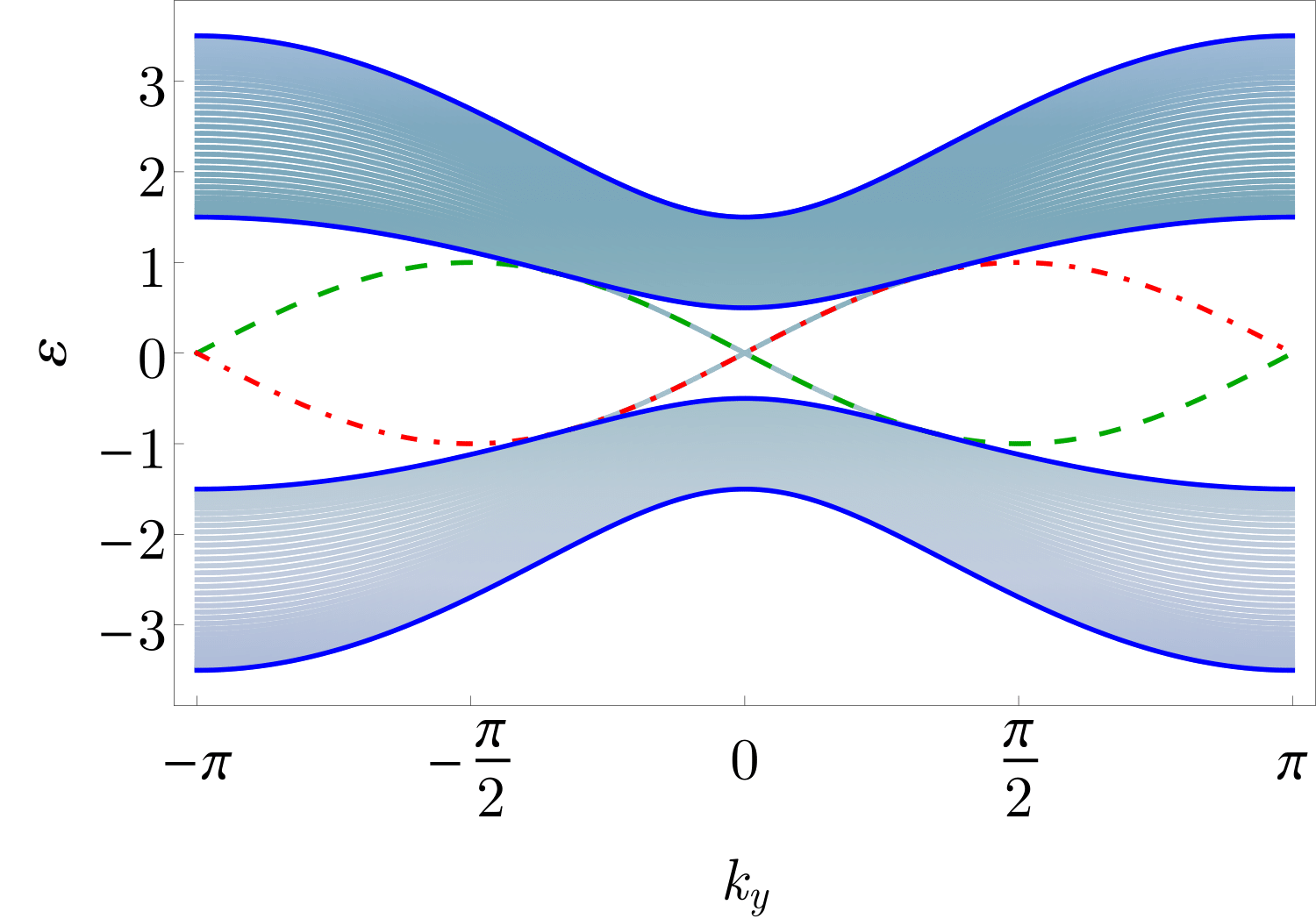} 
  \phantom{blah}
  \includegraphics[width=\columnwidth]{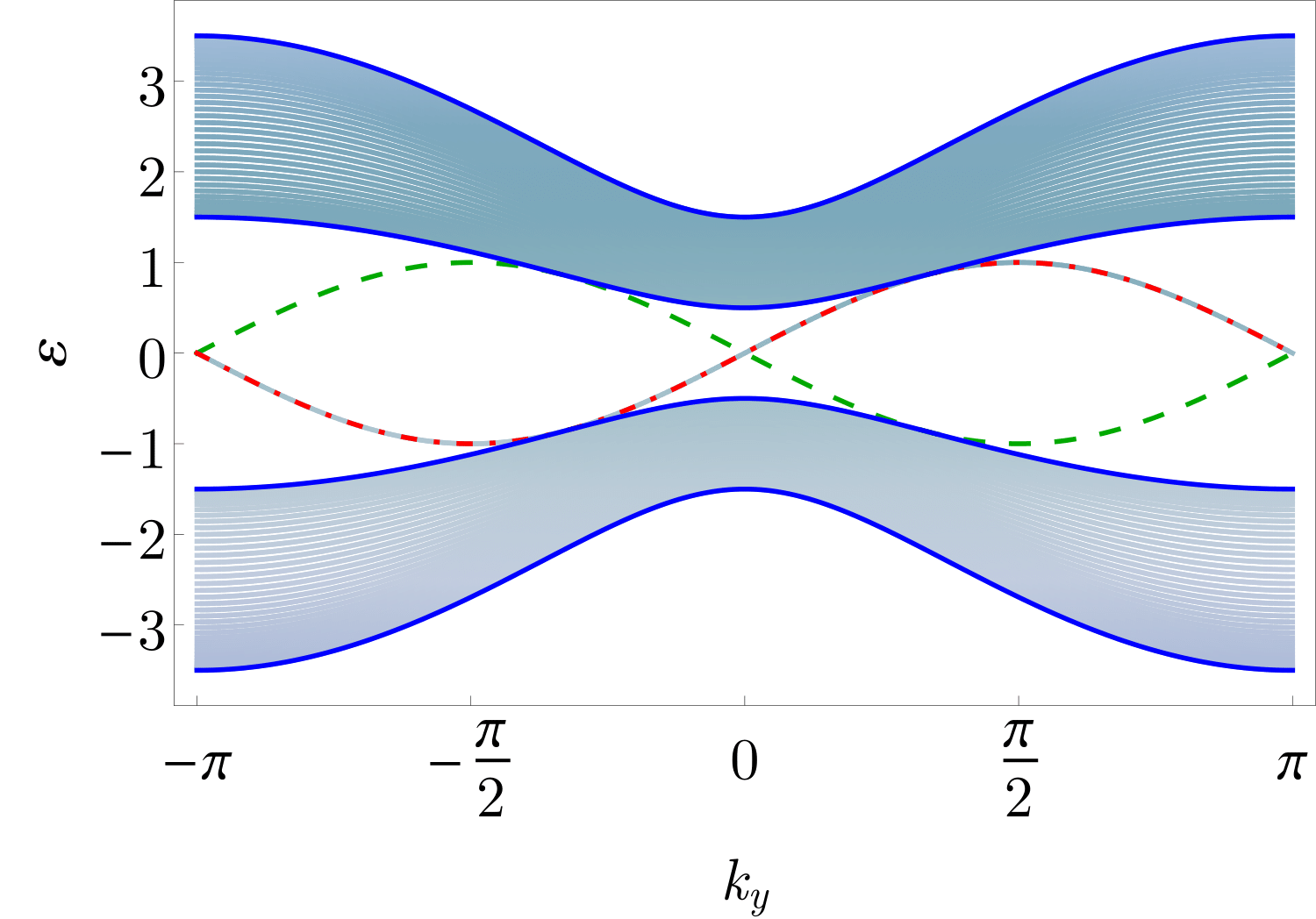}  
  \caption{(color online) The spectrum of Chern insulator for $m = 0.8$, with the band edges (dark blue) computed using the transfer matrix formalism and the left and right edge state dispersion (dashed and dashed-dot) from the Evans equation (\ref{eq:evans}), overlaid on the spectrum computed using exact diagonalization for a (top) commensurate and (bottom) incommensurate system. Note that in the latter case, the edge states seen in exact diagonalization exactly follow the winding right edge state obtained from the transfer matrix.} 
  \label{fig:CI_spectrum}
\end{figure}

We can compute the band edges from this expression, using $\Delta = \pm 2$, to get 
\beq \ve^2 =  \sin^2 k_y + (2 \mp 1 - m - \cos k_y)^2. \eeq 
The bands are symmetric under $\ve \to -\ve$, and stretch between $\ve_{min} < |\ve| < \ve_{max}$, with 
\begin{align}
 \ve_{min} = & \; \sqrt{\sin^2 k_y + (1 - m - \cos k_y)^2 } \nonumber \\ 
 \ve_{max} = & \; \sqrt{\sin^2 k_y + (3 - m - \cos k_y)^2 }.   \label{eq:CI_band_edges}
\end{align}
for $0<m<2$. We can see that this agrees with the spectrum computed using exact diagonalization, as shown in Fig. \ref{fig:CI_spectrum}.

\subsubsection{Unfolding the 1D chain and SSH model}
For the Chern insulator, the unfolding to a 1D chain is particularly neat, as it leads to an alternating bond model, a quintessence of which is the Su-Schrieffer-Hieger(SSH) model\cite{ssh} for polyacetylene. We demonstrate this idea explicitly in the following. We start off with the Hamiltonian 
\beq \hlt = \sin k_x \sigma^x + \sin k_y \sigma^y + (2-m-\cos k_x - \cos k_y) \sigma^z. \eeq
We again identify
\beq J = \frac{1}{2i} \left( \sigma^x - i \sigma^z \right) \equiv  \vv \cdot \vw^\dagger, \eeq 
where
\beq 
\vv = \frac{1}{\sqrt{2}} \left( \begin{array}{c} -i \\ 1 \end{array} \right), \quad \vw = \frac{1}{\sqrt{2}} \left( \begin{array}{c} i \\ 1 \end{array} \right).  \label{eq:vw_CI_def}
\eeq
As $M$ is $2 \times 2$ and hence, by definition, \emph{tridiagonal}, the unfolding requires a unitary operator which takes $J$ to the desired form of \eq{eq:cond_trid}. Take a unitary operator $\uop$, defined as 
\beq \uop = \frac{1}{\sqrt{2}} ( i \id + \sigma ^x), \eeq 
so that $\uop \vv = \uvec_1$ and $\uop \vw = \uvec_2$. The lattice Hamiltonian transforms as $\hlt \to \hlt' =  \uop \hlt \uop^\dagger$, with
\beq  \hlt' = \sin k_x \sigma^x + \sin k_y \sigma^z - (2-m-\cos k_x - \cos k_y) \sigma^y \eeq
We now transform this Hamiltonian to the real space along $x$ to get
\begin{align}
  \hlt'(k_y) = & \; \sum_{n}  \bigg[  \vcd_{n+1} \left( \frac{-i\sigma^x + \sigma^y}{2} \right) \vc_n  \nonumber \\ 
  & -  \vcd_n \left( \frac{i\sigma^x + \sigma^y}{2} \right) \vc_{n+1}  \nonumber \\ 
  & + \vcd_n \left( \sin k_y \sigma^z - \Lambda(k_y) \sigma^y \right) \vc_n \bigg]  
\end{align}
where $\vc_n \equiv (c_n, \bar c_n)^T$. Redefining $\bar c_n = b_{2n}, c_n = b_{2n+1}$ and expanding the products, we get 
\beq \hlt'(k_y) = \sum_{n}  \bigg[ \left( -i \, \tau_n \bd_{n+1} b_n + \text{h.c.} \right) + \mu_n \bd_n b_n \bigg], \eeq
where 
\[ \mu_n =  (-1)^n \sin(k_y), \quad \tau_n = \begin{cases} \Lambda(k_y) & ; n = \text{even}, \\ 1 & ; n = \text{odd}. \end{cases} \]
Hence, by a basis transformation on the Chern insulator, we have obtained the Hamiltonian for a 1D chain with alternating bond strengths $1$ and $\Lambda(k_y)$. This is analogous to the situation with the SSH model, with the addition of an alternating on-site energy term.

\subsubsection{Edge states}   \label{sec:r1_CI_edge}
We can compute the edge spectrum explicitly. For the left edge, applying the Evans condition (\eq{eq:evans}), we get 
\beq 0 = \green_{vv} = \ve + \sin k_y \implies \ve_L(k_y) = -\sin k_y \eeq
while for the right edge, we have 
\beq 0 = \green_{ww} = -(\ve - \sin k_y) \implies \ve_R(k_y) = \sin k_y \eeq
These correspond to the states that satisfy a decay and a boundary condition, but not necessarily the right combination thereof (See Table \ref{tb:edge}). To check that, we will need to compute the eigenvalues of the transfer matrix of $T(\ve_{L,R}(k_y), k_y)$. For $\ve_L(k_y)$,
\beq T(\ve_L(k_y), k_y) =  \frac{1}{|\Lambda(k_y)|} \left( \begin{array}{cc} \Lambda^2(k_y)  \quad \ & -2 \sin k_y \\ 0 \quad & 1 \end{array} \right), \eeq
so that 
\beq T(\ve_L(k_y), k_y) \Phi_1 = |\Lambda(k_y)| \Phi_1, \eeq
where $\Phi_1 = (1, 0)^T$. Hence, for a hard boundary condition, this edge is physical if
\beq |\Lambda(k_y)| = |2 - m - \cos k_y| < 1 \eeq 
which implies that $1 - m < \cos k_y < 3 - m$. Hence, we have edge states for $\cos k_y > 1 - m$ if $m \in (0, 2)$ and $\cos k_y < 3 - m$ if $m \in (2, 4)$. 

Similarly, for the right edge, using a hard boundary condition, we get 
\beq T(\ve_L(k_y), k_y) \Phi_N = \frac{1}{|\Lambda(k_y)|} \Phi_N, \eeq
which is physical if 
\beq \left| \frac{1}{\Lambda(k_y)} \right| > 1 \implies |\Lambda(k_y)| < 1, \eeq 
which is identical to the condition for the left edge state. 

Using the SSH picture, the emergence of edge states is transparent: whenever one opens a boundary, one gets an edge state if the boundary cuts open a strong bond. Scanning as a function of $k_y$, we can see that the edge states vanish when the bonds change their relative strength, i.e, when $\Lambda(k_y) = 1 \implies \cos k_y = 1 - m$, which is what one obtains from more elementary means\cite{hughes-srinidhi_emres} or sees in exact diagonalization.

\begin{figure}
  \centering
  \includegraphics[width=\columnwidth]{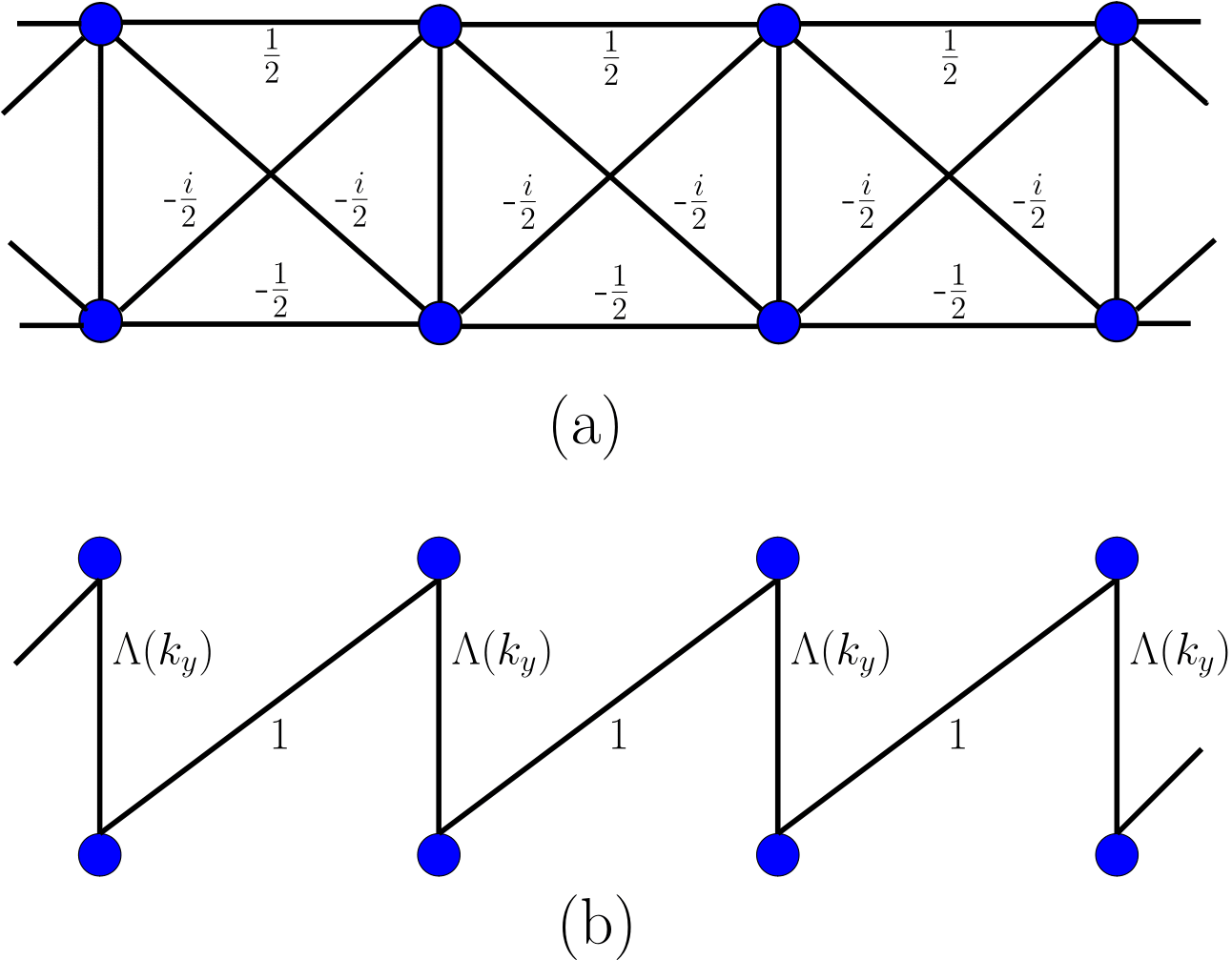}
  \caption{Unfolding the Chern insulator: In (a), we see the Chern insulator in the usual basis, treating the two degrees of freedom as sites. A change of basis in (b) transforms the model to a 1D chain with alternating hopping.}
\end{figure}

Furthermore, in the SSH model, the edge state appears at zero energy\cite{shen_book}. However, for the Chern insulator, we also have an on-site energy term  $\mu_n = (-1)^n \sin(k_y)$. Hence, the spectrum of the edge state is given by $\ve(k_y) = -\sin(k_y)$ for the left edge ($n = 1$) and $\ve(k_y) = \sin(k_y)$ for the right edge ($n = 2 \times \text{number of supercells}$), which is also what we got from a direct computation. 

If the SSH chain has an even number of sites, so that the number of sites is commensurate with the size of the supercell, the edge states always occur in pairs, i.e, either both at the left and right end or not at all. This corresponds to the physical situation, as the aforementioned \emph{sites} correspond to local spin/orbital degrees of freedom and hence always occur in pairs, which explains why the left and the right edges always \emph{switch off} at the same $k_y$ in the computation above. However, if one considers the incommensurate case where the SSH chain has an odd number of sites, there is an edge state for every $k_y$. Hence, if we allow such an (unphysical) boundary condition, we can expose the entire edge state in an exact diagonalization calculation, which is associated with the winding number on the Riemann surface (see fig. \ref{fig:CI_spectrum}).

\subsection{Further examples}    \label{sec:r1_eg}
\begin{figure}
  \centering
  \includegraphics[width=\columnwidth]{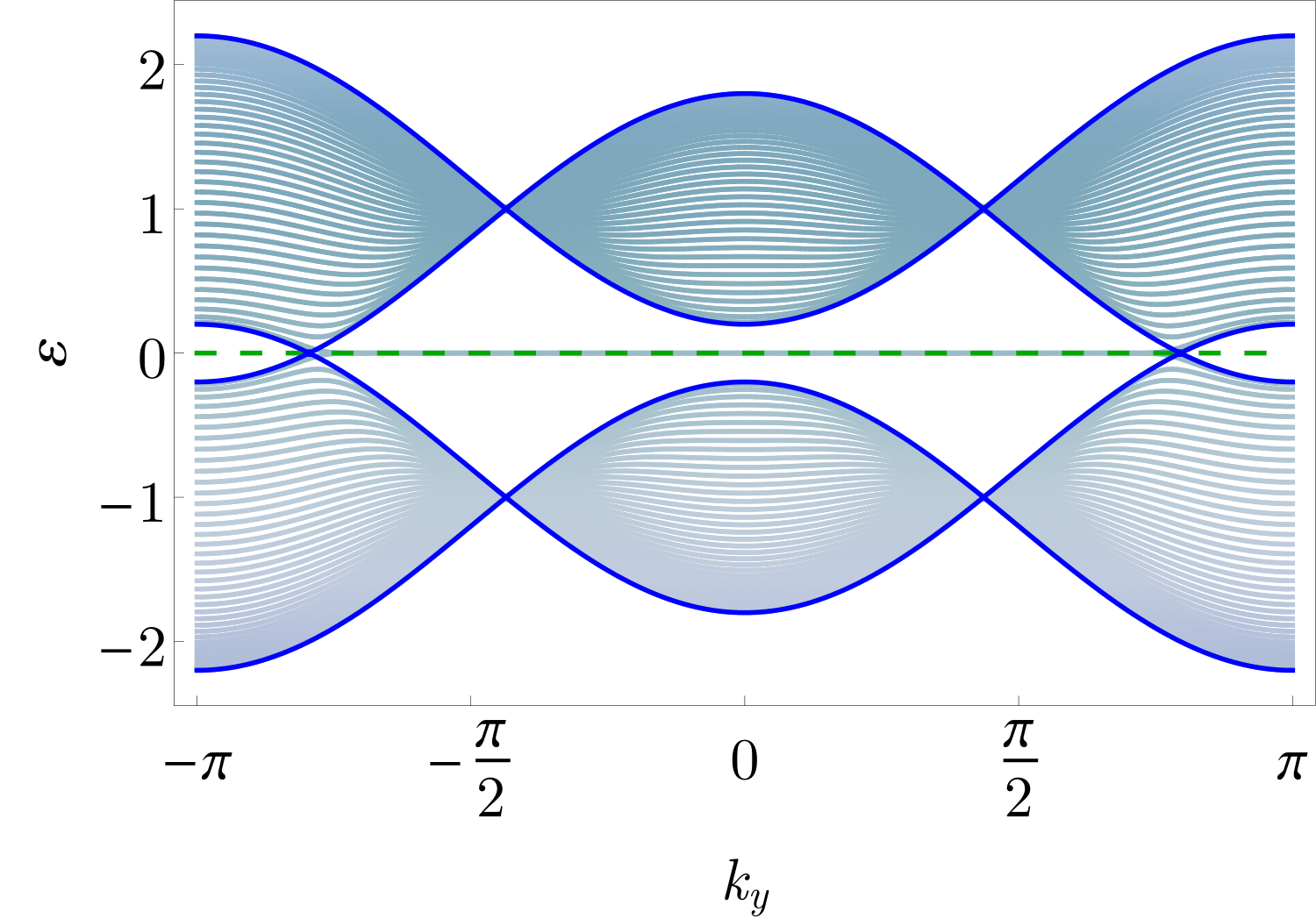}   
  \phantom{blah}
  \includegraphics[width=\columnwidth]{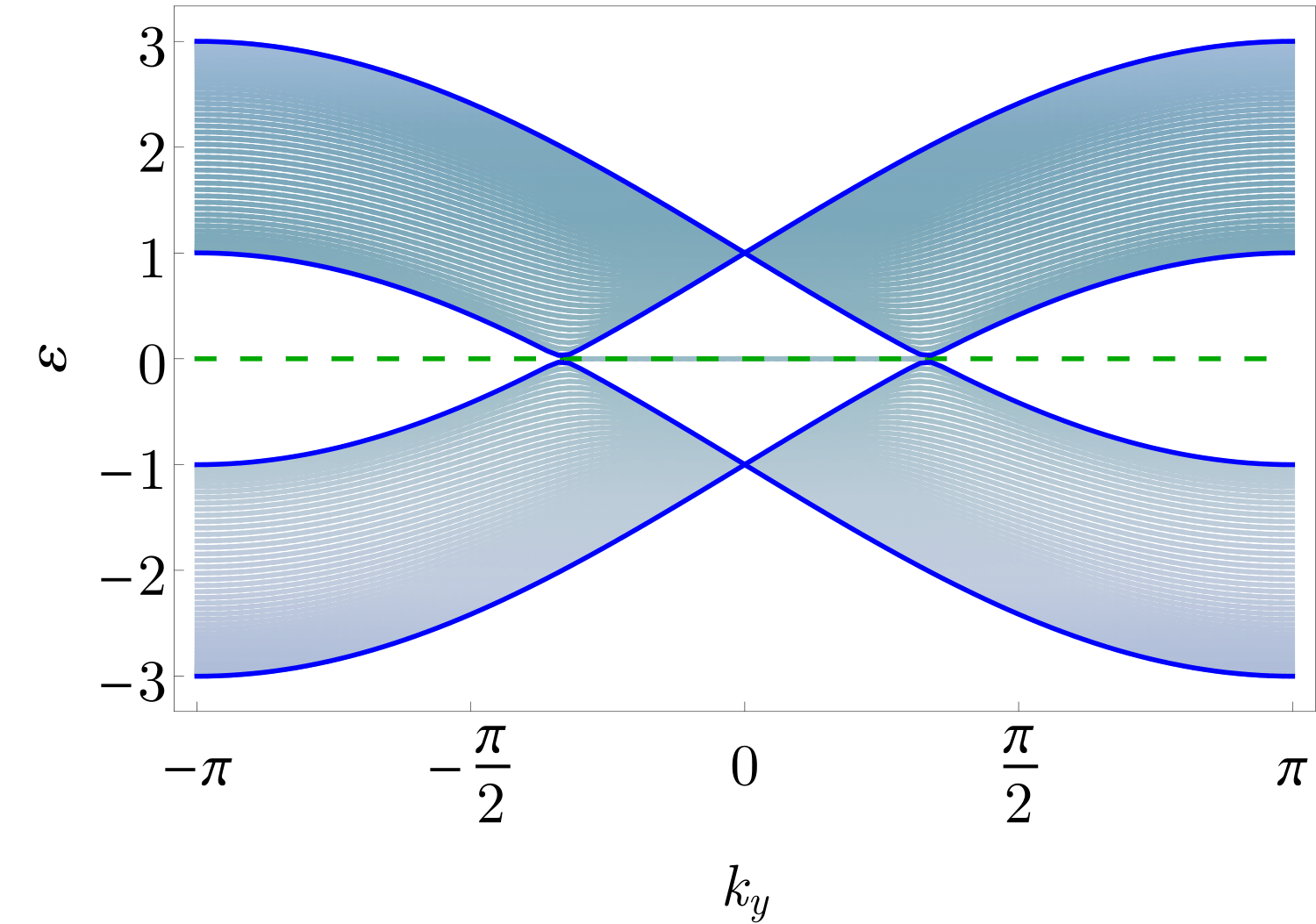} 
  \phantom{blah}
  \includegraphics[width=\columnwidth]{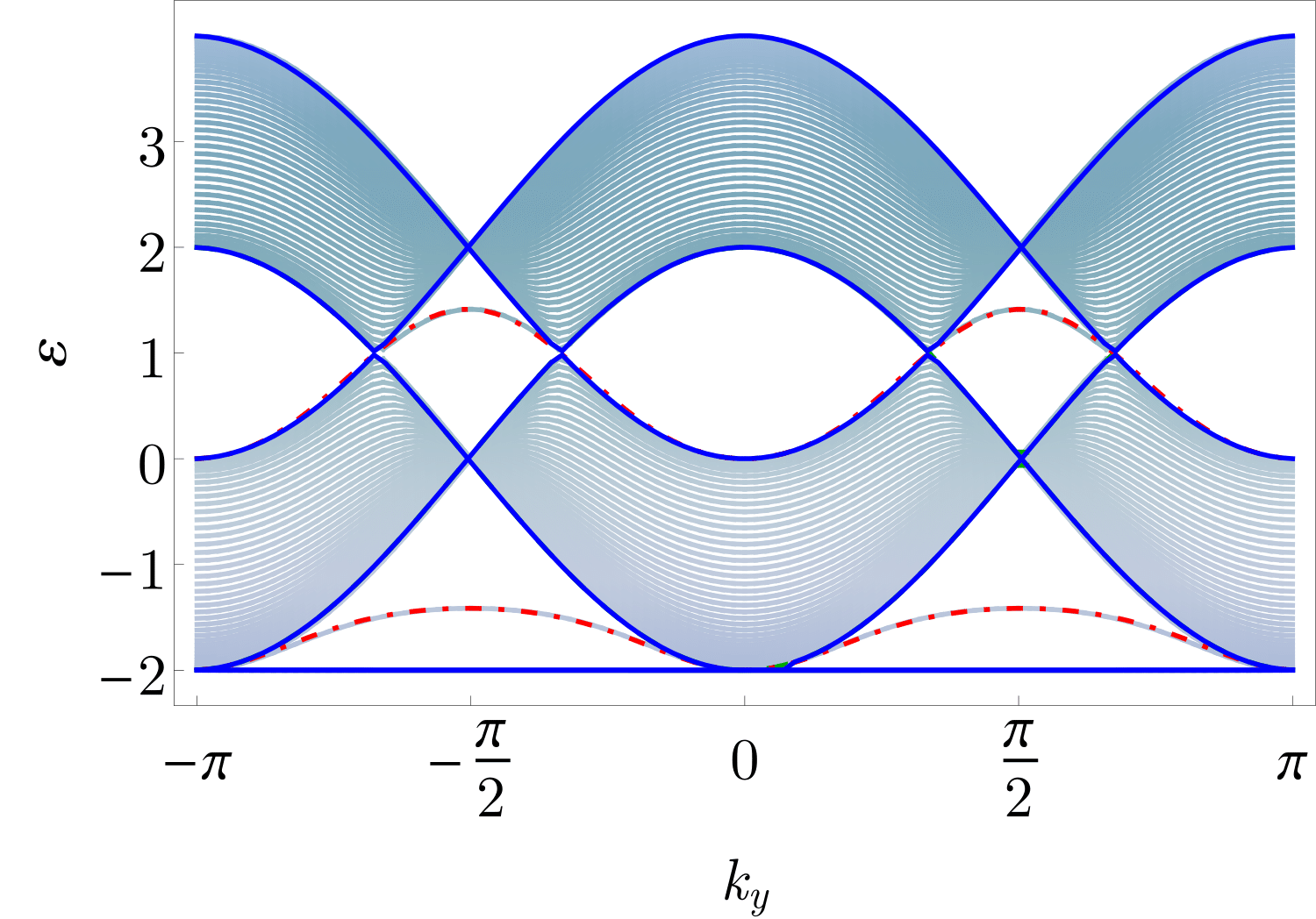} 
  \caption{(color online) The spectrum of (top) Dirac Semimetal,  (middle) Graphene and (bottom) Kagome semimetal. See \S \ref{sec:r1_eg} and table \ref{tb:eg} for details.}     \label{fig:plots}
\end{figure}

Using the transfer matrix construction, the calculation of bulk bands as well as edge states becomes simply a matter of identifying the $J$ and $M$ matrices. We list these matrices for some of the well known topological and semimetal phases in table \ref{tb:eg}. The corresponding band structures and edge states, superimposed over the exact diagonalization result, are collected in Fig. \ref{fig:plots}. 

The parametrization for the case of a square lattice with an edge along a side of the square is straightforward, as we simply identify the direction normal to the edge as $x$ and the other direction as $y$. The same idea works for nonsquare lattices, but it needs some care to define the $J$ and $M$ matrices. In the following, we discuss the identification for kagome semimetal, which had been previously analyzed using less general methods\cite{wang_kagome_edge}. Consider a tight binding model with nearest neighbor hopping on a kagome lattice, which is described by the Bloch Hamiltonian\cite{guo-franz_kagome, wen_kagome}
\beq \hlt(\vk) = 2 \left( \begin{array}{ccc} 
0 & \cos k_1 & \cos k_3 \\ 
\cos k_1 & 0 & \cos k_2 \\
\cos k_3 & \cos k_2 & 0 
\end{array} \right), \eeq 
where $ k_i = \vk \cdot {\bf a}_i$, and
\beq  {\bf a}_1 = \vecenv{1}{0}, \; {\bf a}_2 = \frac{1}{2} \vecenv{-1}{\sqrt{3}}, \; {\bf a}_3 = -\frac{1}{2} \vecenv{1}{\sqrt{3}}, \eeq 
are the lattice vectors corresponding to the three bonds. 

Let us define the $x$ and $y$ directions to be along the unit vectors  
\beq \uvec_x =  \frac{1}{2} \vecenv{\sqrt{3}}{-1}, \quad \uvec_y =  \frac{1}{2} \vecenv{1}{\sqrt{3}}, \eeq 
so that $\vk = k_x \uvec_x + k_y \uvec_y$. We then write $k_i$ in terms of $k_x$ and $k_y$, and decompose the Bloch Hamiltonian as 
\beq \hlt(\vk) = e^{i \sqrt{3} k_x/2} J(k_y) + M(k_y) + e^{- i \sqrt{3} k_x/2} J^\dagger(k_y), \eeq
where $J$ and $M$ are independent of $k_x$. 

\begin{table*}
\setlength{\tabcolsep}{20pt}
\renewcommand{\arraystretch}{4}
 \begin{tabular}{l*2{>{\renewcommand{\arraystretch}{1.2}}c}}
 Model & $J$ & $M$ \\ 
 \hline\hline 
 Chern Insulator & $ \displaystyle \frac{1}{2i} (\sigma^x - i \sigma^z) $  & $ \sin k_y \sigma^y + \Lambda(k_y) \sigma^z $    \\
 \hline
 Dirac Semimetal & $ \displaystyle \frac{1}{2i} (\sigma^x - i \sigma^z) $  & $ \Lambda(k_y) \sigma^z $  \\
 \hline 
 Graphene & $\left( \begin{array}{ccc} 0 & 1 \\ 0 & 0 \end{array} \right)$ & $\left( \begin{array}{cc} 0 & 1 - e^{i k_y} \\ 1 - e^{-i k_y}  & 0 \end{array} \right)$ \vspace{0.1in} \\ 
 \hline 
 Kagome Semimetal & $ 
 \begin{pmatrix}
 0 & 0 & 0 \\
 \mathrm{e}^{i k_y / 2} & 0 & \mathrm{e}^{-i k_y / 2} \\
 0 & 0 & 0
 \end{pmatrix} $ & 
 $\begin{pmatrix}
 	0 & \mathrm{e}^{ i k_y / 2} & 2 \cos k_y \\
 	\mathrm{e}^{-i k_y / 2} & 0 & \mathrm{e}^{i k_y / 2 } \\
 	2 \cos k_y & \mathrm{e}^{-ik_y/2} & 0
 \end{pmatrix}$  \vspace{0.1in} \\
 \hline
\end{tabular}
\caption{A list of $J$ and $M$ matrix for some of the well-known topological and semimetal states. The corresponding spectra are plotted in Fig. \ref{fig:plots}} \label{tb:eg}
\end{table*}
 
\section{Riemann surface and windings}   \label{sec:winding}     
It is well known that the edge states of a topological phase carry topological characteristics dictated by the bulk, which prevents them from being gapped out. Despite the strong dependence of the edge spectrum on the precise boundary condition, the number of (signed) crossings of a given energy level in the band gap is a topological invariant, equal to the bulk Chern number. The proof of this so called \emph{bulk-boundary correspondence} is highly nontrivial\cite{baldes_disorder_IQHE}, and has been worked out in detail for the clean limit only in certain specific cases\cite{xlq-scz_bulk_bdry, vasudha-roger_tm}. However, an alternative perspective, due to Hatsugai, identifies the topological invariants of the edge states as winding numbers of the edge states around certain holes in the (complex) energy Riemann surface. He also provides a proof of this correspondence\cite{hatsugai_cbs_PRL}. 

In this section, following Hatsugai, we describe the geometry associated with the transfer matrices. The central purpose of this analysis is to obtain a better understanding of the topological nature of the edge states.

\subsection{The two complexifications}
In the Bloch analysis of discrete periodic systems, we usually restrict ourselves to real energies and momenta, which correspond to plane wave eigenstates. However, in this section, we shall see that there is much to be gained by allowing them to be complex (``complexifying'' them). In the following, we shall only describe the situation for $r = 1$. Furthermore, we shall restrict ourselves to a system in 2 spatial dimensions, with hard boundary conditions along $x$ and periodic boundary conditions along $y$, so that the transverse momentum is $\vk_\perp = k_y \in S^1$. 

Consider, then, a $2 \times 2$ transfer matrix for a 2-dimensional system, $T(\ve, k_y)$. The eigenvalues of the transfer matrix are 
\beq \rho_\pm = \frac{1}{2} \left[ \Delta \pm \sqrt{\Delta^2 - 4} \right], \quad \Delta = \tr T, \label{eq:tm_soln}\eeq
which satisfy $\rho_+ \rho_- = \det T = 1$. Following the Bloch ansatz, we can put $\rho_+ = e^{ik_x} \implies \rho_- = e^{-ik_x}$, so that $k_x$ is a function of $(\ve, k_y)$. In the standard Bloch theory, $(\ve, k_y) \in \gap$, the band gap, if $|\rho_\pm (\ve, k_y)| \neq 1$, i.e, when $\rho_+ = e^{i k_x}$ has no real solution in $k_x \in \real$. Physically, this simply means that there are no propagating states along $x$ in the gap. 

However, $\rho_\pm (\ve, k_y) = e^{\pm i k_x}$ can always be solved in $\cmplx$, as $\rho_+ \rho_- = 1 \implies \rho_\pm \neq 0$. That is our first complexification. In terms of the Floquet discriminant,
\beq \Delta(\ve, k_y) = 2 \cos k_x. \label{eq:cbs_cond} \eeq 
By solving this equation for $k_x \in \cmplx$, we get the so called \emph{complex band structure} of the system\cite{kohn_cbs, hatsugai_cbs}, which can also be numerically computed and plotted in a 3-dimensional space ($\text{Re}(k_x), \text{Im}(k_x), \ve$) for a given $k_y$\cite{schulman_cbs, tsymbal_TI_edge}. The imaginary part of $k_z$ is interpreted as the inverse penetration depth of the edge modes, with Im$(k_x)$ negative (positive) corresponding to the left (right) edge. 

Now on to the second, and much more interesting, complexification. We note that the expression for the eigenvalues involves $\sqrt{\Delta^2 - 4}$, which is not a genuine function until we choose a branch of the square root. For real $\ve$, the argument of the square root is also real and the two branches are picked for $\rho_\pm$, respectively. However, if we allow $\ve$ to be complex, the square root becomes a genuine function from a two sheeted Riemann surface to the complex plane, with the two sheets corresponding to the two choices for a branch, connected at the branch cuts in the complex plane\cite{kohn_cbs, hatsugai_cbs}. For real eigenvalues, the two sheets correspond to the magnitude of the eigenvalue being greater than (less than) unity. It is this structure that we seek to expose in the following. 

\begin{figure}
  \centering 
  \includegraphics[width=\columnwidth]{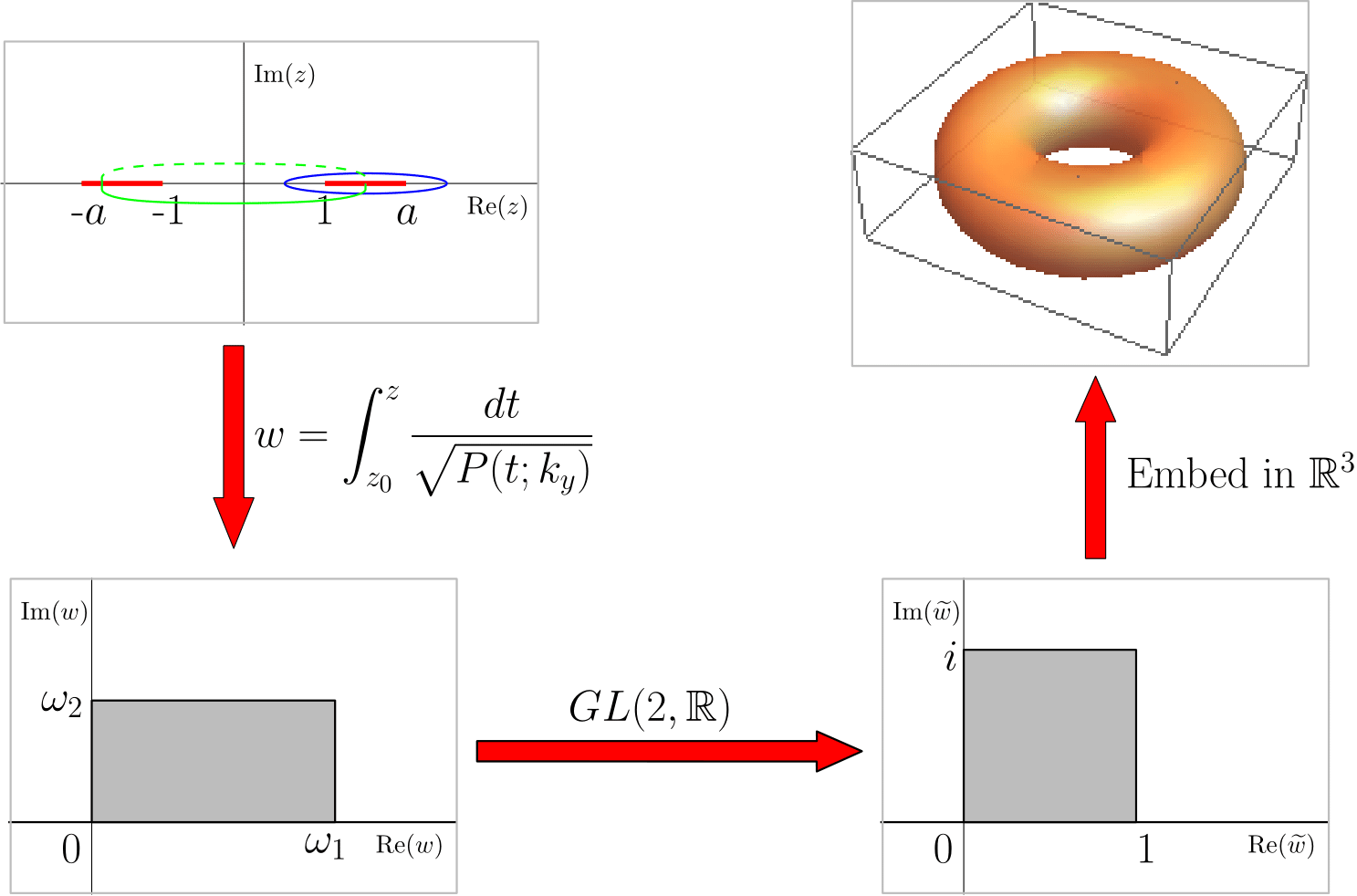}  
  \caption{The schematic for plotting the Riemann sheet corresponding to Chern insulator.}   \label{fig:riemann_sch}    
  \phantom{blah}
  \includegraphics[width=\columnwidth]{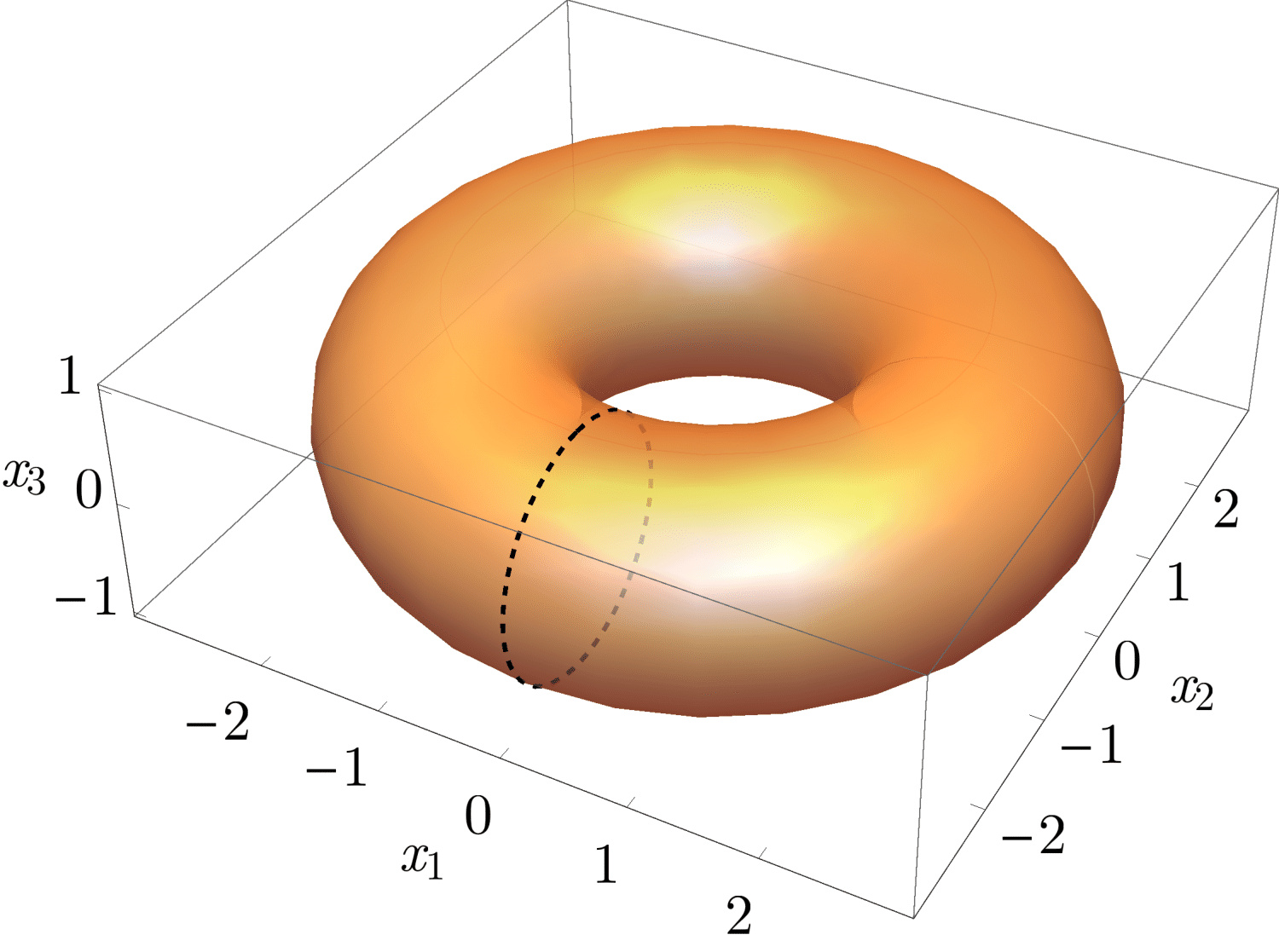}
  \caption{The energy $\ve$-Riemann surface for a Chern insulator, plotted explicitly using Mathematica\texttrademark. The black curve corresponds to an edge state. }   \label{fig:riemann_torus}  
\end{figure}

As remarked earlier, the Floquet discriminant is, in general, a rational function of $\ve$ and $\zeta = e^{ik_y}$. However, we shall restrict ourselves to the cases where it is a polynomial in $\ve$, so that the denominator is independent of $\ve$ (see \S \ref{sec:r1_basis} for relevant conditions for this to happen). Let us, then, define the discriminant of \eq{eq:tm_soln} as 
\beq P(\ve, k_y) = \Delta^2(\ve, k_y) - 4. \eeq 
We shall hereafter simply write $P(\ve)$, tacitly assuming the dependence on $k_y$. For a given system with $\nuc$ degrees of freedom per supercell, the highest power of $\ve$ is that in $\det(\ve \id - M)$, i.e, $\ve^\nuc$, so that $P(\ve)$ is a polynomial of order $2 \nuc$ in $\ve$. 

For a given $k_y$, $P(\ve)$ has $2 \nuc$ real roots, corresponding to the band edges for $\nuc$ bands. Hence, allowing $\ve$ to be complex, we get a $\ve$-Riemann surface with two sheets connected along $\nuc$ branch cuts on the real axis. This corresponds to a surface with genus\cite{mike_book} $\nuc - 1$. 

In the following, we exhibit this structure explicitly for the case of the Chern insulator. Starting with eq. (\ref{eq:CI_trace}), we can write 
\beq P(z) = \ve_{min}^4(z-a) (z-1) (z+1) (z+a) \eeq 
with 
\[ z(k_y) = \frac{\ve}{\ve_{min}}, \quad a(k_y) = \frac{\ve_{max}}{\ve_{min}} > 1, \] 
where $\ve_{min}(k_y)$ and $\ve_{max}(k_y)$ are band edges, as defined in eq. (\ref{eq:CI_band_edges}). 

The prefactor, $\ve_{min}^4$, is nonzero for all $k_y$, except when the parameter $m = 0,2,4$, i.e, at the gapless points. Hence, as far as edge states are concerned, we shall drop it in the subsequent discussion as it does not affect the roots of $P(z)$ and hence the branch-cut structure. On the other hand, for $m = 0, 2, 4$, the system becomes gapless and the topology of the Riemann sheet changes. In fact, for the gapless case, the polynomial can be written as 
\beq P(z) = z^2 (z-\ve_{max}) (z+\ve_{max}), \quad z = \ve, \eeq 
so that the Riemann surface now consists of two sheets connected at the single branch cut running between $-\ve_{max}$ and $\ve_{max}$, which has the topology of a sphere\cite{mike_book}. 

For the gapped case, given $a(k_y)$, we can map the Riemann surface to a torus (or a rectangle in the complex plane with opposite edges identified, to be precise), using the elliptic integral\cite{mike_book}:
\beq w = \int_{z_0}^z \frac{dt}{\sqrt{P(t; k_y)}} \eeq 
where the integral is independent of the path, as long as it does not wind around the branch cuts, corresponding to the two holonomies of the torus. On the other hand, such a winding gives the two periods of the torus, as 
\beq \omega_1(k_y) = \oint_\alpha \frac{dt}{\sqrt{P(t)}}, \quad \omega_2(k_y) = \oint_\beta \frac{dt}{\sqrt{P(t)}}. \eeq 
Hence, the elliptic integral maps the coordinate $z$ on the Riemann sheet to $w$ on the rectangle formed by $0, \omega_1, \omega_1 + \omega_2$ and $\omega_2$ in the complex plane, with the opposite edges identified. We can perform a $\GL(2, \real)$ transform $w \mapsto \widetilde{w}$ to map this rectangle to the square $S$ bounded by $0, 1, 1 + i$ and $i$. Finally, given $\widetilde{w} = \theta + i \phi$, we can embed the torus in 3 dimensional Euclidean space as as 
\begin{align}
 x_1 = & \; (R + \sin\phi) \cos\theta, \nonumber \\ 
 x_2 = & \; (R + \sin\phi) \sin\theta, \nonumber \\ 
 x_3 = & \;  \cos\phi, 
\end{align}
where $(x_1, x_2, x_3) \in \real^3$ and $R > 1$ is a fixed constant. Hence, using the sequence of maps described above, any given curve $\ve(k_y)$ can now be visualized as a curve on a torus. A schematic of this process is depicted in Fig. \ref{fig:riemann_sch}. We also show such a plot in Fig. \ref{fig:riemann_torus}. 

Essentially, what we have is a family of Riemann sheets parametrized by $k_y$, and we used the fact that they all have the same topology independent of $k_y$ to map them all to a single torus. However, we mention in passing that when the topology depends on $k_y$, for instance, if the gap closes for some $k_y$, we can still discuss the family of Riemann sheets in the language of cobordism. We shall not delve into the details of that picture here.

\subsection{Windings on the Riemann surface}
In order to motivate the winding numbers associated with the edge state, we recall that the edge spectrum was computed from the Evans function condition of \eq{eq:evans}, with $\varphi$ equal to the boundary value required by the boundary condition. Given a choice of $\varphi$, the Evans condition can be solved, at least locally, to obtain $\ve$ as a function of $k_y$. As $k_y \in S^1$ winds around the Brillouin zone, $\ve(k_y)$ describes a loop on the $\ve$-Riemann surface which can be associated with a set of winding numbers, one along each of the non-contractible loops of the Riemann surface. 

\begin{figure}
  \centering
  \includegraphics[width=\columnwidth]{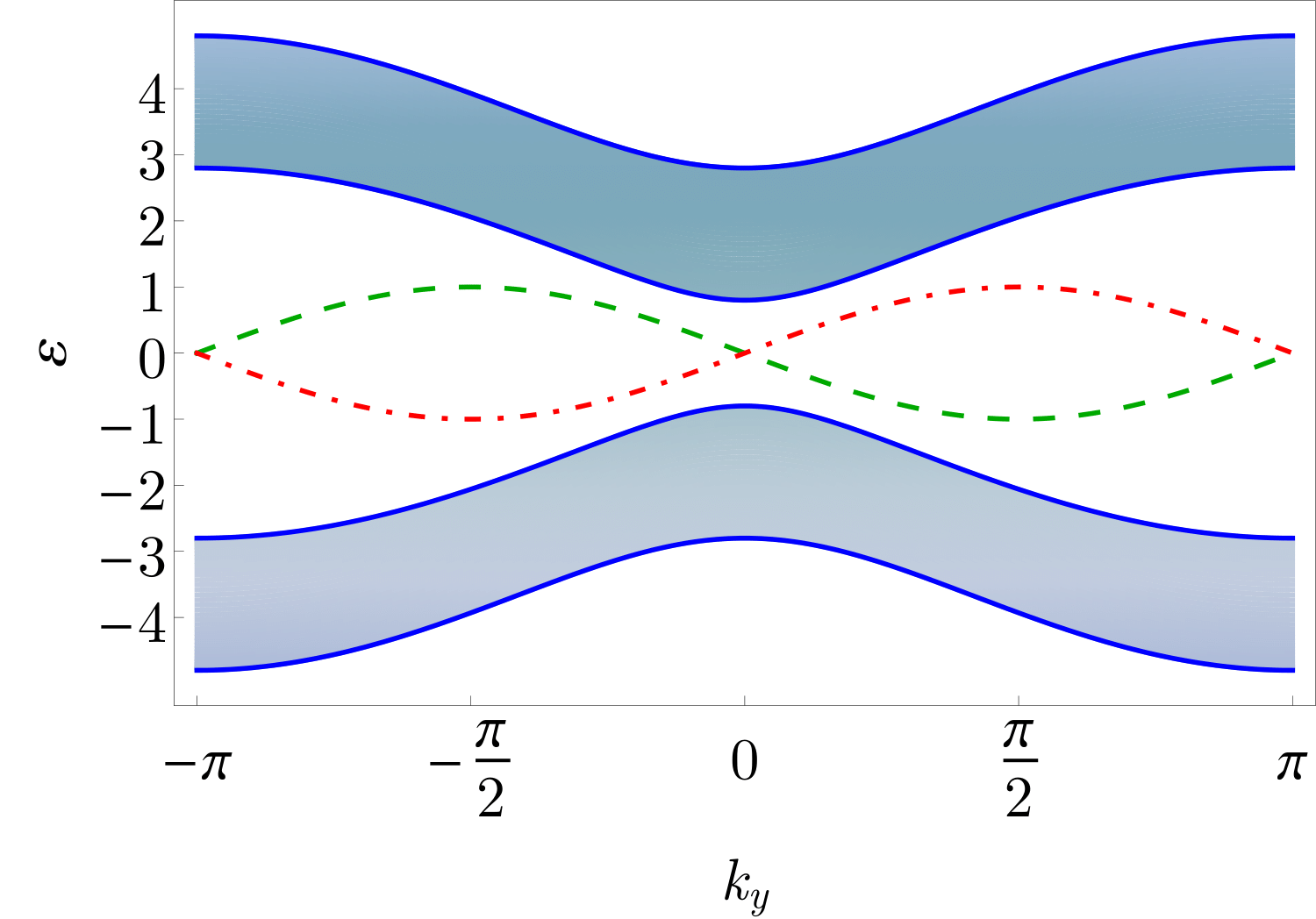}  
  \phantom{blah}
  \includegraphics[width=\columnwidth]{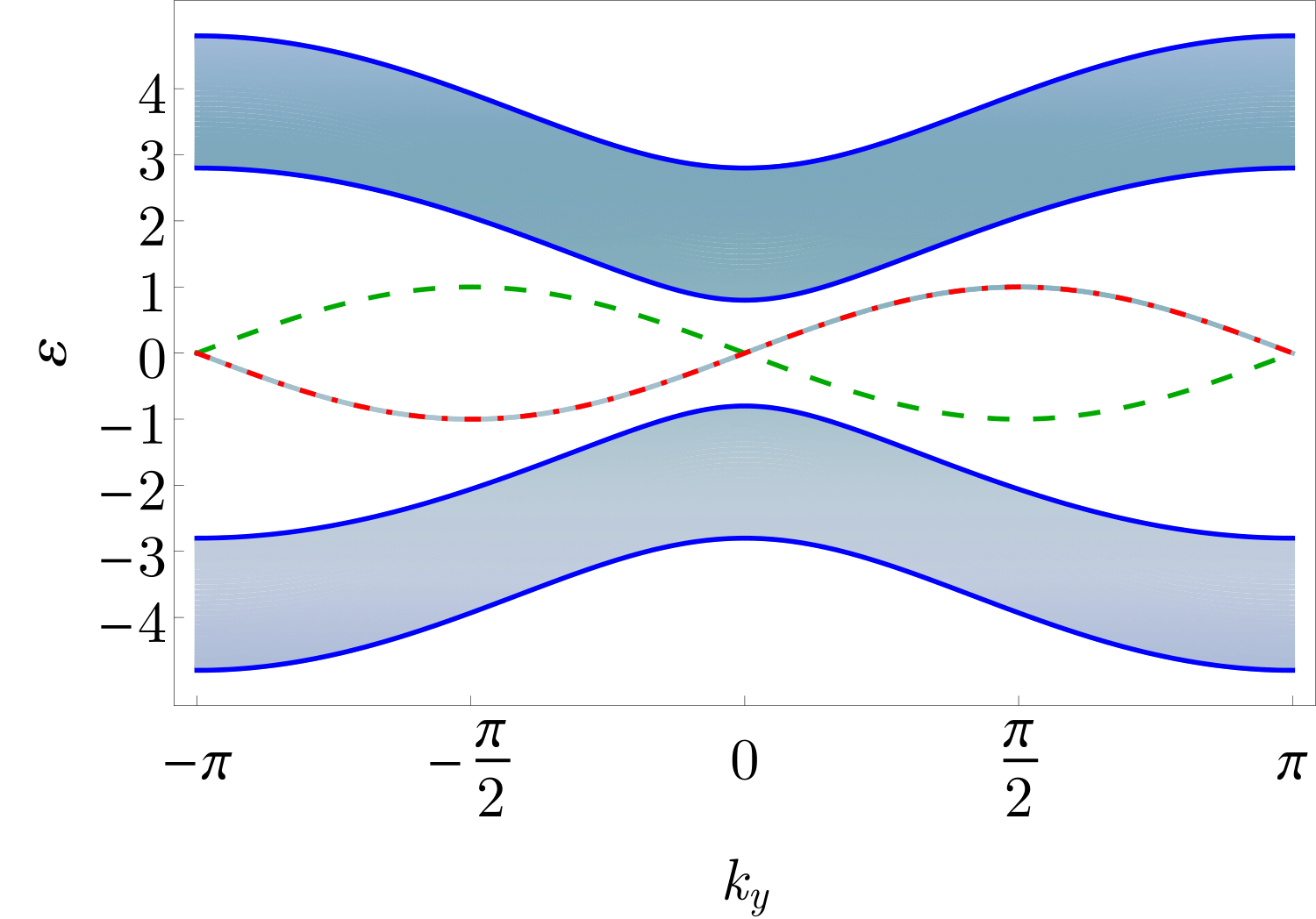}  
  \caption{(color online) The spectrum of Chern insulator for $m = -0.8$, with the band edges (dark blue) computed using the transfer matrix formalism and the left and right edge state dispersion (dashed and dashed-dot) from the Evans equation (\ref{eq:evans}), overlaid on the spectrum computed using exact diagonalization for a (top) commensurate and (bottom) incommensurate system.}    
  \label{fig:CI_trivial}
\end{figure}

Note that as $\ve(k_y)$ is real on this curve, the essential fact of Riemann surface that we are actually using is that it has two copies of the real line, connected at the branch cuts. We could have done that artificially, by gluing together two branches of $\sqrt{P(z)}$ wherever $P(z) = 0$, where the branches yield the same result, but the language of Riemann surface is more familiar and hence less \emph{ad hoc}. The actual maps that we are concerned with are, in essence, $S^1 \to S^1$ such that $k_y \mapsto \ve_L(k_y) \in \real$. This map is associated with just one winding number as $\pi_1(S^1) \cong \intg$, which is not the same as the fundamental group of the Riemann surface, as, for instance, $\pi_1(T^2) \cong \intg \oplus \intg$. 

For concreteness' sake, we plot the curve corresponding to the left edge state for Chern insulator in  Fig. \ref{fig:riemann_torus}. The edge spectrum is $\ve_L(k_y) = -\sin k_y$, as computed in \S \ref{sec:r1_CI_edge}. If the associated curve, $\ve_L(k_y)$, winds around a hole of the Riemann surface, it has to be on both the sheets. But the two sheets correspond to the eigenvalues of $T$ being less than or greater than 1, i.e, for the modes to be decaying as $n \to \infty$ and $n \to -\infty$, respectively. Hence, in order to have a curve with a nontrivial winding, we need both the physical and unphysical states, as defined in Table \ref{tb:edge}. We point out that in Hatsugai's analysis, the winding was obtained using only the \emph{physical} edge states by using a boundary condition such that $\Phi_1 = \Phi_N$ in Table \ref{tb:edge}, so that any given state is physical at at least one of the edges. This corresponds to the incommensurate case in our description.

In the discussion on the $\ve$-Riemann surface, we remarked that its topology changes when the system becomes gapless. In particular, for the Chern insulator at  $m = 0 = k_y$, the Riemann sheet is a 2-sphere, on which all loops are contractible. Hence, as one tunes $m$ across one of these gapless points, the winding number (and hence the Chern number) can change, as the loops that were non-contractible on the torus can be contracted to a point on the sphere. This does not necessarily mean that there are no states anymore that satisfy the boundary and decay conditions; rather, it simply implies that the curves corresponding to such states are now contractible (See Fig \ref{fig:Sp2r_CI}). Furthermore, we can also expose such a state in exact diagonalization by taking an incommensurate system, as shown in Fig. \ref{fig:CI_trivial}. Physically, this indicates that even when the bulk is trivial, there can still be states that decay into the bulk, but they are not topologically protected, and hence can be removed by adding a suitable boundary term.

\subsection{Winding in $\Sp(2, \real)$}
A particularly nice windfall of the $r=1$ systems is that the corresponding transfer matrices $T \in \Sp(2, \real)$, a Lie group which is a 3-dimensional manifold homeomorphic to a solid 2-torus, i.e, $D^2\times S^1$, where $D$ represents the 2-dimensional open disc. In Appendix \ref{app:math_symp}, we describe a particular parametrization of this space.

\begin{figure}
  \centering 
  \includegraphics[width=\columnwidth]{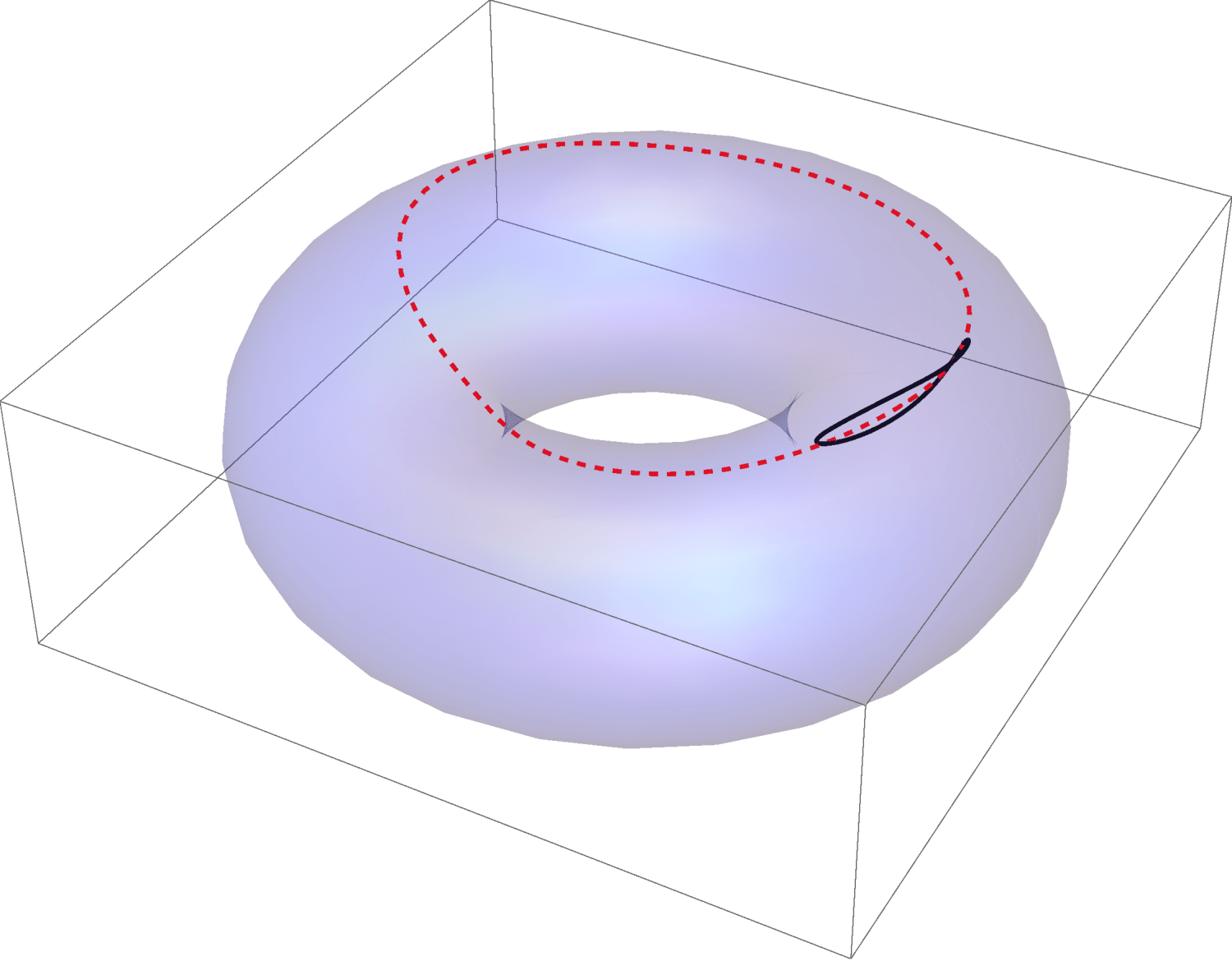}  
  \caption{(color online) The plot of the transfer matrix corresponding to the left edge state for Chern insulator, with $m = +0.8$(red dashed curve) and $m = -0.8$(black solid curve) on the $\Sp(2,\real)$ manifold, which is homeomorphic to a solid torus. See Fig \ref{fig:CI_spectrum} and \ref{fig:CI_trivial}, respectively, for the corresponding spectra.}   \label{fig:Sp2r_CI}    
\end{figure}

Given $\ve(k_y)$ which is a continuous function of $k_y$, consider $T\left(\ve(k_y), k_y\right)$. As $k_y \in S^1$, this describes a curve $\curve$ on $\Sp(2, \real)$, which we can plot explicitly using Mathematica\texttrademark. Now, the Evans condition describes just such a function $\ve(k_y)$, hence, corresponding to every edge state, we have such a curve in $\Sp(2, \real)$. We show an example of such a plot for the Chern insulator in the topological as well as the trivial regime in Fig \ref{fig:Sp2r_CI}. We can clearly see the topological nature of the edge state in the fact that the curve corresponding tot the trivial state is contractible, while the curve corresponding to the topological state is not. 

\begin{figure}
  \begin{center}
   \includegraphics[width=\columnwidth]{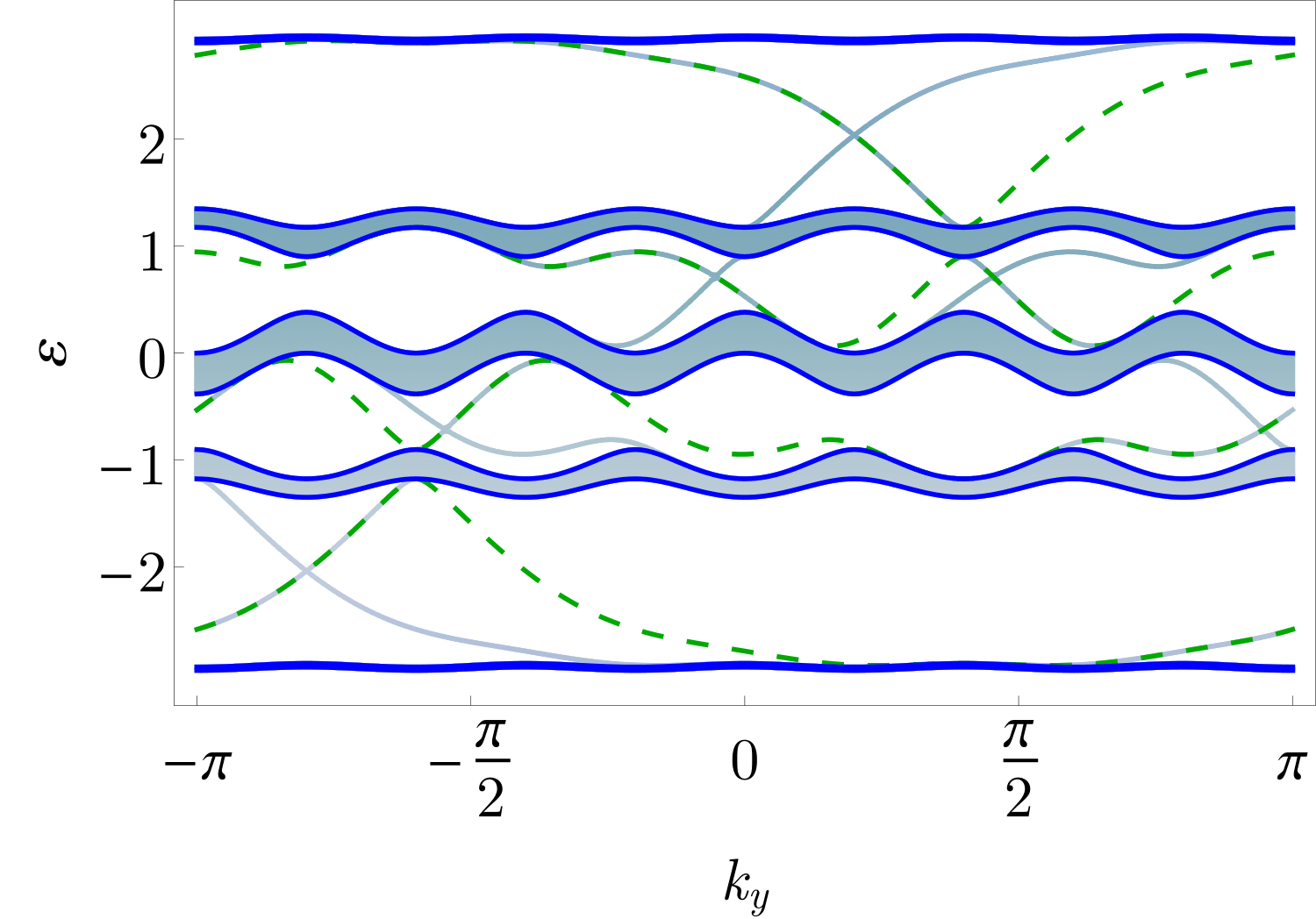}    
  \phantom{blah}
  \includegraphics[width=\columnwidth]{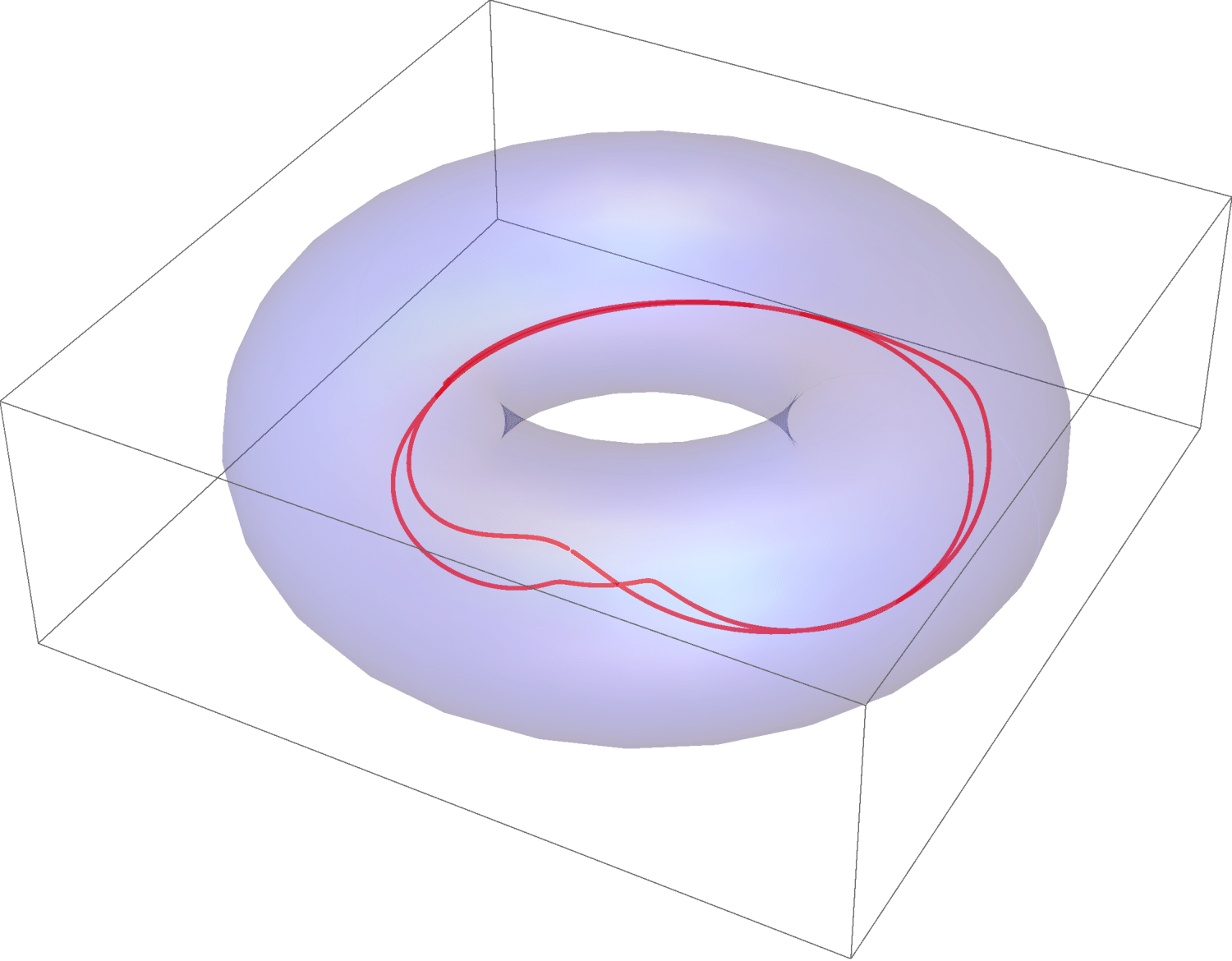}
  \caption{(color online) The spectrum of the Hofstadter model for $\phi = 1/5$ (top), and (bottom) the curve of transfer matrices on $\Sp(2, \real)$ corresponding to the left edge state in the second gap from the bottom. Note that the curve winds around twice in $\Sp(2, \real)$, as expected from the spectrum.}   \label{fig:Sp2r_HS}  
  \end{center}  
  \vspace{-0.3in}
\end{figure}

Note that this computation does not need any of the complexifactions described in the previous sections. Another advantage of plotting these curves in $\Sp(2, \real)$ over the curves on the $\ve$-Riemann surface is that the curves described here are always on a solid torus for all rank 1 systems, as opposed to the Riemann surface, which is a surface whose genus is a function of the number of bands. For instance, we plot the edge state for the Hofstadter model with $\phi = 1/5$ in Fig. \ref{fig:Sp2r_HS}, the Riemann surface corresponding to which has genus 4. The edge state shown has a winding number of 2, a fact that can be easily gleaned from the figure. 

Finally, we show that there exists a winding number associated with $T$ in $\Sp(2, \real)$, which is independent of $\varphi$. To begin with, we note that as the fundamental group $\pi_1\left(\Sp(2, \real)\right) \cong \intg$ (for a proof, see Appendix \ref{app:math_symp}), any curve $\curve$ on $\Sp(2, \real)$ is associated with a winding number (also known as Maslov index\cite{littlejohn_wavepackets, macduff-salamon_symp_top}). Formally, we have a map
\beq \mu :  \loopset(\Sp(2, \real)) \to \intg, \eeq 
which associates a winding number with each loop,  $\curve \in \loopset(\Sp(2, \real))$, where $\loopset(M)$ denotes the set of all closed loops on a smooth manifold $M$. 

Now, the Evans condition for a given $\varphi$ is a continuous function of $k_y$, to which we can associate a curve $\curve_\varphi$, with the corresponding winding number $\mu(\curve_\varphi)$. Hence, for each $\varphi \in \cmplx^2 \backslash \{0\}$, we get a map $\varphi \mapsto \mu(\curve_\varphi) \in \intg$. But as $\mu(\curve_\varphi)$ is an integer, it cannot change continuously under a continuous change of $\varphi$. Thus, $\mu(\curve_\varphi)$ must be independent of $\varphi$'s for a given gap.

So far, we have not shown using our formalism that the winding number of a curve corresponding to a Evans function condition in $\Sp(2, \real)$ should be the same as the winding number of the corresponding curve on the $\ve$-Riemann surface, even though we notice it to be so in all the examples that we checked, and we intuitively expect it to be so. A proof of a similar statement is discussed in Ref. \onlinecite{avila_edge2d} using $K$-theory, but it is rather opaque from the point of view of physicists. Finally, the interpretation of the Chern number as a Maslov index can provide new ways of computing it numerically, as well as analytically\cite{littlejohn-robbins_maslov, avila_edge2d}.


\section{An example for $r = 2$}    \label{sec:r2}
The computation of the transfer matrix naturally becomes more intricate for $r > 1$. However, if the transfer matrix turns out to be symplectic, we can take advantage of the additional structure for exact computations.  Here, we compute the transfer-matrix for a $r=2$ model in closed form and derive exact analytical expressions for its surface spectrum for such a case. The model we study is a topological crystalline insulator (TCI), first introduced by Fu\cite{fu_TCI}, whose topological surface states are protected by crystalline symmetries alongside time reversal symmetry. Moreover, our derived expressions for the topological surface bands correctly capture the closing  of the surface band gap as the model is tuned to its $C_4$ symmetric limit, in agreement with the $k \cdot p$ analysis of Fu\cite{fu_TCI}, as well as the lifting of the degeneracy of the surface bands as we break the $C_4$ symmetry. 

The Fu model is defined on a 3-dimensional tetragonal lattice, with alternating layers of square lattices of $A$ and $B$ type along the $z$ axis. The system has a $C_4$ symmetry in the plane normal to the $z$ axis. The lattice model consists of nearest and next-nearest neighbor hoppings between two orbitals on each site (typically identified as $p_x$ and $p_y$), with the strength of hopping being equal in magnitude but opposite in sign on the A and B sublattices. Thus, the model consists of 4 bands, with 2 orbitals and 2 sublattice degrees of freedom. The Bloch Hamiltonian is given by
\beq 
\hlt(\vk) = \left( \begin{array}{cc} \hlt_A(\vk) & \hlt_{I}(\vk) \\ \hlt_{I}^\dagger(\vk) & \hlt_B(\vk) \end{array} \right)
\eeq 
with the layer Hamiltonian $\hlt_\text{a}, \; \text{a} \in \{A, B\}$ and the inter-layer hopping $\hlt_I$. The $2 \times 2$ blocks are given by 
\begin{align}
 \hlt_\text{a}(\vk) = & \; 2 t_1^\text{a} \left( \begin{array}{cc} \cos k_x & 0 \\ 0 & \cos k_y  \end{array} \right) \nonumber \\ 
  & + 2 t_2^\text{a} \left( \begin{array}{cc} \cos k_x \cos k_y &  \sin k_x \sin k_y \\ \sin k_x \sin k_y & \cos k_x \cos k_y  \end{array} \right) \nonumber \\ 
  = & \;  \left[ t_1^\text{a}(\cos k_x + \cos k_y) + 2 t_2^\text{a} \cos k_x \cos k_y \right] \id_2 \nonumber \\ 
  & + 2 t_2^\text{a} \sin k_x \sin k_y \sigma_x +  t_1^\text{a}(\cos k_x - \cos k_y) \sigma_z \nonumber \\ 
 \hlt_{I}(\vk) = & \; \left[ t_1' + 2 t_2'(\cos k_x + \cos k_y) + t_z' e^{i k_z} \right] \id_2,  \label{eq:hlt_TCI}
\end{align}
where we take $t_i^A = - t_i^B \equiv t_i$ for $i = 1, 2$, so that $\hlt_A = - \hlt_B = \hlt_0$. The system is invariant under $C_4$ rotations, with the $C_4$ action defined by 
\beq  
\mathcal{C}_4 \; \hlt(k_x, k_y, k_z) \; \mathcal{C}_4^{-1} =  \hlt(-k_y, k_x, k_z),
\eeq 
where $\mathcal{C}_4 = i \id_2 \otimes \sigma_y$. Clearly, a cut normal to the $z$ axis preserves the  $C_4$ symmetry. We cut the system along $z$ (as opposed to $x$ in the previous sections but conforming to the notation in Ref. \onlinecite{fu_TCI}). Defining
\beq \hlt_1(\vk_\perp) =  \left[ t_1' + 2 t_2'(\cos k_x + \cos k_y) \right] \id_2, \eeq 
where $\vk_\perp = (k_x, k_y)$, we identify 
\begin{align}
 J = t_z' \left( \begin{array}{cc} 0 & \id_2 \\ 0 & 0  \end{array} \right), \quad M = \left( \begin{array}{cc} \hlt_0 & \hlt_1 \\ \hlt_1 & - \hlt_0 \end{array} \right).
\end{align}
In order to reduce the notational clutter, we set 
\beq \hlt_0 = a \, \id_2 + \vb \cdot \vs, \quad \hlt_1 = m \id_2, \eeq
where we define
\begin{align*}
a = & \; t_1(\cos k_x + \cos k_y) + 2 t_2 \cos k_x \cos k_y,   \\ 
\vb = & \; \left( \begin{array}{ccc} 2 t_2 \sin k_x \sin k_y, & 0, &  t_1(\cos k_x - \cos k_y) \end{array}  \right), \\ 
m = & \; t_1' + 2 t_2'(\cos k_x + \cos k_y),    
\end{align*}
and $b = |\vb|$. We also normalize the parameters of the model so that $t_z' = 1$. 

To compute the transfer matrix, we begin with the SVD of $J$ as $J = V \cdot \Xi \cdot W^\dagger$, with 
\beq V = \vecenv{\id_2}{0}, \quad \Xi =  \id_2, \quad W = \vecenv{0}{\id_2}.  \eeq 
The condition for the transfer matrix being complex-symplectic was that $[\green_{ab}, \Xi] = 0$, which is always true here. Furthermore, as $M$ and $J$ are both real, the transfer matrix will be real. Thus, $T \in \Sp(4, \real)$. 

Next, we need
\begin{align}
 \green = & \; \left( \begin{array}{cc} (\ve - a) \id_2 - \vb \cdot \vs & -m \id_2 \\ -m \id_2 & (\ve + a) \id_2 + \vb \cdot \vs \end{array}  \right)^{-1} \nonumber \\ 
 \equiv & \; \left( \begin{array}{cc} A & B \\ C & D \end{array}  \right)^{-1},
\end{align} 
where $A = \ve \id_2 -  \hlt_0$, $B = C = \hlt_1$ and $D = \ve\id_2 + \hlt_0$. As each block here is invertible for almost all $\ve$, we use the \eq{eq:blk_inv}  from the appendix to get
\beq 
\green = \left( \begin{array}{cc}  \iA + \iA B \schur_{11}^{-1} C \iA   & - \iA B \schur_{11}^{-1} \\  - \schur_{11}^{-1} C \iA  & \schur_{11}^{-1} \end{array}  \right),
\eeq
where $\schur_{11} = \green^{-1}/A = D - C A^{-1} B$. For the definition of $V$ and $W$ as above, the computation of $\green_{ab}, \; a, b \in \{v, w\}$ is simply taking the correct submatrices, viz,
\begin{align}
 \green_{vv} = & \; \iA + \iA B \schur_{11}^{-1} C \iA \nonumber \\ 
 \green_{vw} = & \; -\schur_{11}^{-1} C \iA  \nonumber \\ 
 \green_{wv} = & \; -\iA B \schur_{11}^{-1}  \nonumber \\ 
 \green_{ww} = & \; \schur_{11}^{-1}. 
\end{align}
Using \eq{eq:tmat_def}, the transfer matrix becomes 
\beq 
 T = \left( \begin{array}{cc} B - A C^{-1} D  \quad\quad &  A C^{-1} \\ - C^{-1} D \quad\quad &   C^{-1}  \end{array} \right).
\eeq 
and substituting the blocks, we get 
\beq 
T = \frac{1}{m}  \left( \begin{array}{cc} 
\eta^2 \id_2 - 2 a \vb\cdot\vs  \quad & (a - \ve)\id_2 + \vb\cdot\vs \\ 
(a + \ve)\id_2 + \vb\cdot\vs \quad &   - \id_2  
\end{array} \right),
\eeq
where 
\[ \eta^2 = \ve^2 - a^2 - b^2 - m^2.  \]
As $T$ is symplectic, using results from Appendix \ref{app:tmat} its spectrum is given by
\beq 
\sigma\left[ T \right] = \frac{1}{2} \left( \Delta_\mu\pm \sqrt{\Delta_\mu^2 - 4} \right), \quad \mu = \pm, 
\eeq 
where for $\mu = \pm 1$, 
\begin{align}
\Delta_{\mu} = & \; \frac{1}{2} \left[ \tr T + \mu \sqrt{2 \tr T^2 - (\tr T)^2 + 8} \right] \nonumber \\ 
= & \; \frac{1}{m} \left[  \ve^2 - m^2 - 1  - (a + \mu b)^2  \right].
\end{align}
The band edges are given by 
\beq |\Delta_\mu| = 2 \implies \Delta_\mu = 2\lambda, \quad \lambda = \pm 1, \eeq 
which can be solved to get 
\beq 
\ve = \pm \sqrt{(m + \lambda)^2 + (a + \mu b)^2}, \quad \lambda, \mu = \pm 1.
\eeq

\begin{figure}[t]
  \centering  
  \includegraphics[width=\columnwidth]{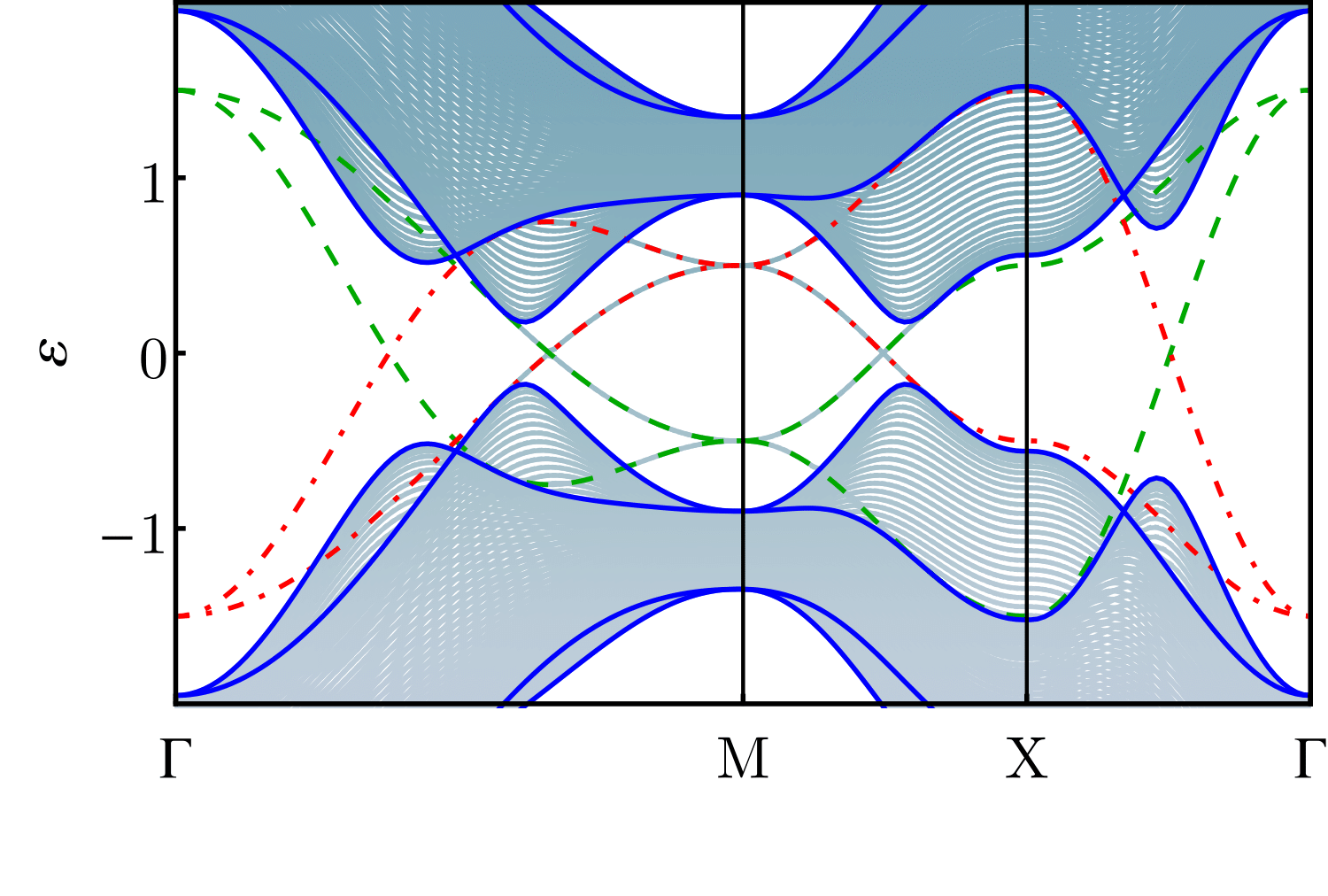}  
  \caption{(color online) The spectrum of the topological crystalline insulator model due Fu\cite{fu_TCI}, with the parameters $t_1 = 0.5$, $t_2 = 0.25$, $t_1' = 1.25$, $t_2' = 0.25$ and $t_z' = 1$ in \eq{eq:hlt_TCI}. The band edges (dark blue) and the left and right edge state dispersion (dashed and dashed-dot) computed using the transfer matrix formalism, overlaid on the spectrum computed using exact diagonalization equivalent to Fig, 2(b) of Ref. \onlinecite{fu_TCI}.}    
  \label{fig:TCI_spectrum}
\end{figure}

For the edge states, given $\Phi = (\bbe, 0)^T$, which satisfies the boundary conditions for the left edge, we demand that $T \Phi$ is in the same subspace as $\Phi$, spanned by $\uvec_1$ and $\uvec_2$. But
\beq 
T \vecenv{\bbe}{0} = \vecenv{\left(  \eta^2 \id_2 - 2 a \vb\cdot\vs \right) \bbe }{\left( (a + \ve)\id_2 + \vb\cdot\vs \right) \bbe}.
\eeq 
Thus, for $\Phi$ to be a left edge state, we demand that 
\beq \left( (a + \ve)\id_2 + \vb\cdot\vs \right) \bbe = 0,  \eeq 
We get a nontrivial solution for $\bbe$ iff the matrix is singular, i.e, iff
\beq \sigma\left[  (a + \ve)\id_2 + \vb\cdot\vs \right] =  (a + \ve)^2 - b^2 = 0. \eeq 
Thus, the left edge spectrum is given by 
\beq \ve_L = -a \pm b, \eeq 
Similarly, the right edge spectrum is given by
\beq \ve_R = a \pm b. \eeq 
Thus, we have analytically obtained explicit expressions for the boundaries of the bulk bands and the edge spectra. We plot them, alongside the spectrum computed from exact diagonalization, in Fig \ref{fig:TCI_spectrum}, overlaid on the band structure obtained by exact diagonalization. 

From the figure, we note that there is a quadratic band touching at the surface near $\vk_M = (\pi,\pi)$, the projection of the $M$ point of the 3D Brillouin zone on a constant $k_z$ plane. Expanding the left edge spectrum in the vicinity of this point as $\vk_\perp = \vk_M + \delta\vk$ upto the second order in $\delta k$, we get 
\begin{align}
 \ve_L  \approx & \; - 2(t_2 - t_1) - \frac{t_1 - 2 t_2}{2} (\delta k)^2  \nonumber \\ 
 & \pm \frac{t_1}{2} \sqrt{(\delta k)^4 + 4 \left[1 - \left( \frac{2 t_2}{t_1} \right)^2 \right] \delta k_x^2 \delta k_y^2 }.
\end{align}
For $t_1 = 2t_2 = t$, we get a radially symmetric quadratic band touching, with the spectrum given by
\beq \ve_L \approx - t \left[ 1 \mp \frac{1}{2} (\delta k)^2 \right]. \eeq 
Thus, we can uncover the fine-tuned nature of this surface quadratic band touching as well as derive the coefficients of a $k \cdot p$ expansion around that point, which was guessed on symmetry grounds in Ref. \onlinecite{fu_TCI}. 

This calculation for the rank 2 TCI reveals reveals some aspects of our formalism that did not come into play in the rank 1 case: 
\begin{itemize} 
 \item The above calculation involved the eigenvalue problem of a $4 \times 4$ transfer matrix, but it was still amenable to analytic calculations leading to explicit closed form expressions for the bulk band edges and edge spectra, owing to the symplectic nature of the transfer matrix. 
 
 \item For $r = 2$, we can potentially have \emph{partial gaps} defined in \S \ref{sec:tmat_appl_bulk}, which corresponds to the $(\ve, \vk_\perp)$ values where a pair of eigenvalues of the transfer matrix lie on the unit circle and the other pair off it. The edge states always touch one of the band edges, but sometimes they can mean the edge to a partial gap, so that for a given $(\ve, \vk_\perp)$, there is an edge state as well as a bulk band state. This is clearly seen in Fig \ref{fig:TCI_spectrum}.
 
 \item The closed form expression of the surface spectrum can be used to analytically track the lifting of degeneracy of the surface states at the high symmetry points on the addition of a $C_4$-breaking term. For instance, we can add a term $\delta \hlt = \mu \sigma_z \otimes \sigma_z$ to the Hamiltonian, corresponding to breaking the degeneracy of the $p_x$ and $p_y$ orbitals. Then, at the $M$ point, 
 \[ a = 2(t_2 - t_1), \quad b = (0,0,\mu),  \] 
 so that the left edge spectrum becomes 
 \beq \ve_L = 2(t_1 - t_2) \pm \mu. \eeq 
 The gap is clearly proportional to $\mu$, the strength of the $C_4$ breaking term. 
\end{itemize}


\section{Application to Disordered Systems}  \label{sec:dis}

The generalized transfer matrix formalism can be used to directly investigate tight binding models in presence of disorder and their metal-insulator transitions. The standard approach\cite{mackinnon1983scaling,kramer1993localization,evers2008anderson} of determining the scaling properties of the longest localization length of a quasi 1D representative of a $d$-dimensional model, in the form of either a cylinder or a strip, may be employed without modification to the generalized transfer matrix formalism. In this section we will demonstrate how this is to be achieved. 

Consider again a generic $d$-dimensional tight binding lattice model with $\ndof$ degrees of freedom per unit cell. The simplest model of disorder is a diagonal (or the Anderson type) disorder, which explicitly breaks the translational symmetry, so that the transverse momentum, $\vk_\perp$ is not a good quantum number anymore. Instead, we consider the system on a \emph{strip geometry}, i.e, infinite along $x$ and finite along all the transverse directions, with open or periodic boundary conditions. 

Thus, we write our system in the position basis and construct the supercells from the sites corresponding to a constant $x$. For instance, for a 2 dimensional strip of width $L_y$, with sites indexed by $m = 1, \dots L_y$ and internal degrees of freedom at each site by $\alpha=1, \ldots, \ndof$. The disorder corresponds to the Hamiltonian 
\beq 
V_n =  \sum_{m=1}^{L_y} \sum_{\alpha=1}^{\ndof}  V_{nm\alpha} c^\dagger_{nm\alpha} c_{nm\alpha}   
\eeq 
where $\{V_{nm\alpha}\}$ are iid. real random variables, taken from a uniform distribution around 0 with width $W$. Our supercells now consist of the $\nuc = \ndof L_y$ degrees of freedom. Using the method described in \S \ref{sec:tmat}, we can identify the hopping matrix $J$ and the on-site matrix $M_n$, where only $M$ depends on $n$ as the disorder is diagonal. We can construct the transfer matrix as a function of $n$, i.e, $\Phi_{n+1} = T_n \Phi_n$, where $T_n$ now depends on the disorder realization. Thus, for a system with $N$ sites along the $x$ axis, we define the total transfer matrix as the product ${\bf T}_N \equiv \prod^N_{n=1} T_n$. 

To investigate the existence of topological edge states, we note that for a strip geometry in 2-dimensions, there are edge states localized at $m = 1, L_y$ along the $y$ axis and strongly delocalized along $x$, even in the presence of disorder. Thus, we need to look for an eigenvalue $\rho$ of the total transfer matrix which lies on the unit circle. Alternatively, we look for the vanishing of a Lyapunov exponent, defined as $\lambda =  \ln |\rho|$. In the next subsection, we describe a recipe to compute the Lyapunov exponents numerically for a given disorder realization.

\subsection{Lyapunov Exponents and Localization Lengths}
The conventional approaches\cite{mackinnon1983scaling,evers2008anderson,kramer1993localization} to studying bulk phases of disordered non-interacting models and their Anderson transitions rely on obtaining the smallest Lyapunov exponent (in magnitude), or equivalently, the longest localization length in the $x$ direction for a fixed energy $\varepsilon$. When the Fermi energy is set to $\varepsilon$, a further finite size scaling analysis of the longest localization length in the transverse directions discriminates between conducting and insulating phases of the bulk. Thus, to observe the quasi- $(d-1)$ dimensional metallic edge modes in a $d$-dimensional disordered topological phase, it is desirable to compute the multiset of all Lyapunov exponents, hereafter termed the Lyapunov spectrum.

For a clean system, the eigenvalues $\rho_i$ of the transfer matrix determine the growth/decay rate of the corresponding eigenstates, so that we can identify the Lyapunov exponents, or alternatively, the inverse localization length, as $\lambda_i = 1/l_i = \ln |\rho_i|$. Alternatively, we can define ${\bf \Lambda} = (T^\dagger T)^{1/2}$ with eigenvalues $\Lambda_i = |\rho_i|$, so that $\lambda_i = \ln \Lambda_i$. For the disordered case, the transfer matrices depend on $n$, so that we define
\begin{align}
{\bf \Lambda} = \lim_{N\rightarrow \infty}\left[ {\bf T}^\dagger_N {\bf T}^{\phantom{\dagger}}_N  \right]^{1/(2N)}; \quad {\bf T}_N \equiv \prod^N_{n=1} T_n.   \label{eq:lyap_def}
\end{align}
The fact that such a finite valued matrix exists is guaranteed by Oseledec's theorem\cite{eckmann1985ergodic}. The Lyapunov exponents are again given by $\lambda_i = \ln \Lambda_i$, where $\Lambda_i \in \real$ are the eigenvalue of ${\bf \Lambda}$. When ${\bf T}_N$ is regarded as the evolution map of a dynamical system in time $N$, the metallic states  correspond to stable limit cycles as $N \to \infty$.

In principle, given the transfer matrix, one could directly compute the matrix product in \eq{eq:lyap_def}, and hence the Lyapunov exponents, as a function of $N$. However, in practice, such a numerical matrix multiplication and diagonalization is usually plagued by numerical rounding and overflow errors, associated with the finite precision of the floating point representation of real numbers. In order to circumvent these issues, we follow the method described in Ref \onlinecite{eckmann1985ergodic}. The key idea is to perform a QR decomposition\cite{kramer1993localization} after every step involving a matrix multiplication.

Explicitly, we begin by performing a $QR$ decomposition of the first transfer matrix in the sequence as $T_1 = Q_1 R_1$, where $Q_1$ is unitary and $R_1$ is upper triangular with real, positive diagonal entries, sorted in descending order. Iterating, we get 
\begin{align}
{\bf T}_N &= 
\left(\prod^N_{n=3} T_n\right)T_2 T_1
=  \left(\prod^N_{n=3} T_n\right)T_2 (Q_1 R_1) \nonumber \\
&= \left(\prod^N_{n=3} T_n\right) T'_2 R_1 
= \left(\prod^N_{n=4}T_n\right)T_3 (Q_2 R_2) R_1 \nonumber \\
&= \ldots  = T'_N \prod_{m=1}^N R_m = Q_N \prod_{m=1}^N R_m,
\end{align}
where we have defined $T_{n+1}' \equiv T_n Q_n$ and carried out its $QR$ decomposition as $T_{n+1}' = Q_{n+1} R_{n+1}$ at each iteration. As $Q^\dagger Q = \id$ and $R^\dagger_m R^{\phantom{\dagger}}_m = S_m$ is diagonal with the diagonal entries $S_{m,ii} = \left( R_{m, ii} \right)^2$, we simply get 
\beq 
{\bf \Lambda} = \left[ \prod_{m=1}^N S_m \right]^{\frac{1}{2N}} = \text{diag} \left\{ \left( \prod_{m=1}^N R_{m,ii} \right)^{\frac{1}{N}} \right\}_{i} 
\eeq 
As $N \to \infty$, the Lyapunov exponents converge to 
\begin{align}
\lambda_i = \lim_{N\rightarrow \infty} \frac{1}{N} \sum_{m=1}^N \ln [(R_m)_{ii}].   \label{eqn:lambda_i}
\end{align}
Hence, only the diagonal elements of $R_m$ are needed at each iteration, thereby avoiding the accumulation of numerical error. Convergence to the true Lyapunov exponents can also be ascertained by studying the statistical fluctuations of the average on the right hand side of \eq{eqn:lambda_i}.

\subsection{Disordered Chern Insulator} \label{sec:dis_CI}
We now specialize to the case of a Chern insulator with diagonal disorder. For a clean system, the on-site matrix $M$ is given by
\begin{align}
M &=\frac{1}{2}\sum_{m=1}^{L_\text{max}} 
\left( \uvec_{m+1} \cdot \uvec_m^\dagger \right)\otimes (i \sigma^y - \sigma^z) +\text{h.c.} \nonumber \\
&+(2-m) \id_{L_y}\otimes\sigma^z 
\end{align}
and the on-site Green's function is 
\begin{align}
\mathcal{G}_n &=  \left( \varepsilon \id_L \otimes \id_2 - M - V_n\right)^{-1},
\end{align}
where $L_\text{max} = L_y$ in the case of PBC and $L_\text{max}=L_y-1$ for the open boundary condition. It is worth remarking that for fixed $\varepsilon$ and $M$, $\mathcal{G}_n^{-1}$ is non-invertible only for a set of measure $V_n$ realizations, i.e, almost everywhere, and so we shall side step questions of its singularity. 

The inter-layer coupling matrix $J$ remains unchanged for this ensemble of disorder and takes the form of a $2{L_y}\times 2{L_y}$ matrix
\begin{align}
J = \frac{1}{2i} \id_{L_y} \otimes (\sigma^x -i \sigma^z),
\end{align}
which, however, remains singular, with rank $r = L_y$. This conforms with the expectation of $L_y$ independent channels in the non-disordered limit, which are explicitly coupled by disorder. The SVD for $J$ remains virtually unchanged:
\begin{align}
&J = \id_{L_y} \otimes ({\bf v} \cdot {\bf w}^\dagger) = \sum_{y=1}^{L_y} {\bf V}_y \cdot {\bf W}^\dagger_y,  
\end{align}
with $\vv$ and $\vw$ defined as in \eq{eq:vw_CI_def}, and we have defined the channels ${\bf V}_y := {\bf e}_y \otimes {\bf v}$ and ${\bf W}_y := {\bf e}_y \otimes {\bf w}$; $\{{\bf e}_y \}_{y=1}^{L_y}$ being the standard basis of $\mathbb{C}^{L_y}$. Also, $\Xi=\id_{L_y}$, which implies that the transfer matrix is symplectic, following \eq{eq:symp_cond}. 

For each $n$, the transfer matrix $T_n$ can now be computed using \eq{eq:tmat_def}, which can be used to further compute the Lyapunov exponents using \eq{eqn:lambda_i}. As the transfer matrix is symplectic, the eigenvalues occur in reciprocal pairs, so that the Lyapunov spectrum will always be symmetric about zero. We seek a localized, potentially topological mode, with a Lyapunov exponent zero (within numerical error). 

\begin{figure}
	\includegraphics[width=0.5\textwidth]{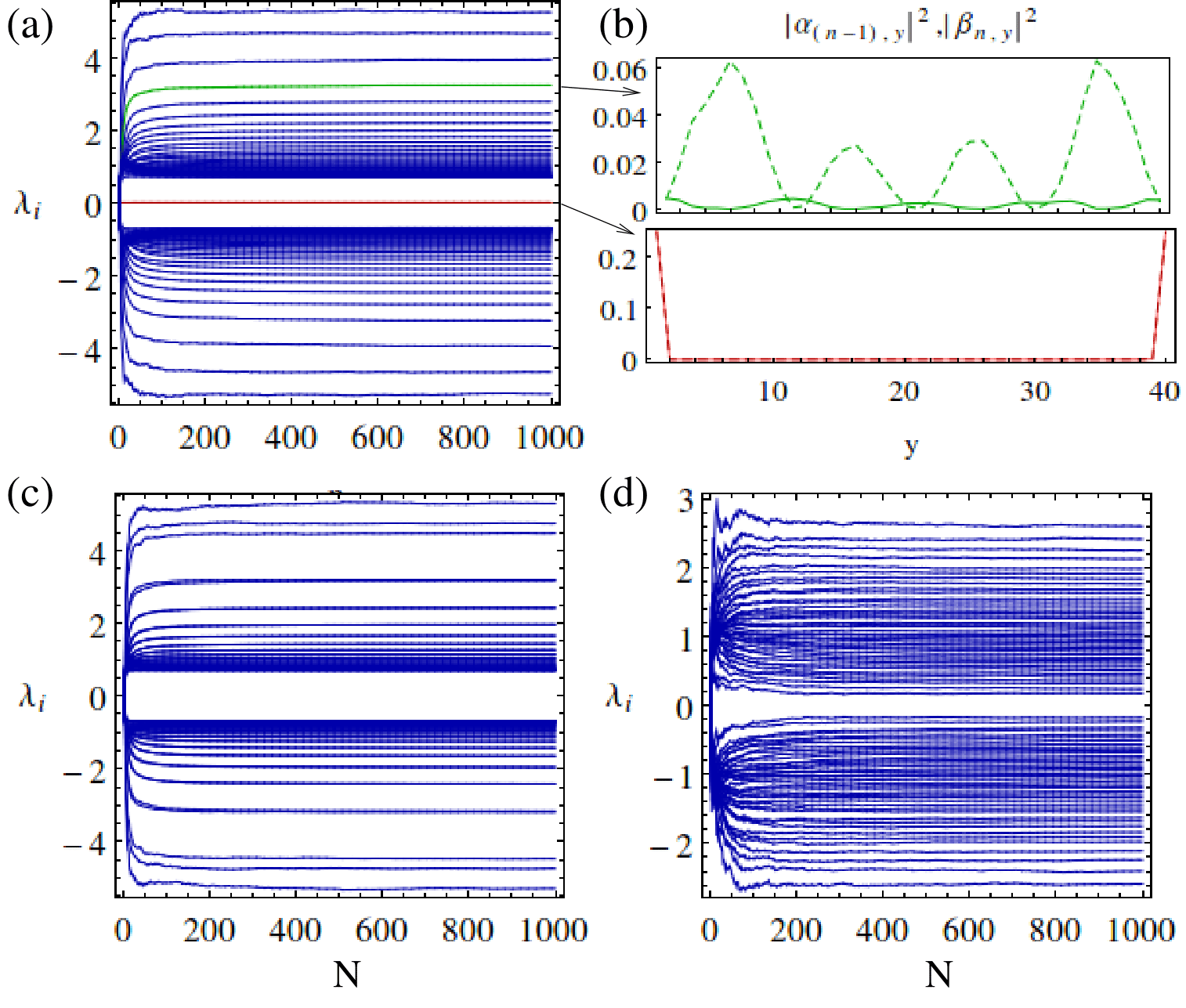}
	\caption{(color online) Numerically computed estimates of Lyapunov exponents as a function of system length $N$ for Chern insulator on a strip geometry with width $L_y = 40$ and parameters $m = 1.0$, $\ve = 0$. For large $N$, the estimates converge to the Lyapunov exponents $\{\lambda_i\}$.  (a) Open boundaries along the vertical and disorder $W=0.1$ shows robust metallic edge modes (red trace) with $\lambda_i=0$ in this scale. Also highlighted  is an insulating bulk mode (green trace) with $\lambda_i\approx 3$. (b) Vertical spatial profiles of the eigenstates for $N = 10^3$  with components ${({\alpha}_{N-1},\beta_N)}$ for the modes highlighted in (a), where the top is the bulk insulating mode [green trace] and the bottom the metallic edge state [red trace] which is strongly localized at the vertical boundaries. Arrows mark the position of these eigenmodes in the Lyapunov spectrum. (c) The same system as in (a) with periodic boundary conditions which shows no metallic edge states. (d) Strongly disordered case ($W=5.0$) with open boundaries and absent metallic states.}
	\label{fig:disoder}
\end{figure}

In Fig. \ref{fig:disoder}, we show the Lyapunov spectrum as a function of $N$ for the Chern insulator in the topological phase, with $m=1$ and strip width $L_y=40$. We limit ourselves to the energy $\varepsilon=0$, corresponding to the center of the band gap. For a weak disorder ($W = 0.1$) and open boundary conditions along $y$, there are two quasi-1D metallic modes with $\lambda_i \approx 0$ at the center of the spectrum, highlighted in red in Fig.\ref{fig:disoder}(a). Numerically, the relevant exponents are never zero to machine precision, but are comparatively small ($|\lambda_i|<10^{-5}$ at $N=10^4$) and systematically decrease (as a power law) with increasing strip length $N$. To confirm the identification of these modes as topological edge states, we plot their spatial profile in Fig.\ref{fig:disoder}(b)(top), which clearly shows localization at the edge, in contrast to an insulating localized mode with $\lambda_i \approx 3$. Furthermore, for the same parameters but with closed periodic boundaries, no metallic modes are observed, as shown in Fig.\ref{fig:disoder}(c). We note that tuning the mass parameter $m$ to the topologically trivial range or moving $\varepsilon$ into the center of the bulk band also removes these metallic modes. 

Finally, for a strong enough disorder ($W$ = 5.0), the metallic modes are also absent, as shown in Fig.13(d). We observe that the “lifting” of the metallic edge modes from the asymptotic value $\lambda_i = 0$ occurs continuously with changes of tuning parameters, in agreement with the theory of continuous Anderson transitions\cite{evers2008anderson}. However, further work is needed to verify that the scaling exponents $\{\nu_i\}$ corresponding to the divergence of the localization lengths $\{l_i\}$ at the metal-insulator transition agrees with the expectations for the Integer Quantum Hall transition\cite{slevin2009critical,obuse2010conformal}. We leave such numerical investigations for future work.


\section{Conclusions and Discussion}    \label{sec:conc} 

In Bloch theory, the band structure is derived for Bloch states indexed by a quasi-momentum $\vk$, which can be thought of as the analogues of plane wave states for a system with a periodic potential. However, a plane wave state is strictly defined only for an infinite periodic system (or equivalently, a system with PBC), which is manifestly violated in the presence of a boundary. Nevertheless the presence of such a boundary is often crucial to make manifest the topological non-triviality of the bulk band structure. In this sense, the transfer matrix approach, being partly a real-space method, is ideally suited to expose this physics. Nonetheless, the usual form of the transfer matrix approach only works in the cases were the hopping operator ($J$ in our notation) is invertible, which has often limited its applicability. 

In this work, we have presented a general construction of the transfer matrix for tight-binding models with physical edges. Crucially, this generalization of the transfer matrix works even in the cases where the hopping matrix element is not invertible. Previously such systems could only be tackled using either numerical diagonalization and/or Green's functions methods. However, the linearity of the transfer matrix equation and the ability to work in the infinite bulk limit gives it many advantages over the other methods. For instance, localization studies in disordered systems using the transfer matrix method will benefit from our generalized formalism by allowing the study of a wider class of tight-binding lattice models and not just their representative Chalker-Coddington network models\cite{evers2008anderson}.

We have also applied our formalism to several important tight-binding models, many of which exhibit an integer quantum Hall effect and topological chiral edge states. The seminal works of Hatsugai\cite{hatsugai_cbs,hatsugai_cbs_PRL} were re-examined using our methods and formalism. When applied to the simple 2D Chern insulator and other $r=1$ systems, our transfer matrix approach simplifies considerably and yields a close relationship with symplectic geometry and has direct analogues with dynamical systems through the Evans function. We have presented -- as Hatsugai has done for the Hofstader model -- the energy-Riemann surface of the Chern insulator and shown it to be the simplest topological integer quantum Hall system. 

Often, the application of transfer matrices is envisioned to be applicable only to square lattices, where the hoppings in the direction normal and parallel to the edge are independent. However, we have also applied our construction to systems on nonsquare lattice, for instance, the honeycomb(graphene) and kagome lattices, to obtain bands and edge states that agree with exact diagonalization. In essence, the singular value decomposition in our construction identifies the relevant modes that hop across the neighboring blocks, thereby mechanizing the process that would otherwise takes significant amount of care to keep track of. 

An interesting connection that we discovered on the side is the mapping of the Chern insulator to the SSH model, an insight which makes the origin of the edge states very clear. Despite not being completely general, this connection hints at the possibility of decomposing other relatively complicated tight binding Hamiltonians as 1-dimensional chains by a suitable basis transformation of the Hamiltonian, which can be gleaned off from the tridiagonalization of $J$ and $M$ in our notation. 

We remark that all the interesting simplifications that let us compute things analytically for the rank 1 case followed simply from the fact that the transfer matrix was symplectic. There is a wealth of interesting results for symplectic matrices that can be used to further study these cases\cite{baldes_fredholm}. We did describe the mathematical conditions for the transfer matrix of systems of higher rank to be symplectic, however, the physical interpretation and implications of those conditions require further study. 

One obvious next step would be a \emph{rigorous} proof of the bulk-boundary correspondence between bulk band Chern number and the count of the chiral edge modes within our generalized transfer matrices with little or no assumptions on the underlying lattice. As the transfer matrix encodes the data about the bulk bands and their wavefunctions, as well as the edge states and their windings, we expect it to be of importance in studying the bulk-boundary correspondence. Optimistically, one can hope for a proof based on elementary algebraic methods. 

It would be interesting to study the effect of discrete symmetries of the Hamiltonian, i.e, the symmetries associated with our $J$ and $M$ matrices, on the corresponding transfer matrix. For instance, transfer matrices can be used to compute the $\intg_2$ invariant and show the bulk-boundary correspondence for systems that preserve time-reversal symmetry\cite{avila_edge2d}. The interplay between the Altland-Zirnbauer classes for tight-binding Hamiltonian and our transfer matrices may shed new light on the classification scheme.

Furthermore, the interpretation of the singular values of the hopping matrix, $J$, as the strength of various \emph{channels} that the system can be decomposed in provides a potential scheme of controlled approximation for the transfer matrix and hence the band structure of the system. This can be particularly useful for a disordered system, where the transfer matrices between consecutive layers would be quite big, but the SVD can extract the \emph{important} degrees of freedom to compute the approximate band structure. 

Over the past century, band theory has continued to prove itself to be a treasure of interesting results, and we hope that this approach will shed further light and offer insights in the ongoing \emph{excavations} in that area, as well as help uncover the associated geometrical and topological structures. We also hope that this work would help bridge at least some of the gap between similar works in condensed matter physics and the relevant mathematics literature.

\begin{acknowledgments}
VD and VC acknowledge the super useful conversations with and moral support from M Stone, J Bronski, S Ramamurthy and A Narayan. VC was supported by the Gordon and Betty Moore Foundation’s EPiQS Initiative through Grant GBMF4305 and VD was supported by the National Science Foundation through Grant NSF DMR 13-06011, and they are both eternally grateful. 
\end{acknowledgments}


\appendix
\section{Mathematical preliminaries}     \label{app:math}

\subsection{Block matrix manipulations}    \label{app:math_matrix}
In this section, we describe a few well known results to do with operations on partitioned matrices with square blocks\cite{bernstein_book}. Consider a square matrix of dimensions $2n \times 2n$, consisting of blocks of dimensions $n \times n$:
\beq M = \left( \begin{array}{cc} A & B \\ C & D \end{array}  \right). \eeq 
We seek formulae relating the properties of $M$ to those of $A, B, C, D$. The starting point is a decomposition of $M$ in terms of triangular matrices,
\beq M = \left( \begin{array}{cc} A & 0 \\ C & \id \end{array}  \right) \left( \begin{array}{cc} \id & \iA B \\ 0 & D - C \iA B \end{array}  \right), \eeq
or, alternatively, 
\beq M =  \left( \begin{array}{cc} \id & B \\ 0 & D \end{array}  \right) \left( \begin{array}{cc} A - B D^{-1} C & 0 \\ D^{-1} C & \id \end{array}  \right), \eeq
which can be verified by a direct computation. 

Using this, we can compute the determinant of $M$ as 
\begin{align}
 \det M = & \; \det (A) \det (D - C \iA B) \nonumber \\ 
 = & \; \det (D) \det (A - B D^{-1} C) \label{eq:det_block}
\end{align}
The quantities of the form $A - B D^{-1} C$ that appear in these expressions are known as Schur complements, usually denoted by 
\beq M/D = A - B D^{-1} C, \eeq
where the order of the matrices in the second term is clockwise in $M$. 

Now the inverse. For a lower triangular matrix with nonsingular $A$ and $D$, the inverse can be computed as 
\beq \left( \begin{array}{cc} A & 0 \\ C & D \end{array}  \right)^{-1} = \left( \begin{array}{cc} \iA & 0 \\ -D^{-1} C \iA  & D^{-1} \end{array}  \right). \eeq
Similarly, for an upper triangular matrix, 
\beq \left( \begin{array}{cc} A & B \\ 0 & D \end{array}  \right)^{-1} = \left( \begin{array}{cc} \iA & -\iA B D^{-1}  \\ 0 & D^{-1} \end{array}  \right). \eeq
An expression for inverse of $M$ is 
\begin{align}
 & M^{-1} =   \left( \begin{array}{cc} \id & \iA B \\ 0 & M/A \end{array}  \right)^{-1} \left( \begin{array}{cc} A & 0 \\ C & \id \end{array}  \right)^{-1} \nonumber \\ 
 & = \left( \begin{array}{cc}  \iA + \iA B (M/A)^{-1} C \iA   & - \iA B (M/A)^{-1} \\  (M/A)^{-1} C \iA  & (M/A)^{-1} \end{array}  \right). \label{eq:blk_inv}
\end{align}
This expression illustrates the principle of decomposing a block matrix into a product of upper-triangular and lower-triangular matrices and computing the inverses individually, using the expressions above. 

\subsection{Symplectic groups and winding}   \label{app:math_symp}
We seek to parametrize $\Sp(2, \real)$, and show that it is homeomorphic to a solid 2-torus\cite{lang_sl2R, conrad_sl2R}. This can be shown using an Iwasawa decomposition\cite{iwasawa}. Explicitly, let us consider a matrix $S \in \Sp(2, \real)$, parametrized as 
\beq S = \left( \begin{array}{cc} a+b & c-d \\ c+d & a-b \end{array} \right), \eeq 
with $(a, b, c, d) \in \real^4$. The determinant condition, $\det S = 1$, demands that 
\beq (a^2 + d^2) - (b^2 + c^2) = 1. \eeq 
Hence, $\Sp(2, \real)$ corresponds to a submanifold of $\real^4$ of codimension 1, which can be thought of as a 4-dimensional analogue of a hyperbola. We reparametrize
\begin{align*}
 a = & \; \cosh \eta \cos \theta_1 \\ 
 b = & \; \sinh \eta \cos \theta_2 \\ 
 c = & \; \sinh \eta \sin \theta_2 \\
 d = & \; \cosh \eta \sin \theta_1 
\end{align*}
where $\eta \in \real$ and $\theta_i \in [0, 2\pi)$. This makes $\Sp(2, \real)$ homeomorphic to $\real \times S^1 \times S^1 \cong \real \times T^2$. Finally, define 
\beq \chi = \frac{1}{2}\left( 1 + \tanh \eta \right) \in (0, 1), \eeq 
so that $\Sp(2, \real) \cong D \times S^1$. Finally, it is straightforward to embed the torus formed by $(\chi, \theta_1, \theta_2)$ in $\real^3$. 

This parametrization also provides a particularly simple proof of the fact that $\pi_1(\Sp(2n, \real)) \cong \intg$ for the $n=1$ case.  Generally, the proof involves the fact\cite{littlejohn_wavepackets, macduff-salamon_symp_top} that $U(n) \subset \Sp(2n, \real)$ is its maximally compact subgroup, so that $\Sp(2n, \real)$ has $U(n)$ as its strong deformation retract. Furthermore, $\pi_1(U(n)) \cong \intg$, which can be seen by the determinant map for $\mathcal{U} \in U(n)$ as $\mathcal{U} \mapsto \det \mathcal{U} \in S^1$, and $\pi_1(S^1) \cong \intg$.   

For $\Sp(2, \real)$, consider the deformation retract
\beq S_t = S(\eta t, \theta_1, \theta_2) : [0, 1] \to \Sp(2n, \real). \eeq
For $t = 1$, we recover $S$, while for $\theta = 0$, we get 
\[ a_0 = \cos \theta_1, \quad d_0 = \sin \theta_1, \quad b_0 = c_0 = 0 \]
so that $S_0$ is parametrized simply by $\theta_1 \in S^1$, which implies that $S^1$ is a deformation retract of $\Sp(2n, \real)$, which proves our result.

\subsection{Jordan canonical form}      \label{app:math_op}
The Jordan canonical form of a matrix $T \in \cmplx^{2r \times 2r}$ can be expressed as\cite{kato_book} 
\beq T = \sum_{s} [\rho_s \proj_s + \nilp_s], \eeq 
where the sum extends over all generalized eigenvalues $\rho_s$, $\proj_s$ are idempotent projectors that project in the eigenspace of $\rho_s$ and $\nilp_s$ are nilpotent operators of order equal to the algebraic multiplicity of the eigenvalue. These operators satisfy
\beq \proj_{s'} \nilp_s = \nilp_s \proj_{s'} = \nilp_s \delta_{ss'}.  \eeq
The generalized eigenvalue equations are then
\begin{align}
(T-\rho_s\mathbb{I})^k \varphi & = 0, \\
\phi^\dagger(T-\rho_s\mathbb{I})^k  & = 0,
\end{align}
for left($\varphi$) and right ($\phi$) eigenvectors, and where $k=\text{rank}(\proj_s)$ is the multiplicity of the eigenvalue $\rho_s$.  	
We define the projector to the decaying subspace as 
\beq \proj_< \equiv \sum_{|\rho_s|<1} \proj_s \eeq 
and construct 
\beq T_< \equiv T \proj_< = \proj_< T = \sum_{|\rho_s|<1} [\rho_s \proj_s + \nilp_s] \eeq 
which, by construction, has generalized eigenvalues satisfying $|\rho_s|<1$, i.e,
\beq \underset{n}{\lim \sup} \| T_< ^n \|^{1/n} = \rho(T_<) < 1 \eeq
by the spectral radius formula \cite{rudin_book}, where $\|.\|$ is the operator norm over $\cmplx^{2r \times 2r}$. Now, given a $\Phi \in \cmplx^{2r}$, a sufficient condition for $|T^n \Phi|$ to decay as $n \to \infty$ is $\proj_< \Phi = \Phi$, as 
\begin{align}
 | T^n \Phi| = & \; | T^n \proj_< \Phi | = | (T_<)^n \Phi | \nonumber \\ 
 \leq  & \; \| T_<^n \| \cdot |\Phi | \rightarrow 0
\end{align} 
as $n\rightarrow \infty$, which proves our assertion.

\section{Transfer matrix and block size}    \label{app:sqr}
The recursion relation is given by 
\beq J \Psi_{n+1} + J^\dagger \Psi_{n-1}  = (\ve \id - M) \Psi_n. \eeq 
Taking $m$ copies of this equation for $n = nm, nm-1, \dots nm-m+1$ and defining 
\beq \wt{\Psi}_n = \left( \Psi_{mn}, \Psi_{mn-1}, \dots \Psi_{mn-m+1} \right)^T,\eeq 
we have a recursion relation for $\wt{\Psi}_n$ as 
\beq \wt{J} \wt{\Psi}_{n+1} + \wt{J}^\dagger \wt{\Psi}_{n-1} = \left( \ve \id - \wt{M} \right) \wt{\Psi}_n, \eeq
where $\wt{J}$ and $\wt{M}$ can be written in terms of $J$ and $M$ as 
\beq \wt{J} = \left( \begin{array}{cccc} 
0 & 0 & \ldots & J \\ 
0 & 0 & \ldots & 0 \\
\vdots & \vdots & \ddots & \vdots \\ 
0 & 0 & \ldots & 0 
\end{array} \right), \quad 
\wt{M} = \left( \begin{array}{cccc} 
M & J^\dagger & \ldots & 0 \\ 
J & M & \ldots & 0 \\
\vdots & \vdots & \ddots & \vdots \\ 
0 & 0 & \ldots & M
\end{array} \right).  \eeq 
We have $\rank(\wt{J}) = \rank(J) = r$, and the reduced SVD of $\wt{J}$ is
\beq \wt{J} = \wt{V} \cdot \Xi \cdot \wt{W}^\dagger, \eeq
where
\beq \wt{V} = \left( \begin{array}{c} V \\ \vdots \\ 0 \end{array} \right)_{\nuc m \times r}, \quad 
     \wt{W} = \left( \begin{array}{c} 0 \\ \vdots \\  W \end{array} \right)_{\nuc m \times r},  \eeq
and the singular values, $\Xi$, are same as those of $J$. 

Now, following the calculation in \S \ref{sec:tmat_general}, we compute the recursion relations for $\wt{\bal}_n$ and $\wt{\bbe}_n$, the coefficients of $\wt{\Psi}_n$ along $\wt{V}$ and $\wt{W}$, and construct a $2r \times 2r$ transfer matrix $\wt{T}$, so that
\beq \wt{\Phi}_{n+1} = \wt{T} \wt{\Phi}_n, \quad \wt{\Phi}_n \equiv \left( \begin{array}{c} \wt{\bbe}_{n} \\ \wt{\bal}_{n-1} \end{array} \right). \eeq
But using the definition of $\wt{\Psi}_n$, we get 
\begin{align}
 \wt{\bal}_n = & \; \wt{V}^\dagger \wt{\Psi}_n = V^\dagger \Psi_{nm} = \bal_{nm}, \nonumber \\
 \wt{\bbe}_n = & \; \wt{W}^\dagger \wt{\Psi}_n = W^\dagger \Psi_{nm-m+1} = \bbe_{(n-1)m+1} , 
\end{align}
so that 
\beq \wt{\Phi}_n  = \left( \begin{array}{c} \bbe_{(n-1)m+1} \\ \bal_{(n-1)m} \end{array} \right) = \Phi_{(n-1)m+1}. \eeq 
Using the old transfer matrix, $T$, we also have 
\beq \wt{\Phi}_{n+1} = \Phi_{nm+1} = T^m \Phi_{(n-1)m+1} = T^m\wt{\Phi}_{n}, \eeq 
so that the action of $\wt{T}$ is identical to the action of $T^m$ on any arbitrary wavefunction. We conclude that $\wt{T} = T^m$.

\section{Transfer matrices and Floquet discriminants}   \label{app:tmat}
Given a $N \times N$ real square matrix $A$, finding its eigenvalues is equivalent to finding the roots of its characteristic polynomial, a polynomial of degree $N$ with real coefficients. Generally, the solution cannot be obtained in a closed form if $N \geq 4$. However, if the matrix is symplectic, we can often find all of its eigenvalues analytically. We discuss the procedure in the following. 

Recall that a matrix $A$ is symplectic if $A \in \real^{2r \times 2r}$ and
\beq A^T \symp A = \symp, \quad \symp = \left( \begin{array}{cc} 0 & \id \\ -\id & 0 \end{array}  \right). \eeq 
An immediate consequence\cite{macduff-salamon_symp_top} of this result is $\det A = 1$, from which we deduce that all eigenvalues of $A$ are nonzero.  
 
Let the characteristic polynomial of $A$ be 
\beq P(\rho) = \det (\rho \id - A ) = \sum_{n=0}^{2r} a_n \rho^n. \eeq 
We are interested in the eigenvalues of $A$, i.e, the zeros of $P(\rho)$. We begin by noting that if $\rho$ is an eigenvalue of $A$, so is $\rho^{-1}$. To see that, take $A \varphi = \rho \varphi$. Then
\[ \symp \varphi = A^T \symp A \varphi = A^T \symp \rho \varphi \implies A^T \left( \symp \varphi \right) = \rho^{-1}  \left( \symp \varphi \right), \]
and as $A$ and $A^T$ have the same spectrum, we conclude that $\rho^{-1}$ is an eigenvalue of $A$. Hence, $P(\rho)=0$ implies that
\beq 0 = P(\rho^{-1}) = \sum_{n=0}^{2r} a_n \rho^{-n} = \rho^{-2r} \left( \sum_{n=1}^{2r} a_{2r-n} \rho^n \right). \eeq 
As $\rho \neq 0$, we conclude that $a_n = a_{2r-n}$, i.e, the characteristic polynomial is \emph{palindromic}\cite{littlejohn_wavepackets,jared_mod_ins}. Thus, we can rewrite the eigenvalue condition as 
\beq 0 = P(\rho) = \rho^r \left( a_r + \sum_{n=1}^{r} a_{r-n} \left( \rho^n + \rho^{-n} \right) \right). \eeq 
Defining $\Delta = \rho + \rho^{-1}$, we can express $\rho^n + \rho^{-n}$ as Chebyshev polynomials of the first kind in $\Delta$. To see this, define $\rho = e^{i\theta}$, so that
\beq  T_n (\cos\theta) = \cos(n\theta) = \frac{1}{2} \left( \rho^n + \rho^{-n} \right). \eeq
But as $\Delta = 2 \cos\theta$, we get 
\beq \rho^n + \rho^{-n} = 2 \; T_n \left( \frac{\Delta}{2} \right). \eeq 
Explicitly,
\begin{align*}
 \rho^2 + \rho^{-2} = & \; \Delta^2 - 2, \\ 
 \rho^3 + \rho^{-3} = & \; \Delta^3 - 3\Delta, \; \text{etc.} 
\end{align*} 
The eigenvalue problem then becomes the problem of finding the zeros of a polynomial of order $r$ in $\Delta$, which can be written explicitly as 
\beq a_r + 2 \sum_{n=1}^{r} a_{r-n} T_n \left( \frac{\Delta}{2} \right) = 0 \eeq 
Once we solve for the $r$ roots $\Delta_1, \dots \Delta_r \in \cmplx$, we can solve for $\rho$ as 
\beq \rho + \rho^{-1} = \Delta_n \implies \rho = \frac{1}{2} \left[ \Delta_n \pm \sqrt{\Delta_n^2 - 4} \right],  \eeq 
where $n = 1, 2 \dots r$.

In the following, we work out the case of $r=2$ explicitly. The eigenvalue condition becomes 
\[ a_0 (\rho^2 + \rho^{-2}) + a_1 (\rho + \rho^{-1}) + a_2 = 0, \]
with 
\[ a_0 = 1, \quad a_1 = - \tr A, \quad a_2 = \frac{1}{2} \left( (\tr A)^2 - \tr A^2 \right). \]
In terms of $\Delta$, we get 
\[ a_0 (\Delta^2 - 2) + a_1 \Delta  + a_2 = 0 \]
which implies that
\[ \Delta_\pm = \frac{1}{2a_0} \left[ -a_1 \pm \sqrt{a_1^2 - 4 a_0(a_2 - 2a_0)} \right]. \] 
Substituting $a_n$'s, we get the Floquet discriminants as
\beq \Delta_{\pm} = \frac{1}{2} \left[ \tr A \pm \sqrt{2 \tr A^2 - (\tr A)^2 + 8} \right].  \eeq

\section{Transfer matrix for Chern insulator}     \label{app:chern}
The Chern insulator is a 2-dimensional lattice model described by the lattice Hamiltonian\cite{bernevig-hughes_book}
\beq \hlt = a \sin k_x \sigma^x + a \sin k_y \sigma^y + b (2-m-\cos k_x - \cos k_y) \sigma^z \eeq
The system is gapped in the bulk, except for $m = 0, 2, 4$, when the gap closes. It is topological for $0<m<2$ with edge states around $k = 0$ and and for $2<m<4$ with edge states around $k = \pi$. 

Let us put the Chern insulator on a cylinder which is periodic along $y$ and finite along $x$. Then, we need to inverse Fourier transform along $x$ (as $k_x$ is not well-defined for a finite system) and write the Hamiltonian as
\begin{align}
\hlt(k_y) = & \sum_{n = 0}^N  \bigg[ \frac{a}{2i} \left( \cd_{n+1} \sigma^x c_n - \cd_n \sigma^x c_{n+1} \right)  \nonumber \\ 
& - \frac{b}{2} \left( \cd_{n+1} \sigma^z c_n + \cd_n \sigma^z c_{n+1} \right)  \nonumber \\
&  \frac{}{} +  \cd_n \left(  \sin k_y \sigma^y + b \Lambda(k_y) \sigma^z \right) c_n \bigg]
\end{align}
where $c_n(k_y)$ is a row vector, corresponding to the annihilation operator for the two degrees of freedom on each lattice site and $\Lambda(k_y) = 2 - m - \cos k_y$. Here, $k_y \; (\equiv \vk_\perp)$ just acts as a parameter in the Hamiltonian. We are only concerned with a topological state for $n\geq 0$.

The corresponding recursion relation is
\begin{align}
  \frac{1}{2i} \left( a \sigma^x  - i b \sigma^z \right) \psi_{n+1} -  \frac{1}{2i} \left( a \sigma^x  + i b \sigma^z \right) \psi_{n-1} \nonumber \\ 
 =  \;  \left( \ve \id - a \sin k_y \sigma^y - b \Lambda(k_y)  \sigma^z \right)\psi_n 
\end{align}
We identify the hopping matrix
\beq J = \frac{1}{2i} \left( a \sigma^x  - i b \sigma^z \right) \eeq
which has eigenvalues
\beq \sigma(J) = \pm \frac{1}{2} \sqrt{b^2 - a^2}. \eeq
Hence, $J$ becomes singular when $a = b$, which is precisely the case that we are interested in. For the subsequent calculations, we set $a = b = 1$. Hence,
\beq J =  \frac{1}{2i} \left(  \sigma^x  - i \sigma^z \right) = -\frac{1}{2} \left( \begin{array}{cc} 1 & i \\ i & -1 \end{array} \right) \eeq
and ker$(J)$ is spanned by $\vv = (1, i)^T$, while ker$(J^\dagger)$ is spanned by  $\vw = (1, -i)^T$. The crucial fact, that helps us compute the transfer matrix, is that $\vv$ and $\vw$ are orthogonal, i.e, $\langle \vv, \vw \rangle = 0$.

To see that explicitly, we write out $\psi_n = (\psi_n^1, \psi_n^2)^T$, and the recursion relation as
\begin{align}
\left( \begin{array}{cc} 1 & i \\ i & -1 \end{array} \right) \left( \begin{array}{c} \psi_{n+1}^1 \\ \psi_{n+1}^2 \end{array} \right) - \left( \begin{array}{cc} -1 & i \\ i & 1 \end{array} \right) \left( \begin{array}{c} \psi_{n-1}^1 \\ \psi_{n-1}^2 \end{array} \right) \nonumber \\ 
= - 2 \left( \begin{array}{cc} \ve - \Lambda(k_y)  & i \sin k_y \\ -i \sin k_y & \ve + \Lambda(k_y)  \end{array} \right) \left( \begin{array}{c} \psi_{n}^1 \\ \psi_{n}^2 \end{array} \right)  \label{eq:CI_recur}
\end{align}

We now premultiply the above expression by $(1,i)$ and $(1, -i)$ to get two recursion relations, one excluding $\psi_{n+1}$ and one excluding $\psi_{n-1}$. We can simplify these expressions greatly by defining
\beq \phi_n = \psi_n^2 + i \psi_n^2, \quad \bar\phi_n =   \psi_n^2 - i \psi_n^2. \eeq
Notice that these are not complex conjugates, as $\psi_n^i$'s are in general complex. In terms $\phi$'s, we get
\begin{align}
(\ve + \sin k_y) \phi_n - \Lambda(k_y)  \bar\phi_n + \bar\phi_{n-1} = & \; 0 \nonumber \\
\phi_{n+1} - \Lambda(k_y)  \phi_n + (\ve - \sin k_y) \bar\phi_n = & \; 0
\end{align}
Replacing $n \to n+1$ in the former and reorganizing the terms, we get
\beq \left( \begin{array}{c} \bar\phi_{n+1} \\ \phi_{n+1} \end{array} \right) =  \left( \begin{array}{cc}  \frac{1 - \ve^2 + \sin^2 k_y}{\Lambda(k_y) } & \ve + \sin k_y  \\  - (\ve - \sin k_y) & \Lambda(k_y)   \end{array} \right) \left( \begin{array}{c} \bar\phi_{n} \\ \phi_{n} \end{array} \right). \label{eq:CI_tmat}  \eeq
Hence, we have managed to compute the transfer matrix, acting as
\beq  \Phi_{n+1} = T \Phi_n, \quad \Phi_n =  \left( \begin{array}{c} \bar\phi_{n} \\ \phi_{n} \end{array} \right) \eeq
We can explicitly check that $\det(T) = 1$. The other useful quantity is the trace,
\beq \Delta(\ve, k_y) = \frac{1 - \ve^2 + \Lambda^2(k_y) + \sin^2 k_y}{\Lambda(k_y)}. \eeq
This is equal to the trace obtained by using the formal construction in \eq{eq:CI_trace}. Finally, we can compute the band edges and edge states, as described in \S \ref{sec:tmat_appl}.

\bibliography{transfer_matrix}

\end{document}

%% file: tm_defcommands.tex
\newcommand{\beq}{\begin{equation}}
\newcommand{\eeq}{\end{equation}}

\newcommand{\eq}[1]{eq. (\ref{#1})}

\newcommand{\tr}{\text{tr}}
\newcommand{\Tr}[1]{\text{Tr}\left( #1 \right)}

\newcommand{\rank}{\text{rank}}

\newcommand{\vecenv}[2]{\left( \begin{array}{c} #1 \\ #2 \end{array} \right)}

\newcommand{\cmplx}{\mathbb{C}}
\newcommand{\intg}{\mathbb{Z}}
\newcommand{\real}{\mathbb{R}}

\newcommand{\uvec}{\mathbf{e}}

\newcommand{\id}{\mathbbm{1}}
\newcommand{\symp}{\mathcal{J}}
\newcommand{\proj}{\mathcal{P}}
\newcommand{\nilp}{\mathcal{D}}
\newcommand{\uop}{\mathcal{U}}

\newcommand{\curve}{\mathcal{C}}

\newcommand{\U}{\mathrm{U}}

\newcommand{\Sp}{\mathrm{Sp}}
\newcommand{\GL}{\mathrm{GL}}

\newcommand{\hlt}{\mathcal{H}}

\newcommand{\green}{\mathcal{G}}
\newcommand{\gralt}{\mathscr{G}}

\newcommand{\ket}[1]{| #1 \rangle}

\newcommand{\bd}{b^\dagger}
\newcommand{\cd}{c^\dagger}
\newcommand{\ve}{\varepsilon}

\newcommand{\vc}{\mathbf{c}}
\newcommand{\vcd}{\mathbf{c}^\dagger}
\newcommand{\vn}{\mathbf{n}}
\newcommand{\vk}{\mathbf{k}}

\newcommand{\vv}{\mathbf{v}}
\newcommand{\vw}{\mathbf{w}}
\newcommand{\vx}{\mathbf{x}}

\newcommand{\bal}{{\boldsymbol{\alpha}}}
\newcommand{\bbe}{{\boldsymbol{\beta}}}
\newcommand{\bga}{{\boldsymbol{\gamma}}}
\newcommand{\bze}{{\boldsymbol{\zeta}}}

\newcommand{\ndof}{q}
\newcommand{\nuc}{\mathscr{N}}
\newcommand{\iA}{A^{-1}}

\newcommand{\proja}{\mathcal{Q}_\bal}
\newcommand{\projb}{\mathcal{Q}_\bbe}

\newcommand{\tmat}{\mathcal{T}}
\newcommand{\torus}{\mathbb{T}}
\newcommand{\gap}{\mathscr{G}}
\newcommand{\pgap}{\mathscr{P}}
\newcommand{\band}{\mathscr{B}}
\newcommand{\basis}{\boldsymbol{\xi}}